\documentclass[12pt, reqno]{amsart}
\usepackage{amsmath,amssymb,pdflscape,bm,xr}
\usepackage{amsfonts,natbib}
\usepackage{tikz}
\usepackage{tikz-network}
\usepackage{graphicx,epsfig,setspace}
\usepackage[margin=1in]{geometry}

\numberwithin{equation}{section}
\DeclareMathOperator*{\cart}{\times}

\begin{document}

\newcommand{\cov}{\textnormal{Cov}}
\newcommand{\var}{\textnormal{Var}}
\newcommand{\diag}{\textnormal{diag}}
\newcommand{\plim}{\textnormal{plim}_n}
\newcommand{\dum}{1\hspace{-2.5pt}\textnormal{l}}
\newcommand{\ind}{\bot\hspace{-6pt}\bot}
\newcommand{\co}{\textnormal{co}}
\newcommand{\tr}{\textnormal{tr }}
\newcommand{\fsgn}{\textnormal{\footnotesize sgn}}
\newcommand{\sgn}{\textnormal{sgn}}
\newcommand{\fatb}{\mathbf{b}}
\newcommand{\fatp}{\mathbf{p}}
\newcommand{\trace}{\textnormal{trace}}
\newcommand{\Eta}{\textnormal{H}}

\newtheorem{dfn}{Definition}[section]
\newtheorem{rem}{Remark}[section]
\newtheorem{cor}{Corollary}[section]
\newtheorem{thm}{Theorem}[section]
\newtheorem{lem}{Lemma}[section]
\newtheorem{notn}{Notation}[section]
\newtheorem{con}{Condition}[section]
\newtheorem{prp}{Proposition}[section]
\newtheorem{pty}{Property}[section]
\newtheorem{ass}{Assumption}[section]
\newtheorem{ex}{Example}[section]
\newtheorem*{cst1}{Constraint S}
\newtheorem*{cst2}{Constraint U}
\newtheorem{qn}{Question}[section]


\onehalfspacing

\title[Network CLT]{Central Limit Theory for Models of Strategic Network Formation}
\author[Konrad Menzel]{Konrad Menzel\\New York University}
\date{March 2021 The author gratefully acknowledges support from the NSF (SES-1459686).}

\begin{abstract}
We provide asymptotic approximations to the distribution of statistics that are obtained from network data for limiting sequences that let the number of nodes (agents) in the network grow large. Network formation is permitted to be strategic in that agents' incentives for link formation may depend on the ego and alter's positions in that endogenous network. Our framework does not limit the strength of these interaction effects, but assumes that the network is sparse. We show that the model can be approximated by a sampling experiment in which subnetworks are generated independently from a common equilibrium distribution, and any dependence across subnetworks is captured by state variables at the level of the entire network. Under many-player asymptotics, the leading term of the approximation error to the limiting model established in \cite{Men16} is shown to be Gaussian, with an asymptotic bias and variance that can be estimated consistently from a single network.\\[4pt]

\noindent\textbf{JEL Classification:} C1, C12, C23, C33\\
\textbf{Keywords:} Network Data, Large-Network Asymptotics, Multiple Equlibria, Exchangeability
\end{abstract}

\maketitle

\section{Introduction}


We develop an asymptotic theory for structural models of network formation where agents (nodes) may be pursuing strategic motives in deciding whether or not to form a link (edge). In economic contexts, the language of networks has been extremely useful to describe collections of transactions, contracts, or other relationships, where the various agreements among pairs or small groups of economic agents may be interrelated.

For example if a network transmits information, then the benefit from a link to another node varies with how centrally that node is located in the network, and the sign of that effect generally depends on whether information is rival or not (\cite{CAr04} and \cite{CAJ04}). Third parties may be used to screen, monitor, or otherwise secure risky transactions, providing an incentive to form more tightly connected clustered or cliques of agents (\cite{JRT12}, \cite{AMS14}, and \cite{GG16}). If bargaining takes place on a network of longer-term relationships, node pairs at critical bottlenecks may be able to extract greater surplus than more marginal nodes (\cite{LFo13} and \cite{Man18}). \cite{CJP09} and \cite{CJP10} analyze a friendship formation model with homophilous preferences and endogenous search effort that predicts an effect of relative group size on friendship networks (``relative homophily") in addition to assortatitivity on the homomphilous attributes.

Stylized models for these problems predict that networks formed subject to these strategic motives depart in systematic ways from  an idealized benchmark (e.g. Erd\H{o}s-R\'enyi random graphs). However more comprehensive empirical models that would allow to separate those patterns from other confounding effects in order to validate these predictions empirically remain extremely difficult to solve and estimate. With strategic interdependencies in link payoffs, links between agent pairs can no longer be modeled independently, and standard solution concepts may fail to predict a unique outcome. While these challenges closely resemble known problems for simultaneous discrete choice and game-theoretic models (see e.g. \cite{Hec78}, \cite{BVu84}, \cite{BRe91}, and \cite{Tam03}), they manifest themselves at a much greater level of complexity in the context of networks. Real-world examples of networks typically involve a fairly large number of agents and the relevant ``action space" (corresponding to the sets of other nodes that an agent can link to) also grows in dimension with the number of nodes in the network.

\subsection*{Contribution}

This paper develops tractable asymptotic approximations to the distribution of network moments, where we take limits of observable quantities along sequences of networks consisting of a finite, but increasing number of nodes drawn at random from a common population. As an approximation device, this thought experiment yields probability limits for observable features of the network that can be characterized directly in terms of payoff parameters, as well as a central limit theorem to describe the asymptotic behavior of the approximation error. These approximations can be used for estimation and inference regarding payoff parameters via likelihood- or moment-based methods.


\cite{Men16} already derived the first-order limit of conditional link formation frequencies for the setup considered in this paper and showed how these can be used to obtain a pseudo likelihood or other moment conditions in terms of structural payoff parameters. For a range of scenarios, expressions for these limits are available in closed form given a vector of aggregate state variables that solve a population fixed-point condition. For estimation \cite{Men16} proposed a pseudo maximum likelihood estimator that is conditional on the observable aggregate states, and treats the unobserved aggregate states as nuisance parameters which satisfy the aggregate constraints implied by the theory. The structure of the limiting problem is therefore very similar to computational problems that are commonly encountered in empirical industrial organization, including stationary discrete dynamic programming problems with a value function defined by a Bellman iteration, for which reliable and efficient computational solutions already exist (see e.g. \cite{Rus87}, \cite{HMi93}, \cite{AMi07}, \cite{SJu12}). This paper derives a (many player) asymptotic theory, including weak convergence and the asymptotic distribution of network moments, that can be used to formalize this particular approach for estimation and inference. In addition, the results in this paper can also be used when the researcher estimates parameters using the limiting approximation to other moment conditions for the network data which may take the form of equality or inequality restrictions.

Most importantly, this paper quantifies the impact of the approximation errors under many-player asymptotics for estimation and inference, where we establish a characterization of asymptotic bias and a central limit theorem for the leading stochastic component of the many-player limiting approximation. Since that leading term of the approximation error results from sampling uncertainty regarding the nodes forming the network, the distributional theory can be used for inferential statements regarding payoff parameters that are based on a finite number of large networks, possibly a single realization of the network. One key qualitative feature of our asymptotic analysis is that we consider sparse network sequences in which each node connects to a number of alters that remains stochastically bounded as the size of the network grows, which we think of as a reasonable qualitative approximation to many real-world economic networks with a large number of agents.

The key idea behind our approach is to exploit exchangeability and invariance properties that are built into a typical empirical model. At the center of this approximation is a representation of the network formation model as a sample of conditionally independent subnetworks. In that sampling representation, the endogenous network neighborhood for each subnetwork is drawn independently from a common distribution of exogenous and (endogenous) network attributes for the nodes that may potentially accept a link to one or several nodes in that subnetwork.

Since the value of network attributes for these neighboring nodes are in part determined by the subnetwork itself, we introduce the notion of potential values for network attributes, accounting for the equilibrium response in the network to conjectural local changes. Furthermore, in addition to the ``local" endogeneity modeled as potential values, the distribution generating potential network neighbors is determined by ``global" equilibrium conditions in the network. We show that, to the first order of approximation, this endogeneity is captured by aggregate state variables, which are determined in equilibrium by an aggregate fixed-point condition. This ``flattening" of the network graph then allows to evaluate probabilities for subgraph configurations among a finite subset of nodes entirely in terms of subnetworks of a fixed size.

\subsection*{Literature}


This paper builds on the approach in \cite{Men16} who derived the first-order limit of the network formation model but provided no asymptotic theory for stochastic convergence of network moments to their asymptotic analogs. The law of large numbers and central limit theory in this paper provide a sampling theory, allowing to apply his results for estimation or inference of structural parameters. The general approach in this paper, exploiting symmetry and identifying aggregate state variables has conceptual parallels with \cite{Men12}'s framework of many-player asymptotics for discrete games, however the present setting is fundamentally different due to the non-anonymous nature of interactions in a network formation problem among identifiable agents. We also build on insights regarding the asymptotic distribution of multi-way dependent arrays in \cite{Men17}, where we find that for the sparse network sequence considered in this paper,  non-Gaussian components of the distribution are dominated in the limit. A similar phenomenon was found by \cite{Gra20} for sparse dyadic arrays.


A different set of asymptotic results for large network formation models is available from \cite{Leu16} and \cite{LMo19} who give conditions on link preferences under which the network separates into bounded, non-overlapping components that do not interact strategically. Our approach differs from theirs in that we do not constrain the strength of strategic interaction effects, allowing for preference cycles of arbitrary length. The other key difference is in the role of node ``locations" in the attribute space. Our framework allows for distance (or other functions of node positions) to have an effect on link preferences that remains at the same order of magnitude as unobserved taste shocks and strategic effects, whereas their framework requires the effect of distance in homophilous attributes to dominate as the network grows. Earlier work by \cite{Leu15} also considered an asymptotic theory for network models based on weak dependence under large-domain asymptotics.

Our approach also differs from a more established literature on random networks in probability and statistics that derives asymptotic properties of large networks based on independence or exchangeability assumptions, starting with the classical \cite{ERe59} random graph (RGM) model and including more recent work by \cite{Lov12}, \cite{BCL11}, \cite{BBi15} and others. In economics, this approach has been generalized substantially by \cite{ChJ16} who propose a model in which not only edges but also larger subgraphs are formed at random. Among other advantages this allows to approximate patterns of clustering that theoretical models of strategic network formation would predict. However for the settings of strategic interactions considered in this paper, a pairwise stable graph among finitely many nodes is usually not jointly exchangeable. Furthermore our main objective is to interpret observable features of the graph in terms of economic primitives, rather than purely descriptive, ``reduced-form" parameters which may not be interpretable in terms of the strategic motives that gave rise to the network or even constitute stable features of the model especially when a pairwise stable network is not unique.

The results in this paper are also complementary to recent advances in understanding identification of network formation models, including \cite{Mel17}, \cite{DRT18}, \cite{She20},\cite{Gra17}, \cite{Dze14}, and \cite{Gra12}. This work clarifies identification and provides moment conditions relating observable quantities to economic primitives, and the results in this paper can be used for inference based on those identification strategies when the data comes from a small number of large networks. \cite{GdP20} also provides a summary of recent developments in the econometric literature on network data.

The remainder of the paper is structured as follows: we first describe our general framework for estimation, then two separate sections introduce the main two concepts for our analysis - ``local" potential values for endogenous network variables, and aggregate states characterizing a ``global" equilibrium - together with key intermediate results. Section 5 then states the main formal results, a law of large numbers and the asymptotic distribution for the class of network moments considered in this paper. Section 6 concludes.

\section{Setup}

We assume that the researcher observes data from a network among $n$ agents (``nodes"/``vertices") where $n$ is large. Network connections are represented by the $n\times n$ adjacency matrix $\mathbf{L}=(L_{ij})_{i,j}$, with the $(i,j)$th entry corresponding to the value of the edge from node $i$ to $j$. For our purposes, links between agents are assumed to be undirected, $L_{ij}=L_{ji}$ and unweighted, $L_{ij}\in\{0,1\}$ for all $i,j$, and we rule out the possibility of self-links, $L_{ii}=0$. Each node is associated with a vector $x_i$ of attributes, with $\mathbf{X}:=[x_1,\dots,x_n]'$ denoting the stacked attribute vectors for the entire network. We also let $\mathbf{L}-\{ij\}$ be the network resulting from deleting the edge $ij$ from $\mathbf{L}$, that is from setting $L_{ij}=L_{ji}=0$. Similarly, $\mathbf{L}+\{ij\}$ denotes the network resulting from adding the edge $ij$ to $\mathbf{L}$.

For pedagogical purposes, we will also primarily discuss the case in which the data consists of the matrices $\mathbf{L},\mathbf{X}$ for a single network. Since our theory concerns the asymptotic behavior of the network formation model, it is not essential to our results which nodes and attributes are observed by the researcher. Our theoretical arguments can also be easily extended to multiple large networks.

\subsection{Network Moments}

Our main results concern $D$-adic moments that are a function of the attributes of the nodes in the $D$-ad interacted with indicators for events regarding the subgraph (edges) among those nodes.

Specifically, let $\mathbf{L}_{|i_1,\dots,i_D}$ denote the subnetwork among the nodes $i_1,\dots,i_D$, as represented by the submatrix of $\mathbf{L}$ corresponding to the rows and columns indexed with $\mathbf{i}:=(i_1,\dots,i_D)'$, and $\mathbf{x}_{\mathbf{i}}\equiv \mathbf{x}_{i_1,\dots,i_D}:=(\mathbf{x}_{i_1,\dots,i_D})'$ the attributes of the nodes forming the $D$-ad.

Let $A_{\mathbf{i}}\subset \{0,1\}^{D(D-1)/2}$ denote an event regarding undirected subnetworks among the $D$ nodes. We then consider moment functions that are of the form
\[m_{\mathbf{i}}(\mathbf{L},\mathbf{X};\theta) = \dum\left\{\mathbf{L}_{|\mathbf{i}}\in A_{\mathbf{i}}\right\}\mathbf{h}(\mathbf{x}_{\mathbf{i}};\theta)\]
for some function (``instrument") $\mathbf{h}:\mathcal{X}^D\times\Theta \rightarrow\mathbb{R}^q$ of node attributes and the parameter $\theta$. Our asymptotic theory concerns $D$-adic moments of the form
\begin{equation}\label{moment_fct}\hat{m}_n(\theta) = \left[\binom{n}{D}p_n\right]^{-1}\sum_{i_1,\dots,i_D} m_{i_1\dots i_D}(\mathbf{L},\mathbf{X};\theta)\end{equation}
These generalized subgraph counts are appropriately normalized averages over all subnetworks consisting of $D$ distinct nodes. The focus of this paper is on sparse network sequences under which link probabilities at the level of a dyad go to zero, so that the normalizing sequence $p_n$ will be chosen to ensure that $p_n^{-1}\mathbb{E}[m_{\mathbf{i}}(\mathbf{L},\mathbf{X};\theta)]$ converges to a nontrivial limit.

For expositional purposes we focus on the case of moments based on a single network event $A_{\mathbf{i}}$, but more generally our arguments easily generalize to stacked moments corresponding to different events $A_{\mathbf{i}}^{(1)},\dots,A_{\mathbf{i}}^{(Q)}$ with respective instruments $\mathbf{h}^{(1)}(\mathbf{x}_{\mathbf{i}};\theta),\dots,\mathbf{h}^{(Q)}(\mathbf{x}_{\mathbf{i}};\theta)$ and normalizing sequences $p_n^{(1)},\dots,p_n^{(Q)}$.

\begin{ass}\label{moment_ass}\textbf{(Moment Functions)} (a) The function $\mathbf{h}:\mathcal{X}^D\times\Theta \rightarrow\mathbb{R}^q$, $(\mathbf{x},\theta)\mapsto\mathbf{h}(\mathbf{x};\theta)$ is bounded and continuous in $\theta$. Given the asymptotic sequence of networks, the normalizing sequence $p_n$ is such that $p_n^{-1}\mathbb{P}_n\left(\mathbf{L}_{|\mathbf{i}}\in A_{\mathbf{i}}\right)$ is bounded as $n$ increases.
\end{ass}

Denoting
\[\pi_n(A_{\mathbf{i}}|\mathbf{x}_{1,\dots,D}):=
\mathbb{P}_n\left(\left.\mathbf{L}_{|\mathbf{i}}\in A_{\mathbf{i}}\right|
\mathbf{x}_{\mathbf{i}}=\mathbf{x}_{1,\dots,D}\right)\]
the conditional expectation of the moment function given $\mathbf{x}_{\mathbf{i}}$ is
\[\mathbb{E}_n\left[\left.m_{\mathbf{i}}(\mathbf{L},\mathbf{X})\right|\mathbf{x}_{\mathbf{i}}=\mathbf{x}_{1,\dots,D}\right]
=\pi_n(A_{\mathbf{i}}|\mathbf{x}_{1,\dots,D})h(\mathbf{x}_{1,\dots,D};\theta)\]
in any finite network, so that under Assumption \ref{moment_ass}, the expectation of $\hat{m}_n(\theta)$ is bounded. Similarly, the conditional variance
\[\var_n\left[\left.m_{\mathbf{i}}(\mathbf{L},\mathbf{X})\right|\mathbf{x}_{\mathbf{i}}=\mathbf{x}_{1,\dots,D}\right]
=\pi_n(A_{\mathbf{i}}|\mathbf{x}_{1,\dots,D})\left(1-\pi_n(A_{\mathbf{i}}|\mathbf{x}_{1,\dots,D})\right)
h(\mathbf{x}_{1,\dots,D};\theta)^2\]
If $A_{\mathbf{i}}$ does not include the empty graph $\mathbf{L}_{|\mathbf{i}}=0$, then for a sparse network $\pi_n(A_{\mathbf{i}}|\mathbf{x}_{1,\dots,D})\rightarrow 0$ for all values of $\mathbf{x}_{1,\dots,D}$ so that the expectation and the variance of the same order of magnitude. We also denote the unconditional probability with
\[p_n:=\mathbb{E}\left[\pi_n(A_{\mathbf{i}}|\mathbf{x}_{1,\dots,D})\right] \]
which we will use as the normalizing sequence in (\ref{moment_fct}).

\subsubsection*{Many-Player Asymptotics}

Following the approach in \cite{Men16}, we derive a limiting model, where we consider limits of the form
\[\pi_0(A_{\mathbf{i}}|\mathbf{x}_{1,\dots,D}):=\lim_{n}p_n^{-1}\pi_n(A_{\mathbf{i}}|\mathbf{x}_{1,\dots,D})\]
as the number of nodes in the network grows to infinity. Then the conditional expectation of the moment function given $\mathbf{x}_{\mathbf{i}}$ under the limiting model is
\[\mathbb{E}_0\left[\left.m_{\mathbf{i}}(\mathbf{L},\mathbf{X};\theta)\right|\mathbf{x}_{\mathbf{i}}=\mathbf{x}_{1,\dots,D}\right]
\equiv \pi_0(A_{\mathbf{i}}|\mathbf{x}_{1,\dots,D})h(\mathbf{x}_{1,\dots,D};\theta)\]
and the limit of the moment is given by
\[m_0(\theta):=\mathbb{E}_0\left[\pi_0(A_{\mathbf{i}}|\mathbf{x}_{1,\dots,D})h(\mathbf{x}_{1,\dots,D};\theta)\right]\]
which is also bounded under Assumption \ref{moment_ass}.


The main purpose of our analysis is to provide asymptotic approximations to the distribution of network moments $\hat{m}_n(\theta)$ that are of the form (\ref{moment_fct}), assuming that the number of agents (nodes) in the network is large. A previous paper by \cite{Men16} derived a limit model that is a distribution (or set of distributions) for the network variables of interest. This limit model implies a set of limiting values (possibly singleton) for the network moment $\hat{m}_n(\theta)$. In practice, this limiting model can be used to obtain bounds or other moment conditions for identification and estimation of payoff parameters.

For data from an economic or social network with only finitely many agents, these moment conditions only hold up to an approximation error which vanishes along the asymptotic sequence. The objective of this paper is to characterize the leading terms of the approximation error from evaluating a moment of the data under the limiting distribution instead of the finite-player distribution. We obtain an approximation of the form
\begin{equation}\label{stochastic_expansion} \hat{m}_n(\theta)=m_0(\theta) + n^{-1/2}Z_n(\theta) + o_P(n^{-1/2})
\end{equation}
where the focus of this paper is on estimating the distribution of the random variable $Z_n(\theta)$ which is asymptotically tight, but not necessarily zero-mean.

The subgraph counts $\hat{m}_n(\theta)$ share some qualitative features with U-statistics of order $D$ whose convergence rate and asymptotic distribution generally depend on its order of degeneracy and whether the network is sparse. We find that under the assumptions of this paper, $\hat{m}_n(\theta)$ converges in probability to an element of a set of possible limit values $m_0(\theta)$ which in some cases can be characterized in closed form. Furthermore, the leading term of the asymptotic approximation error is Gaussian,
\[V_n(\theta)^{-1/2}(Z_n(\theta)-B_n(\theta))\stackrel{d}{\rightarrow}N(\mathbf{0},\mathbf{I})\]
for sequences $(B_n(\theta),V_n(\theta))_n$ converging to the asymptotic bias and the asymptotic variance matrix. Our main formal results below are the law of large numbers in Theorem \ref{primitive_lln_thm}, and the central limit theorem in Theorem \ref{asy_Gauss_thm}.

Since these limits are obtained using a statistical theory that treats the nodes of the endogenously formed network as random draws from a distribution of types, the network moment is subject to sampling uncertainty, which is captured by the stochastic term $Z_n(\theta)$. Theorem \ref{asy_Gauss_thm} will therefore serve as a basis for inference that properly accounts for simultaneity and statistical dependence in network outcomes, and a future version of this paper will give additional practical results concerning estimation of the asymptotic distribution of $Z_n(\theta)$.

\subsection{Structural Model}

\label{sec:struct_model}

Player $i$'s incentives for link formation will be given by a payoff function
\[\Pi_i(\mathbf{L},\mathbf{X}) = B_i(\mathbf{L},\mathbf{X}) - C_i(\mathbf{L},\mathbf{X})\]
which distinguishes a gross benefit $B_i(\mathbf{L},\mathbf{X})$ to $i$ from the network structure, and a total cost $C_i(\mathbf{L},\mathbf{X})$ of maintaining links. The parametrization of the model will be in terms of the \emph{incremental} benefit of adding a link $ij$ to the network $\mathbf{L}$,
\[U_{ij}(\mathbf{L},,\mathbf{X}):=B_i(\mathbf{L}+\{ij\},\mathbf{X}) - B_i(\mathbf{L}-\{ij\},\mathbf{X})\]
and the cost increment of adding that same link,
\[MC_{ij}(\mathbf{L},\mathbf{X}):= C_i(\mathbf{L}+\{ij\},\mathbf{X}) - C_i(\mathbf{L}-\{ij\},\mathbf{X})\]
We refer to $U_{ij}(\mathbf{L},,\mathbf{X})$ as the marginal benefit, and $MC_{ij}(\mathbf{L},,\mathbf{X}))$ as the marginal cost of the link $ij$ to agent $i$.

We assume that marginal benefits are of the form
\begin{equation}\label{mu_model}U_{ij}(\mathbf{L},\mathbf{X}) = U_{ij}^*(\mathbf{L},\mathbf{X}) + \sigma\varepsilon_{ij}\end{equation}
where $U_{ij}^*(\mathbf{L},\mathbf{X})$ is a deterministic function of attributes $x_1,\dots,x_n$ and the adjacency matrix $\mathbf{L}$, and will be referred to as the \emph{systematic part} of the marginal benefit function. The idiosyncratic taste shifters $\varepsilon_{ij}$ are assumed to be independent of $x_i$ and $x_j$ and distributed according to a continuous c.d.f. $G(\cdot)$, and $\sigma>0$ is a scale parameter. In principle, this formulation would allow for arbitrary strategic interference effects among links in the network, however we will restrict dependence of marginal benefits through dependence on $\mathbf{L}$ further below.

The marginal cost of the link $ij$ to $i$ is assumed to be
\begin{equation}\label{mc_model}MC_{ij}(\mathbf{L},\mathbf{X}):=\max_{k=1,\dots,J} \sigma\varepsilon_{i0,k}\end{equation}
where $\varepsilon_{i0,k}$ are independent of $x_i$ and across draws $k=1,2,\dots$, and the choice of the number of draws $J$ will be discussed below. In particular, we let $J$ to grow as $n$ increases in order for the resulting network to be sparse. In this formulation, marginal costs do not depend on the network structure, so that in the following we denote marginal cost of the link $ij$ by $MC_i$ without explicit reference to $j$ or the network $\mathbf{L}$. Note that in the absence of further restrictions on the systematic parts of benefits $U_{ij}^*(\mathbf{L},\mathbf{X})$, this is only a normalization.



Our framework allows for various types of interaction effects on the marginal benefit function. The marginal benefit from adding the link from $i$ to $j$ may depend on agent $i$ and $j$'s exogenous attributes $x_i$ and $x_j$, and the structure of the network through vector-valued statistics $S_i,S_j,T_{ij}$ that summarize the payoff-relevant features of the network $\mathbf{L}$,
\begin{equation}\label{reference_model}U_{ij}^*(\mathbf{L},\mathbf{X}) \equiv U^*(x_i,x_j;S_i,S_j,T_{ij}) \end{equation}
Specifically, the marginal benefit of a link may directly depend on node $i$ and $j$'s exogenous attributes, $x_i$ and $x_j$, respectively, as well as interaction effects between the two. $U_{ij}^*(\mathbf{L},\mathbf{X})$ may vary in $x_i$, e.g. because some node attributes may make $i$ attach more value to any additional links. Dependence on $x_j$ allows for target nodes with certain attributes to be generally more attractive as partners. Finally, a non-zero cross-derivative between components of $x_i$ and $x_j$ could represent economic complementarities, or a preference for being linked to nodes with similar (homophily) or different (heterophily) attributes.

In addition to preferences for exogenous attributes, the propensity of agent $i$ forming an additional link, and the attractiveness of a link to agent $j$, may depend on the absolute position of either node $i$ and $j$ in the network. To account for effects of this type, we can include \emph{node-specific network statistics} of the form
\begin{equation}\label{node_stat_def}S_i:=S(\mathbf{L},\mathbf{X};i)\end{equation}
where we assume that the function $S(\cdot)$ is invariant to permutation of player indices.\footnote{Formally, we assume that for any one-to-one map $\tau:\{1,\dots,n\}\rightarrow\{1,\dots,n\}$ and $i=1,\dots,n$ we have $S(\mathbf{L}^{\tau},\mathbf{X}^{\tau};\tau(i))=S(\mathbf{L},\mathbf{X};i)$, where the matrices $\mathbf{X}^{\tau}$ and $\mathbf{L}^{\tau}$ are obtained from $\mathbf{X}$ and $\mathbf{L}$ by permuting the rows (rows and columns, respectively) of the matrix according to $\tau$.}

\begin{ex}\textbf{(Degree and Composition)}
Node specific network statistics include the network degree (number of direct neighbors),
\[S_1(\mathbf{L},\mathbf{X};i):=\sum_{j\neq i} L_{ij}\]
Alternatively we could measure the share of $i$'s direct neighbors that are of a given exogenous type,
\[S_2(\mathbf{L},\mathbf{X};i):=\frac{\sum_{j\neq i} L_{ij}\dum\{x_{jk}=\bar{x}_k\}}{\sum_{j\neq i} L_{ij}}\]
where the $k$th component of $x_j$ may be e.g. gender or race, and $\bar{x}_k$ the value corresponding to the category in question (e.g. with respect to gender or race).
\end{ex}

The network degree of a node plays a special role in the description of the link frequency distribution. In the remainder of the paper, we therefore take the first component of $s_i$ to denote the network degree of node $i$, $s_{1i}:=\sum_{j=1}^nL_{ij}$.

Payoffs may also depend on the relative position of the node $i$ with respect to $j$ in the network. Specifically, the researcher may also want to include \emph{edge-specific network statistics} of the form
\begin{equation}\label{edge_stat_def}T_{ij}:=T(\mathbf{L},\mathbf{X};i,j)\end{equation}
where $T(\cdot)$ may again be vector-valued, and we assume that the function $T(\cdot)$ is invariant to permutations of player indices.\footnote{That is, we assume that for any permutation $\tau:\{1,\dots,n\}\rightarrow\{1,\dots,n\}$ and $i,j=1,\dots,n$ we have $T(\mathbf{L}^{\tau},\mathbf{X}^{\tau};\tau(i),\tau(j))=T(\mathbf{L},\mathbf{X};i,j)$.} In the following, we also assume that the statistic is symmetric, $T_{ij} = T_{ji}$.\footnote{In order to accommodate the general case of asymmetric edge-specific statistics, it would be possible an additional argument in the marginal benefit function, and the technical results would continue to go through without substantive modifications.} In our description of preferences regarding $T_{ij}$ we will occasionally use $t_0$ to denote an arbitrarily chosen ``default" value for the statistic. 

\begin{ex}\textbf{(Transitive Triads)} A preference for closure of transitive triads can be expressed using statistics of the form
\[T_1(\mathbf{L},\mathbf{X};i,j) = \sum_{k\neq i,j}L_{ik}L_{jk},\hspace{0.3cm}\textnormal{or }T_2(\mathbf{L},\mathbf{X};i,j) = \max\left\{L_{ik}L_{jk}:k\neq i,j\right\}\]
Here, $T_{1ij}$ counts the number of immediate neighbors that both $i$ and $j$ have in common, and $T_{2ij}$ is an indicator whether $i$ and $j$ have any common neighbor. More generally, $T_{ij}$ could include other measures of the distance between agents $i$ and $j$ in the absence of a direct link, or indicators for potential ``cliques" of larger sizes.
\end{ex}

Patterns of transitivity may emerge for example in economic models of social capital where supporting links to common neighbors may enhance the value or viability of a connection between an agent pair, see e.g. \cite{JRT12} or \cite{GG16}. Transitivity may also reflect a biased search process where agents may be more likely to ``meet" through common neighbors.

Some of our results concern special cases in which the network statistics $S_i,S_j$, and $T_{ij}$ only depend on nodes at up to a finite network distance from $i$ and $j$, respectively. Specifically, we say that $S_i$ is \emph{a function of the network neighborhood of radius $r_S$ around $i$} if $S(\mathbf{L},\mathbf{X};i)=S(\tilde{\mathbf{L}},\mathbf{X};i)$ for any networks $\mathbf{L},\tilde{\mathbf{L}}$ such that $\tilde{\mathbf{L}}_{kl}=L_{kl}$ whenever the network distance between $i$ and $k$ is less than or equal to $r_S$. Similarly, we say that $T_{ij}$ \emph{is a function of the network neighborhood of radius $r_T$ around $i$ and $j$} if $T(\mathbf{L},\mathbf{X};i,j)=T(\tilde{\mathbf{L}},\mathbf{X};i,j)$ for any networks $\mathbf{L},\tilde{\mathbf{L}}$ such that $\tilde{\mathbf{L}}_{kl}=L_{kl}$ whenever the network distance of $k$ to $i$ or $j$ is less than $r_T$.

In contrast to node attributes $x_i,x_j$, the variables $S_i,S_j,$ and $T_{ij}$ are endogenous to the network formation process, and the characterization of the limiting model therefore must include equilibrium conditions for the joint distribution of types $x_i$ and network statistics $S_i$ and $T_{ij}$. Following the previous literature on social interactions models we refer to the payoff contribution of the exogenous attributes $x_i,x_j$ as \emph{exogenous interaction effects}, and the contribution of the endogenous network characteristics $s_i,s_j,t_{ij}$ as \emph{endogenous interaction effects}.



In order to avoid technical challenges from characterizing equilibrium conditions in infinite-dimensional vector spaces, the main results of this paper focus on the case of discrete attributes and network statistics:

\begin{ass}\textbf{(Network Statistics)}\label{finite_supp_ass} 
(i) The supports of the payoff-relevant network statistics, $\mathcal{S}$ and $\mathcal{T}$, and the type space $\mathcal{X}$ are finite. (ii) The values of $S(\mathbf{L},\mathbf{X};i)$ and $T(\mathbf{L},\mathbf{X};i,j)$ are determined by a network neighborhood of radius $r<\infty$ around $i$ ($i,j$, respectively). That is for any $\tilde{L}_{kl}=L_{kl}$ and $\tilde{x}_k=\tilde{x}_k$ whenever $(\mathbf{L}^r)_{ik}=(\mathbf{L}^r)_{jk}=0$ we have
\[S(\mathbf{L},\mathbf{X};i)=S(\tilde{\mathbf{L}},\tilde{\mathbf{X}};i)\hspace{0.5cm}\textnormal{and } T(\mathbf{L},\mathbf{X};i,j)=T(\tilde{\mathbf{L}},\tilde{\mathbf{X}};i,j)\]
\end{ass}

The finite support assumption for network statistics is restrictive for some leading cases, for example for preferences depending on degree centrality, the effective support has to be truncated at some finite value above which preferences become insensitive to a further increase of $s_{1i}$. For most of the results, we posit that this assumption could be replaced with a compact support assumption and adequate regularity conditions on the payoff functions, however we leave such a generalization for future research. Part (iii) is primarily for notational convenience, especially in connection with the assumption of a discrete type space for $x_i$.

\begin{ass}\textbf{(Systematic Part of Payoffs)}\label{surplus_bd_ass} Node attributes $x_i$ are i.i.d. draws from a distribution on $\mathcal{X}$.  
(i) The systematic parts of payoffs are uniformly bounded in absolute value for some value of $t=t_0$, $|U^*(x,x',s,s',t_0)|\leq\bar{U}<\infty$. Furthermore, (ii) at all values of $s,s'$, the function $U^*(x,x',s,s',t_0)$ is $p\geq1$ times differentiable in $x$ with uniformly bounded partial derivatives. (iii) The supports of the payoff-relevant network statistics, $\mathcal{S}$ and $\mathcal{T}$, and the type space $\mathcal{X}$ are  compact sets. 
\end{ass}

We next state our assumptions on the distribution of unobserved taste shifters. Most importantly, we impose sufficient conditions for the distribution of $\varepsilon_{ij}$ to belong to the domain of attraction of the extreme-value type I (Gumbel) distribution. Following \cite{Res87}, we say that the upper tail of the distribution $G(\varepsilon)$ is of type I if there exists an auxiliary function $a(s)\geq0$ such that the c.d.f. satisfies
\[\lim_{s\rightarrow\infty}\frac{1-G(s + a(s)v)}{1-G(s)} = e^{-v}\]
for all $v\in\mathbb{R}$. We are furthermore going to restrict our attention to distributions for which the auxiliary function can be chosen as $a(s):=\frac{1-G(s)}{g(s)}$, where $g(s)$ denotes the density associated with the c.d.f. $G(s)$. This property is shared for most standard specifications of discrete choice models, e.g. if $\varepsilon_{ij}$ follows the extreme-value type I, normal, or Gamma distribution, see \cite{Res87}.

In order to characterize the discrepancy between the tails of the distribution with c.d.f. $G(z)$ and that of the extreme-value distribution of type I with c.d.f. $\Lambda(z):=\exp\left\{- e^{-z}\right\}$, we define the function
\begin{equation}
\label{tail_h_fct_def} h(z):=G^{-1}(\Lambda(z))
\end{equation}
Clearly, extreme-value convergence to $\Lambda(z)$ requires that $\lim_{z\rightarrow\infty} h(z)=0$. Unfortunately, convergence may be very slow for some distributions so that the error from an extreme-value approximation may dominate the limiting distribution. We therefore need to strengthen that requirement to control that rate and narrow our focus on distributions for $\varepsilon_{ij}$ for which convergence is sufficiently fast. Specifically, we make the following assumption:


\begin{ass}\label{set_unobs_ass}\textbf{(Idiosyncratic Part of Payoffs)} $\varepsilon_{ij}$ and $\varepsilon_{i0,k}$ are i.i.d. draws from the distribution $G(s)$, and are independent of $x_i,x_j$, where (i) the c.d.f. $G(s)$ is absolutely continuous with density $g(s)$, and (ii) for the function $h(z):=G^{-1}(\Lambda(z))$, we have $h''(\log n)=o(n^{-1/2})$.
\end{ass}

The second requirement is satisfied when $G(z)$ is the c.d.f. of an exponentially or extreme-value type I (Gumbel) distributed random variable. However it rules out some other common parametric specifications for a random utility model, in particular the Gaussian distribution.

We embed the finite economy into the asymptotic sequence proposed in \cite{Men16}, which is chosen to retain a number of qualitative features in the limit. Most importantly, that sequence will serve only as an approximation device rather than a factual description on how the network may evolve if additional nodes were added. Formally, we assume the following:

\begin{ass}\textbf{(Network Size)}\label{asy_rate_ass}(i) The number $n$ of agents in the network grows to infinity, and (ii) the random draws for marginal costs $MC_i$ are governed by the sequence $J=\left[n^{1/2}\right]$, where $[x]$ denotes the value of $x$ rounded to the closest integer. (iii) The scale parameter for the taste shifters $\sigma\equiv \sigma_n = \frac{1}{a(b_n)}$, where $b_n = G^{-1}\left(1-\frac1{\sqrt{n}}\right)$, and $a(s)$ is the auxiliary function specified in Assumption \ref{set_unobs_ass} (ii). Furthermore, (iv) for any values $t_1\neq t_2\in\mathcal{T}$, $|U(x,x',s,s',t_1) - U(x,x',s,s',t_2)|$ may increase with $n$, and there exists a constant $B_T<\infty$ such that for any sequence of pairwise stable networks $\left(\mathbf{L}_n^*\right)_{n\geq2}$, $\sup_{x,x',s,s'}\left(\mathbb{E}\left[\exp\left\{2|U(x,x',s,s',T(\mathbf{L}_n^*,x,x',i,j)) - U(x,x',s,s',t_0)|\right\}\right]\right)^{1/2}\leq \exp\{B_T\}$ for $n$ sufficiently large.
\end{ass}

Most importantly, the network is assumed to be sparse in the sense that each node is connected to a stochastically bounded number of alters, which remains stochastic. Furthermore, the relative scale of the systematic and idiosyncratic parts remains of the same order of magnitude so that differences in the systematic part remain predictive, but do not fully determine the network outcomes. Finally, preferences for certain network features may vary along the asymptotic sequence to ensure non-degenerate distributions for edge-specific network statistics. See also \cite{Men16} for a more detailed justification and discussion of these choices.

\subsection{Solution Concept} Our formal analysis assumes pairwise stability as a solution concept, which was first introduced by \cite{JWo96}. While there are alternative sets of primitive conditions for existence of a pairwise stable network, we make the following high-level assumption on the observed network $L^*$:

\begin{dfn}\textbf{(Pairwise Stable Network)}\label{pw_def}
The undirected graph $\mathbf{L}^*$ is a \textbf{pairwise stable network (PSN)} if for any link $ij$ with $L_{ij}^*=1$,
\[U_{ij}(\mathbf{L}^*,\mathbf{X})\geq MC_i,\hspace{0.3cm}\textnormal{and }U_{ji}(\mathbf{L}^*)\geq MC_j\]
and any link $ij$ with $L_{ij}^*=0$,
\[U_{ij}(\mathbf{L}^*,\mathbf{X})< MC_i,\hspace{0.3cm}\textnormal{or }U_{ji}(\mathbf{L}^*)< MC_j\]
\end{dfn}

In general the pairwise stable network is not unique, but it is possible to give general conditions on the payoffs to ensure existence, see e.g. \cite{She20} for a discussion. Our results on the possible limits of the network moment do not require any auxiliary assumptions regarding equilibrium selection, our derivation of its asymptotic distribution requires that the (possibly stochastic) equilibrium selection mechanism  is stochastically independent of payoffs. Specifically we assume the following:

\begin{ass}\label{solution_ass}\textbf{Solution Concept} (i) The observed network $\mathbf{L}^*$ satisfies the payoff conditions for pairwise stability in Definition \ref{pw_def}. (ii) In the presence of multiple pairwise stable networks, agents coordinate on a pairwise stable network via a public signal $\mathbf{v}$ that is measurable with respect to a sigma field $\mathcal{F}$ but need not be observable to the researcher. (iii) Selection is independent of payoffs, that is the payoff-relevant variables $x_i,MC_i,\varepsilon_{ij}$ are drawn i.i.d. conditional on $\mathcal{F}$.
\end{ass}

Our results describing the set of possible network outcomes, including the law of large numbers in Theorem \ref{primitive_lln_thm}, will not rely on any assumptions on the selection mechanism, however parts (ii) and (iii) will be assumed for the CLT in Theorem \ref{asy_Gauss_thm}. In particular, while the set of pairwise stable networks certainly depends on realized payoffs, our distribution theory will assume that the selection rule can be described in terms of a randomization device that is stochastically independent of link payoffs. For example, a stochastic selection mechanism could be specified in terms of a random initial condition and sequence for updating links in a t\^atonnement process - under appropriate conditions t\^atonnement can induce a distribution over the pairwise stable networks as the stationary points of that process.

\subsection{Additional Restrictions}

Our main objective in this paper is to introduce key theoretical ideas and demonstrate their practical use for a sufficiently broad set of applications. We therefore do not aim for the greatest generality but maintain some additional restrictions for key formal results to simplify the exposition. To preview the main restrictions, some key results will only be stated for the case of node-specific interaction effects, $T_{ij}:=\{\}$. The implications of edge-specific interaction effects for our approach are discussed separately in Appendix \ref{sec:edge_spec_samp_app}.

Furthermore, we also restrict the subnetwork events $\mathbf{A}_{i_1\dots i_D}$ in the construction of network moments in order to avoid some complications from the multiplicity of pairwise stable networks. Specifically, we only consider composite events $\mathbf{A}_{i_1\dots i_D}$ such that the existence of a pairwise stable network supporting each elementary subnetwork outcome are mutually exclusive events. Dropping this requirement does not fundamentally change our theoretical analysis, but will pose additional practical problems for implementation as we discuss in Section \ref{subsec:samp_rep_sec}.

Finally, for the asymptotic distribution of the network moment $\hat{m}_n(\theta)$ we only consider the case in which the probability for the subnetwork events of interest is uniquely determined by the (possibly non-unique) equilibrium values of certain aggregate quantities for the entire network. This can result either from the nature of strategic interaction effects or from additional constraints on equilibrium selection. Our analysis could in principle be extended to the case without uniqueness of local network outcomes, however the asymptotic distribution of network moments could only be characterized indirectly in terms of which bounds on aggregate state variables are simultaneously binding at any point of interest.

\section{Potential Values and Random Network Neighborhoods}

\label{sec:potential_values_sec}


For our analysis we need to characterize the probability of events regarding the links to a node $i$ and the values of the payoff-relevant network statistics for another set of nodes and statistics.

To this end, we first define potential values for certain network outcomes as an equilibrium response to exogenous constraints on some of the link formation decisions. Specifically given an arbitrary network $\mathbf{L}$, we define $D_{ij}:=\dum\left\{U_{ij}(\mathbf{L},\mathbf{X})\geq MC_i\right\}$ as an indicator whether $i$ would agree to form a link to $j$ given that network $\mathbf{L}$, and let $D_{ij}^*:=\dum\left\{U_{ij}(\mathbf{L}^*,\mathbf{X})\geq MC_i\right\}$ denote the corresponding indicator given the pairwise stable network $\mathbf{L}^*$. We also let $\mathbf{D}$ and $\mathbf{D^*}$ denote the matrices with entries $D_{ij}$ and $D_{ij}^*$, respectively, and refer to $\mathbf{D}$ as a directed \emph{proposal network}. These indicators are directed, i.e. in general $D_{ij}$ need not be equal to $D_{ji}$, and from the definition of a pairwise stable network, $L_{ij}^*\equiv D_{ij}^*D_{ji}^*$, or
\[\mathbf{L}^* \equiv \mathbf{D}^*\odot(\mathbf{D}^*)'\]
where $\mathbf{A}\odot\mathbf{B}$ denotes the Hadamard (element wise) product of two matrices $\mathbf{A},\mathbf{B}$ of equal size. ssTo characterize restrictions on $\mathbf{D}$, we denote a collection of $r$ directed edges $(i_1,j_1),\dots,(i_r,j_r)$ with the $2\times r$ matrix
\[\mathbf{E}:=\left[\begin{array}{ccc}i_1&\cdots&i_r\\j_1&\cdots&j_r\end{array}\right].\]
We denote the corresponding entries of $\mathbf{D}$ with the $r$-dimensional vector $\mathbf{D}_{\mathbf{E}}=(D_{i_1,j_1},\dots,D_{i_r,j_r})$ with elements $D_{\mathbf{E},i_1j_1}$.

We now consider the pairwise stable networks that may arise from holding link acceptance decisions at the directed edges in $\mathbf{E_1}$ fixed at $\mathbf{D}_{\mathbf{E}}$: Given random payoffs and node attributes, we let
\[\mathcal{D}^*(\mathbf{E_1},\mathbf{D}_{\mathbf{E_1}}):=\left\{\mathbf{D}^*:D_{ij}^*=\left\{\begin{array}{lcl} D_{\mathbf{E_1},ij}&\hspace{0.5cm}&\textnormal{if }(i,j)\in\mathbf{E_1}\\
\dum\{U_{ij}(\mathbf{D}^*\odot(\mathbf{D}^*)',\mathbf{X})\geq MC_i\}&&\textnormal{otherwise}\end{array}\right.\right\}\]
be the collection of link acceptance indicators that are supported by a pairwise stable network after fixing the proposals for the edges in $\{\mathbf{E_1}\}$ at $\mathbf{D}_{\mathbf{E_1}}$, where we let $\{\mathbf{E}\}:=\left\{(E_{11},E_{21}),\dots,(E_{1r},E_{2r})\right\}$ for the set of nodes corresponding to the edges in $\mathbf{E}$. We also denote a typical entry of a matrix $\mathbf{D^*}\in\mathcal{D}^*(\mathbf{E_1},\mathbf{D}_{\mathbf{E_1}})$ with $D_{ij}^*(\mathbf{E}_1,\mathbf{D}_{\mathbf{E}_1})$.


In order to characterize the values of network statistics $s_j^*$ that may result from local changes to the subnetwork on $\mathbf{E_1}$, we need to account both for the direct (``mechanical") effect on the position of node $j$ from changing links in the network, as well as the indirect (``equilibrium") effects from adjustments outside $\mathbf{E_1}$ that may be needed for the network on the complement of $\mathbf{E_1}$ to be pairwise stable. That distinction is important for our analysis. Here, the payoff inequalities characterizing pairwise stability only concern unilateral link-by-link decisions given the status quo where agents only need to consider direct ``mechanical" effects. In contrast, probabilities for the resulting subnetwork events have to incorporate indirect ``general equilibrium" effects as well.

We therefore evaluate \emph{potential values} for the network statistic $s_j$ under changes to two, potentially separate, edge sets $\mathbf{E}_1,\mathbf{E}_2$ and proposal subnetworks on those edges, $\mathbf{D}_{\mathbf{E}_1},\mathbf{D}_{\mathbf{E}_2}$. As a shorthand notation, we denote the subset of non-zero edges in $(\mathbf{E},\mathbf{D_E})$ with
\[\mathbf{E}(\mathbf{D}_{\mathbf{E}}):=\left\{(E_{1i},E_{2j}):D_{ij}=1\right\}\]
We then let
\[\mathbf{L}(\mathbf{D}^*,\mathbf{E}_2,\mathbf{D}_{\mathbf{E}_2}):=
\mathbf{D}^*\odot(\mathbf{D}^*+\{\mathbf{E}(\mathbf{D}_{\mathbf{E}_2})\}-\left\{\mathbf{E}(1-\mathbf{D}_{\mathbf{E}_2})\right\})'\]
denote the network resulting from $\mathbf{D}^*$ after setting the edges in the set $\mathbf{E}$ equal to the values specified in $\mathbf{D}_{\mathbf{E}}$, which need not be pairwise stable. We then define the potential values for the node-level network statistic $s_j$ at the proposal subnetwork  $\mathbf{D}_{\mathbf{E}_2}$ and given proposals $\mathbf{D}_{\mathbf{E_1}}$ as the set
\[\mathcal{S}_{j}^*(\mathbf{E}_1,\mathbf{D}_{\mathbf{E}_1},\mathbf{E}_2,\mathbf{D}_{\mathbf{E}_2}):=\left\{ S(\mathbf{L}(\mathbf{D}^*,\mathbf{E}_2,\mathbf{D}_{\mathbf{E}_2});j): \mathbf{D}^*\in\mathcal{D}^*(\mathbf{E}_1,\mathbf{D}_{\mathbf{E}_1})\right\}\]
That is, $\mathcal{S}_{j}^*(\mathbf{E}_1,\mathbf{D}_{\mathbf{E}_1},\mathbf{E}_2,\mathbf{D}_{\mathbf{E}_2})$ is the set of values for $s_j^*$ from changing the proposal subnetwork on $\mathbf{E_2}$ to $\mathbf{D_{E_2}}$, that are supported by a pairwise stable proposal network after holding the subnetwork on $\mathbf{E}_1$ fixed at $\mathbf{L_{E_1}}$. In the same fashion we can define the potential values for the edge-level statistic $t_{ij}$ as
\[\mathcal{T}_{ij}^*(\mathbf{E}_1,\mathbf{D}_{\mathbf{E}_1},\mathbf{E}_2,\mathbf{D}_{\mathbf{E}_2}):=\left\{ T(\mathbf{L}(\mathbf{D}^*,\mathbf{E}_2,\mathbf{D}_{\mathbf{E}_2});i,j): \mathbf{D}^*\in\mathcal{D}^*(\mathbf{E}_1,\mathbf{D}_{\mathbf{E}_1})\right\}\]
Note again that in this notation, the crucial difference in the role of the subnetworks $(\mathbf{E_1},\mathbf{D}_{\mathbf{E_1}})$ and $(\mathbf{E_2},\mathbf{D}_{\mathbf{E_2}})$ is that potential values of $s_j,t_{ij}$ account for direct \emph{and} indirect effects from exogenously fixing the edges on $\mathbf{E_1}$, but only for mechanical effects from changing edges in $\mathbf{E_2}$ from the pairwise stable network to the values $\mathbf{D_{E_2}}$.

\begin{figure}
\SetCoordinates[xAngle=-10,yAngle=90,zAngle=10,yLength=1,xLength=1,zLength=1]
\begin{tikzpicture}[multilayer=3d]
\SetLayerDistance{4.5}
\Plane[x=-1,y=-3,width=5,height=7,color=blue,opacity=0]
\Plane[x=-1,y=-3,width=5,height=7,color=blue,opacity=0,layer=2]
\Plane[x=-1,y=-12,width=5,height=7,color=blue,opacity=0,layer=2]
\Plane[x=-1,y=-3,width=5,height=7,color=blue,,opacity=0,layer=3]
\Vertices{potential_vertices_ml.csv}
\Edges{potential_edges_ml.csv}
\end{tikzpicture}
\caption{Potential values for various configurations $\mathbf{L}_{E_F}$ around node $F$, and realized outcome in a pairwise stable network.}
\end{figure}
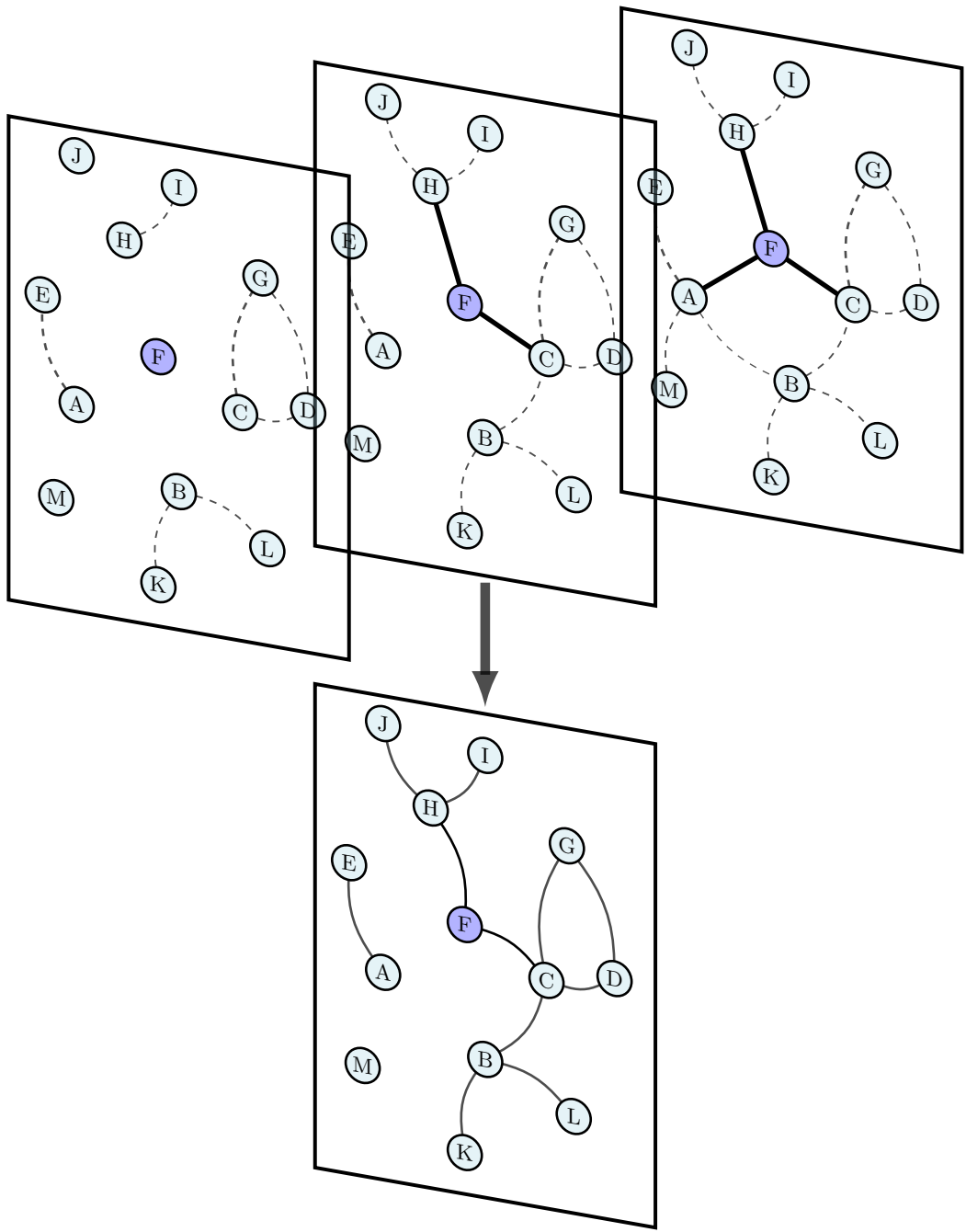

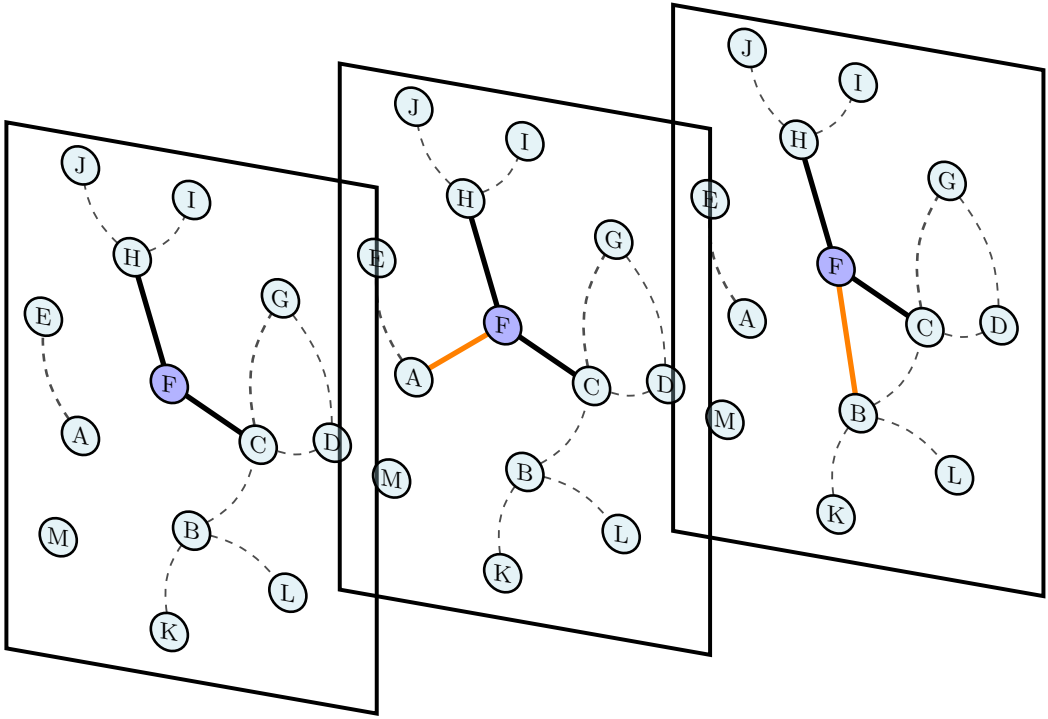
\begin{figure}
\SetCoordinates[xAngle=-10,yAngle=90,zAngle=10,yLength=1,xLength=1,zLength=1]
\begin{tikzpicture}[multilayer=3d]
\SetLayerDistance{4.5}
\Plane[x=-1,y=-3,width=5,height=7,color=blue,opacity=0]
\Plane[x=-1,y=-3,width=5,height=7,color=blue,opacity=0,layer=2]
\Plane[x=-1,y=-3,width=5,height=7,color=blue,,opacity=0,layer=3]
\Vertices{potential_vertices_ml2.csv}
\Edges{potential_edges_ml2.csv}
\end{tikzpicture}
\caption{Potential values for $\mathbf{E}_1$ with $\mathbf{L_{E_2}}=1$ for $\mathbf{E}_2=()$, $\mathbf{E}_2=(A,F)$, and $\mathbf{E}_2=(B,F)$, respectively.}
\end{figure}

\begin{ex}\textbf{(Potential Values)} To fix ideas consider an example with three nodes $\{1,2,3\}$ with marginal utilities
\[U_{ij} = s_j -1 + \varepsilon_{ij}\]
where $s_i = \sum_{j\neq i}L_{ij}$. Let the edge set $\mathbf{E}_1:=\left[\begin{array}{cc}1&1\\2&3\end{array}\right]$ be the directed edges from node $1$ to each of the remaining nodes, and consider potential outcomes for $s_2$. If $D_{12}=0$, then $s_2=1$ iff $U_{23}\geq MC_2$ and $U_{32}\geq MC_3$, and zero otherwise, noting that $U_{23}$ also depends on $s_3$, and therefore also on $D_{13}$. If $D_{12}=1$, then $s_2=2$ iff $U_{23}\geq MC_2$ and $U_{32}\geq MC_3$, and $U_{21}\geq MC_2$, $s_2=0$ if $U_{21}\geq MC_2$ and $U_{23}\geq MC_2$ or $U_{32}\geq MC_3$, and otherwise $s_2=1$. The distribution for these potential values then derives from the probabilities for these events.
\end{ex}

\subsection{Potential Values and Pairwise Stability}

The distribution of network moments (\ref{moment_fct}) primarily depends on the probability of network events of the form $A_{\mathbf{i}}$ for a $D$-ad $\mathbf{i}=(i_1,\dots,i_D)$, or the joint probabilities over finitely many nodes. Clearly the events $A_{\mathbf{i}}$ and $A_{\mathbf{j}}$ may be dependent between $D$-ads $\mathbf{i}$ and $\mathbf{j}$, especially (but not only) when they have nodes in common. For example, the expectation of $\hat{m}_n(\theta)$ is determined by probabilities
\[\pi_n(A_{\mathbf{i}}|\mathbf{x}_{\mathbf{i}}):= \mathbb{P}\left(\mathbf{L}_{|\mathbf{i}}^*\in A_{\mathbf{i}}|\mathbf{x}_{\mathbf{i}}\right)\]
and the variance by the probabilities
\[\pi_n(A_{\mathbf{i}}\cap A_{\mathbf{j}}|\mathbf{x}_{\mathbf{i}},\mathbf{x}_{\mathbf{j}}):= \mathbb{P}\left(\mathbf{L}_{|\mathbf{i}}^*\in A_{\mathbf{i}}\cap \mathbf{L}_{|\mathbf{j}}^*\in A_{\mathbf{j}}|
\mathbf{x}_{\mathbf{i}},\mathbf{x}_{\mathbf{j}}\right)\]
for pairs of $D$-ads $\mathbf{i},\mathbf{j}$.

Since links and network statistics are determined simultaneously, the probability $\pi_n(A_i|\mathbf{x}_i)$ is difficult to evaluate. However we argue that the problem becomes more tractable by separating it into generating the relevant potential values for network statistics in a first step, and then evaluating the conditional probability of the network event $A_i$ given those potential values.

We formalize this step as a Lemma for which we need some additional notation: for the $2\times r$ matrix of directed edges, $\mathbf{E_1}$, we denote the vector of indices corresponding to the source nodes with $\mathbf{E_{1,1\cdot}}:=(E_{1,11},\dots,E_{1,1r})$, and that corresponding to the the target nodes with $\mathbf{E_{1,2\cdot}}:=(E_{1,21},\dots,E_{1,2r})$. $\mathbf{E_{1,1\cdot}}$ and $\mathbf{E_{1,2\cdot}}$ also corresponds to the first row and second row, respectively, of the matrix $\mathbf{E_{1}}$. We also let $\mathbf{s}_{\mathbf{E_{1,2\cdot}}}:=(s_{E_{1,12}},\dots,s_{E_{1,r2}})'$ denote a matrix containing the network statistics for the nodes indexed by $\mathbf{E_{1,2\cdot}}$. Furthermore, we let the random mapping $\mathcal{D}_{\mathbf{E_1}}^*(\mathbf{s_{E_{1,2\cdot}}})$ denote the set of proposals on the edge set $\mathbf{E_1}$ that are pairwise stable given the network statistics $\mathbf{s_{E_{1,2\cdot}}}$, and $\mathcal{S}_{\mathbf{E_{1,2\cdot}}}^*(\mathbf{D}_{\mathbf{E_1}})$ be the set of potential values for the network statistics $\mathbf{s}_{\mathbf{E_{1,2\cdot}}}$ given the proposals $\mathbf{D}_{\mathbf{E_1}}$.

The next Lemma then follows immediately from the definition of potential values for network statistics:
\begin{lem}\label{psn_potential_val_lem}
The proposals $\mathbf{D_{E_1}}$ and the network statistics $\mathbf{s_{E_{1,2\cdot}}}$ are jointly supported by a pairwise stable network if and only if $\mathbf{D_{E_1}}\in\mathcal{D}_{\mathbf{E_1}}^*(\mathbf{s_{E_{1,2\cdot}}})$ and $\mathbf{s_{E_{1,2\cdot}}}\in
\mathcal{S}_{\mathbf{E_{1,2\cdot}}}^*(\mathbf{D}_{\mathbf{E_1}})$.
\end{lem}

Sharp upper and lower bounds for $\pi_n(A_i|\mathbf{x}_{\mathbf{i}})$ are generally determined by the probability that certain outcomes, or sets of outcomes, are supported by a pairwise stable network. Given Lemma \ref{psn_potential_val_lem}, that probability corresponds to that of drawing the potential values given the implied relevant overlap times the conditional probability of forming the edges corresponding to the event.

\subsection{Independence}

For our purposes it suffices to consider potential outcomes with respect to edge sets $\mathbf{E_1}$ that contain the outgoing edges from a node or a polyad $\mathbf{i}=(i_1,\dots,i_D)'$. We now show that for that particular structure, certain subsets of potential values are determined by a set of payoff shifters that is disjoint from those determining stability on $\mathbf{E_1}$ given potential values, and therefore independent.

To be specific, we consider edge sets of the form
\[\mathbf{E_1}\equiv\mathbf{E_i}:=\left[\mathbf{E}_{i_1},\dots,\mathbf{E}_{i_D}\right]\]
for
\[\mathbf{E}_{i_d}:=\left[\begin{array}{ccc}i_d&\dots&i_d\\1&\dots&n\end{array}\right]\]
For notational ease, we let $\mathbf{E_1}$ include self links, which were all set to zero under the general assumptions of our framework, so that the number of columns (edges) of $\mathbf{E_1}$ is $r=nD$. Adapting earlier notation we also let $\mathcal{S}_{-\mathbf{i}}^*(\mathbf{D_{E_1}})$ denote the set of potential values $\left(s_j\right){j\notin\{\mathbf{i}\}}$ with respect to the proposal subnetwork $\mathbf{D_{E_1}}$ on $\mathbf{E_1}$.

The random mappings $\mathcal{S}_{-\mathbf{i}}^*:\{0,1\}^r\rightarrow 2^{\mathcal{S}^{n-D}}$ and $\mathcal{D}_{\mathbf{E_1}}^*:\mathcal{S}^{n-D}\rightarrow 2^{\{0,1\}^r}$ are implicitly defined by the attributes $x_1,\dots,x_n$ as well as taste shocks $\varepsilon_{12},\dots,\varepsilon_{nn-1}$ and marginal costs $MC_1,\dots,MC_n$. We now give conditions for the events
\begin{eqnarray}
\nonumber A_D(\mathbf{E_i},\mathbf{D_{E_i}},\mathbf{s_{-\mathbf{i}}})&:=&\left\{\mathbf{D_{E_1}}\in\mathcal{D}_{\mathbf{E_i}}^*(\mathbf{s_{-\mathbf{i}}})\right\}\\
\nonumber
A_s(\mathbf{E_i},\mathbf{D_{E_i}},\mathbf{s_{-\mathbf{i}}})&:=&\left\{\mathbf{s_{-\mathbf{i}}}\in
\mathcal{S}_{-\mathbf{i}}^*(\mathbf{D}_{\mathbf{E_i}})\right\}
\end{eqnarray}
to be conditionally independent given attributes $x_1,\dots,x_n$ and $\mathbf{F}$ for any values of $\mathbf{D_{E_i}},\mathbf{s_{-\mathbf{i}}}$.


For an edge set $\mathbf{E_i}$ of this form, we obtain the following independence result:
\begin{lem}\label{app_independence_lem}\textbf{(Independence)} Suppose Assumption \ref{set_unobs_ass} holds. 
Then for any polyad $\mathbf{i}=(i_1,\dots,i_D)$, $A_D(\mathbf{E_i},\mathbf{D_{E_i}},\mathbf{s_{-\mathbf{i}}})$ and
$A_s(\mathbf{E_i},\mathbf{D_{E_i}},\mathbf{s_{-\mathbf{i}}})$ are conditionally independent given $\mathbf{x_{\mathbf{i}}}$ and $\mathcal{F}$. The analogous conclusion holds after conditioning on taste shocks $\varepsilon_{k_1l_1},MC_{k_1},\dots,\varepsilon_{k_ql_q},MC_{k_q}$ for any $\{k_1,\dots,k_q\}\cap\{i_1,\dots,i_D\}=\emptyset$.
\end{lem}

See the appendix for a proof. Note that this problem is different from other classical settings with triangular or other simultaneous equations, where unobservables are allowed to be arbitrarily correlated. Here, the assumption of i.i.d. taste shocks permits to establish independence relationships as long as there is no overlap in which factors affect either set of potential outcomes. It is also important to point out that we only show that independence holds for a network outcomes across a particular partition of nodes into two disjoint sets which does not imply any stronger conclusions of conditional or unconditional independence across nodes.

\subsection{Invariance}

As a second step, we establish invariance of the distribution of potential values with respect to permutations of elements of $\mathbf{E}$ and $\mathbf{s}$. To compare nodes in the edge set $\{\mathbf{E}\}$ to other nodes in the network, we define the (payoff) relevant overlap between a set of nodes $j_1,\dots,j_D$ and $\mathbf{E}$. In the following, let $\mathbf{L}_{\mathbf{E}}(\mathbf{D}_0,\mathbf{D_E})$ denote the adjacency matrix obtained from $\mathbf{L}(\mathbf{D}_0,\mathbf{D}_{\mathbf{E}})$ by setting all entries outside $\mathbf{E}$ to zero.

\begin{dfn}\textbf{(Relevant Overlap)} The \emph{relevant overlap} between the directed edges $\mathbf{E}$ and the nodes $j_1,\dots,j_D$ is a sufficient statistic for $\mathbf{L}_{\mathbf{E}}(\mathbf{D}_0,\mathbf{D_E})$ with respect to $s_{j_1},\dots,s_{j_D}$ and $t_{kl}$ for each edge in $\mathbf{E}$ that includes a node in $\{j_1,\dots,j_D\}$. That is, we let $R(\mathbf{E},\mathbf{D_E},\mathbf{X};j_1,\dots,j_D)$ be a mapping such that for all $\mathbf{D_0}\in\{0,1\}^{n(n-1)}$ and $\mathbf{\tilde{D}_E}\in\{0,1\}^r$, we have 
$S(\mathbf{L}(\mathbf{D_0},\mathbf{\tilde{D}_E}),\mathbf{X};j) = S(\mathbf{L}(\mathbf{D_0},\mathbf{D_E}),\mathbf{X};j)$ for $j\in\{j_1,\dots,j_D\}$, and
$T(\mathbf{L}(\mathbf{D_0},\mathbf{\tilde{D}_E}),\mathbf{X};k,l) =  T(\mathbf{L}(\mathbf{D_0},\mathbf{D_E}),\mathbf{X};k,l)$ for each edge in $\mathbf{E}$ that includes a node in $\{j_1,\dots,j_D\}$ whenever $R(\mathbf{E},\mathbf{L}_{\mathbf{E}}(\mathbf{D}_0,\mathbf{\tilde{D}_E}),\mathbf{X};j_1,\dots,j_D)
=R(\mathbf{E},\mathbf{L}_{\mathbf{E}}(\mathbf{D}_0,\mathbf{D_E}),\mathbf{X};j_1,\dots,j_D)$.
\end{dfn}

Note that since by Assumption \ref{finite_supp_ass} the network statistics only take finitely many values, we can without loss of generality assume that $R(\cdot)$ only takes finitely many values as well. To fix ideas we next give a few examples, where we let $\tilde{L}_{kl}$ denote the element of $\mathbf{L}_{\mathbf{E}}(\mathbf{D}_0,\mathbf{\tilde{D}_E})$ corresponding to the edge $(kl)\in\mathbf{E}$.

\begin{ex}\textbf{Relevant Overlap - Degree Centrality} Suppose that $\mathbf{E}_i$ is the set of the directed edges $\left((i,j)\right)_{j=1}^n$ and the relevant network statistic $s_j$ depends only on the network neighborhood of radius 1 around $j$, i.e. the set of nodes $k$ such that $L_{jk}=1$. Then the relevant overlap with node $j$, $R(\mathbf{E}_i,\mathbf{L}_{\mathbf{E}}(\mathbf{D}_0,\mathbf{\tilde{D}_E}),\mathbf{X};j)=\tilde{L}_{ij}$. This includes in particular the case of degree centrality, $s_j:=\sum_{k\neq j}L_{ij}.$
\end{ex}

\begin{ex}\textbf{Relevant Overlap - Transitivity} If the payoff relevant network statistic is the indicator whether $i$ and $j$ have a common network neighbor, $t_{ij}:=\max_{k\neq i, j}L_{ik}L_{jk}$, then the relevant overlap for the set $\mathbf{E}_i$ of directed edges $\left((i,j)\right)_{j=1}^n$ with the node pair $j,k$ is $R(\mathbf{E}_i,\mathbf{L}_{\mathbf{E}}(\mathbf{D}_0,\mathbf{\tilde{D}_E}),\mathbf{X};j,k)=\dum\{\tilde{L}_{ij}=\tilde{L}_{ik}=1\}$; the relevant overlap of the set $\mathbf{E}_i\cup\mathbf{E}_k$ of directed edges $\left((i,k),(j,k)\right)_{k=1}^n$ with the node $k$ is
$R(\mathbf{E}_i,\mathbf{L}_{\mathbf{E}}(\mathbf{D}_0,\mathbf{\tilde{D}_E}),\mathbf{X};j,k)=\dum\{\tilde{L}_{ik}=\tilde{L}_{jk}=1\}$. Note that in this second case, the value of $T_{ij}$ is fully determined given the relevant overlap.
\end{ex}

\begin{ex}\textbf{Relevant Overlap - Recursive Statistics} Suppose that the node-level statistic $s_j$ depends on a network neighborhood of radius greater than 1 around $j$ but is defined recursively as
\[s_j = \tilde{S}_2\left(L_{1j}x_1,\dots,L_{nj}x_n,L_{1j}\tilde{S}_1(\mathbf{L},\mathbf{X};1),\dots,L_{nj}\tilde{S}_1(\mathbf{L},\mathbf{X};1)\right)\]
where $\tilde{S}_1(\cdot)$ defines a vector of $m$ node-specific network statistics, and $\tilde{s}_2(\cdot)$ is symmetric in its $n(k+m)$ arguments. That is, $s_j$ is a function of the exogenous attributes and ``intermediate" network positions $\tilde{s}_{k1}:=\tilde{S}_1(\mathbf{L},\mathbf{X};k)$ of $j$'s immediate network neighbors. Then for the set $\mathbf{E}_i$ of directed edges $\left((i,j)\right)_{j=1}^n$, the relevant overlap with node $j$ is
$R(\mathbf{E}_i,\mathbf{L}_{\mathbf{E}}(\mathbf{D}_0,\mathbf{\tilde{D}_E}),\mathbf{X};j)=\left(\tilde{L}_{ij},\tilde{s}_{i1}\right)$,
where $\tilde{s}_{i1}:=\tilde{S}_1(\mathbf{L}_{\mathbf{E}}(\mathbf{D}_0,\mathbf{\tilde{D}_E}),\mathbf{X};k)$.
\end{ex}

We then consider an arbitrary permutation $\tau:\{1,\dots,n\}\rightarrow\{1,\dots,n\}$ which we identify with a one-to-one map of the set of node indices to itself. We also let $\mathbf{E}^{\tau}:=\tau(\mathbf{E})$ where we apply the mapping $\tau$ to each node index in the matrix $\mathbf{E}$, and let ``$\stackrel{d}{=}$" denote equality in distribution. We can now state the main version of our invariance result:

\begin{lem}\textbf{(Invariance)}\label{app_invariance_lem} Suppose Assumptions \ref{finite_supp_ass}-\ref{asy_rate_ass} hold. Then for any $i,j=1,\dots,n$, $\mathbf{E}$, and $\mathbf{D}_{\mathbf{E}}$, and arbitrary permutation $\tau$ of $\{1,\dots,n\}$ we have for the node-level statistics
\[\mathcal{S}_{j}^*(\mathbf{E}_1,\mathbf{D}_{\mathbf{E}_1},\mathbf{E}_2,\mathbf{D}_{\mathbf{E}_2})
\stackrel{d}{=}\mathcal{S}_{\tau(j)}^*(\mathbf{E}_1,\mathbf{D}_{\mathbf{E}_1},\mathbf{E}_2^{\tau},\mathbf{D}_{\mathbf{E}_2})\]
conditional on $R(\mathbf{E}_1^{\tau},\mathbf{L}_{\mathbf{E}_1}(\mathbf{D}_0,\mathbf{D_{E_1}}),\mathbf{X}^{\tau};\tau(j))
=R(\mathbf{E}_1,\mathbf{L}_{\mathbf{E}_1}(\mathbf{D}_0,\mathbf{D_{E_1}}),\mathbf{X};j)$, $x_{\tau(j)}=x_j$, and $\mathcal{F}$.  Furthermore, for edge-level statistics \[\mathcal{T}_{ij}^*(\mathbf{E}_1,\mathbf{D}_{\mathbf{E}_1},\mathbf{E}_2,\mathbf{D}_{\mathbf{E}_2})
\stackrel{d}{=}\mathcal{T}_{\tau(i)\tau(j)}^*(\mathbf{E}_1,\mathbf{D}_{\mathbf{E}_1},\mathbf{E}_2^{\tau},\mathbf{D}_{\mathbf{E}_2})\]
conditional on $R(\mathbf{E_1}^{\tau},\mathbf{L}_{\mathbf{E}_1}(\mathbf{D}_0,\mathbf{D_{E_1}}),\mathbf{X}^{\tau};\tau(i),\tau(j))
=R(\mathbf{E}_1,\mathbf{L}_{\mathbf{E}_1}(\mathbf{D}_0,\mathbf{D_{E_1}}),\mathbf{X};i,j)$, $x_{\tau(i)}=x_i,x_{\tau(j)}=x_j$, and $\mathcal{F}$.
\end{lem}

See the appendix for a proof. Since $\tau$ is a bijection, the lemma also immediately implies that
\[\mathcal{S}_{j}^*(\mathbf{E}_1,\mathbf{D}_{\mathbf{E}_1},\mathbf{E}_2,\mathbf{D}_{\mathbf{E}_2})
\stackrel{d}{=}\mathcal{S}_{j}^*(\mathbf{E}_1^{\tau},\mathbf{D}_{\mathbf{E}_1},\mathbf{E}_2,\mathbf{D}_{\mathbf{E}_2})\]
conditional on relevant overlap. For some steps, we also need to characterize the joint distribution of node-level statistics for a finite number of nodes. That extension is straightforward and is given separately in the appendix.

\subsection{Sampling Representation}

\label{subsec:samp_rep_sec}

Under pairwise stability, the links to and from a polyad $(i_1,\dots,i_D)$ - i.e. the network on the edge set $\mathbf{E_i}:=\mathbf{E}_{i_1}\cup\dots\cup\mathbf{E}_{i_D}$ - are jointly determined with network attributes of other nodes in the network. We now establish a ``local" representation which will allow to approximate probabilities for network events on $\mathbf{E_i}$ without explicitly modeling the remainder of the network. For greater ease of exposition, we state our results in the remainder of the paper only for the case of node-specific interaction effects, $T_{ij}:=\{\}$, and leave a discussion of the general case to Appendix \ref{sec:edge_spec_samp_app}.

Given the payoff shocks $\left(\varepsilon_{kl}\right)_{k,l}$ and $MC_1,\dots,MC_n$, we let
\[D_{ij0}:=\dum\{\bar{U} + \varepsilon_{ij}\geq MC_i\}\textnormal{  and  }D_{ji0}:=\dum\left\{\bar{U} + \varepsilon_{ji}\geq MC_j\right\}.\] Since by Assumption \ref{surplus_bd_ass} $\bar{U}$ is an upper bound for the systematic part of random utility, $D_{ij0}\geq D_{ij}^*$ almost surely. We then define a network neighborhood of node $i$ as the set $\mathcal{N}_i$ of nodes $j$ with $D_{ij0}=D_{ji0}=1$. That set excludes any node $j$ that will not form a pairwise stable link to node $i$ for any realization of exogenous attributes $x_i,x_j$ and network attributes $s_i,s_j,t_{ij}$, and by Assumption \ref{asy_rate_ass} the size of $\mathcal{N}_i$ is asymptotically bounded. We refer to $\bigcup_{d=1}^D\mathcal{N}_{i_d}$ as the network neighborhood of the polyad $(i_1,\dots,i_D)$.

A key challenge in evaluating the probability of an event on the edge set $\mathbf{E}_{\mathbf{i}}:=\bigcup_{d=1}^D\mathbf{E}_{i_d}$ is that network attributes $s_j^*$ and link proposals $D_{ji_r}^*$ for $r=1,\dots,D$ and other nodes $j\in\bigcup_{d=1}^D\mathcal{N}_{i_d}$ are jointly determined with the subnetwork on that edge set. We now propose an alternative sampling experiment which replaces the network neighborhood $\mathcal{N}_i$ for each node in the $(D+1)$ad with a random sample $\tilde{\mathcal{N}}_i$ consisting of a stochastically bounded number of Poisson draws of potential values and taste shocks from the respective marginal distributions. We then justify that representation by showing that it approximates the relevant conditional moments of the distribution generated by the finite-population network formation model.

By invariance in Lemma \ref{app_invariance_lem}, the distribution of potential values $s_j^*$ is invariant to node permutations conditional on
\[R_{\mathbf{i}j}:=R\left(\mathbf{E}_{\mathbf{i}},\mathbf{L}_{\mathbf{E}}(\mathbf{D}_0,\mathbf{\tilde{D}_{E_i}});j\right)\]
where in the following we assume without loss of generality that $R_{\mathbf{i}j}$ is the maximal invariant of $R\left(\mathbf{E}_{\mathbf{i}},\mathbf{L}_{\mathbf{E}}(\mathbf{D}_0,\mathbf{\tilde{D}_{E_i}});j\right)$ under permutations of $\{1,\dots,n\}\backslash\{j\}$.\footnote{In particular the event $R_{\mathbf{i}k}=R_{\mathbf{i}j}$ is equivalent to the existence of a permutation $\tilde{\tau}$ of $\{1,\dots,n\}\backslash\{k\}$ such that $R\left(\mathbf{E}_{\mathbf{i}}^{\tilde{\tau}},\mathbf{L}_{\mathbf{E}}(\mathbf{D}_0,\mathbf{\tilde{D}_{E_i}});k\right)= R_{\mathbf{i}j}$.} We will therefore use the notation $\tilde{s}_k^*(R)$ to denote a draw from a conditional distribution of potential values given $R_{\mathbf{i}k}=R$. Similarly, we use \[\tilde{D}_{ki}^*(s;R):=\dum\left\{U^*(x_k,x_i;\tilde{s}_k(R),s,t)+\tilde{\varepsilon}_{ki}\geq\widetilde{MC}_k\right\}\]
to denote an indicator for a proposal from $k$ to $i$ for draws $\tilde{\varepsilon}_{ki},\widetilde{MC}_k$ and a fixed value of $s$. 

\begin{figure}
\begin{tikzpicture}[scale=0.6]
\Vertices{sampling_rep2_vertices2.csv}
\Edges{sampling_rep2_edges.csv}
\end{tikzpicture}
\caption{Pairwise stable network (left) and sampling representation of subnetworks as typical draws from a distribution of network neighborhoods (right).}
\end{figure}
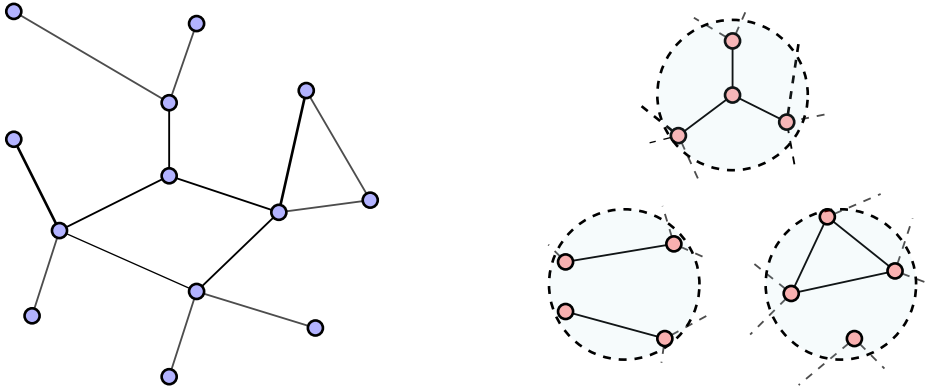

In the absence of edge-specific interaction effects, the approximate representation replaces the network neighborhood $\mathcal{N}_{\mathbf{i}}$ with network neighborhoods $\tilde{\mathcal{N}}_{i_1},\dots,\tilde{\mathcal{N}}_{i_D}$ that are independent realizations of a Poisson (point) process.\footnote{Recall that a point measure on a set $\mathcal{X}$ is of the form $\xi=\sum_{i=1}^q\delta_{X_i}$ for an integer-valued random variable $q$ and a discrete set of random values $\{X_1,\dots,X_q\}\in\mathcal{X}$. Given a finite measure $M$ on $\mathcal{X}$, a Poisson process with intensity (mean measure) $M$ is a random point measure such that for any Borel set $A\subset\mathcal{X}$ and non-negative integer $k$, $P(\xi(F)=k)=\exp\{-M(F)\}M(F)^k/k!$, and for any disjoint sets $F_1,\dots,F_k$, $\xi(F_1),\dots,\xi(F_k)$ are independent, see e.g. \cite{Res87}.} The approximating process is governed by the intensity
\begin{equation}\label{emp_ref_dist}\hat{M}_n(x,s|R):= \frac{\frac1n\sum_{k=1}^n\sum_{\mathbf{\tilde{i}}:k\notin\mathbf{\tilde{i}}}\dum\left\{x_k=x,s_k^*=s,R_{\mathbf{\tilde{i}}k}=R,
D_{k\tilde{i}_10}=D_{\tilde{i}_1k0}=1\right\}}{\frac1{n^2}\sum_{k=1}^n\sum_{\mathbf{\tilde{i}}:k\notin\mathbf{\tilde{i}}}\dum\left\{R_{\mathbf{i}k}=R\right\}} \end{equation}
This empirical (Poisson) rate characterizes at what rate nodes $k$ with attributes $x_k$ and network attributes $s_k^*$ are part of the network neighborhood $\mathcal{N}_{\mathbf{i}}$ of node $i$ conditional on $R_{\mathbf{i}k}=R$. Note also that under Assumption \ref{asy_rate_ass} $\hat{M}_n$ is stochastically bounded as $n$ increases. Below we will refer to the population analog of $\hat{M}_n$ as the \emph{reference distribution} of the network. Clearly, this rate is determined endogenously in the network formation model. However as shown in Lemma \ref{app_independence_lem}, independence of taste shocks across nodes ensures that link payoffs for nodes in $\{i_1,\dots,i_D\}$ are conditionally independent of (appropriately defined) potential values of network statistics for their alters outside of that set so that we can nevertheless approximate probabilities for network events. Moreover we show in the next section that in the limit, $\hat{M}_n$ can be approximated by the solution of a fixed point problem.

We can now consider an event concerning the subnetwork among the nodes included in the multi-index $\mathbf{i}:=(i_1,\dots,i_D)'$, where we define the edge set $\mathbf{E}_{\mathbf{i}}:=\left[\mathbf{E}_{i_1},\dots,\mathbf{E}_{i_D}\right]$. To evaluate the probability of such an event, we now propose the following alternative sampling scheme:
\begin{itemize}
\item For each node $i$ in the multi-index $\mathbf{i}=(i_1,\dots,i_D)'$, a set of nodes $\mathcal{\tilde{N}}_i:=\left\{j_1,\dots,j_r\right\}$ with attributes and potential values $(x_j,s_j^*(R))_{j\in\mathcal{N}_i}$ is generated by a point process with Poisson intensity $\hat{M}_n(x,s|R)$.
\item For each node pair $(i,j)$ with $i\in\{i_1,\dots,i_D\}$ and $j\in\mathcal{N}_i\cup\{i_1,\dots,i_D\}\backslash\{i\}$ we also generate i.i.d. draws $\widetilde{MC}_i$ and $\tilde{\varepsilon}_{ij}$ from their respective marginal distributions.
\item The subnetwork $\mathbf{L_{E_{i}}}$ on $\mathbf{E_i}$ is then supported if for each $d=1,\dots,D$ and $j\in\mathcal{\tilde{N}}_{i_d}$
\[L_{i_dj}=\dum\left\{U^*(x_{i_d},x_{j};s_{i_d},s_{j}^*(R_{\mathbf{i}j}))+\tilde{\varepsilon}_{i_dj}\geq \widetilde{MC}_{i_d}\right\}D_{ji_d}(s_{i_d},R_{\mathbf{i}j}),\]
and $L_{i_dj}=0$ for all $j\notin\mathcal{\tilde{N}}_{i_d}$, where $s_{i_d}:=s(R_{\mathbf{i}i_d})$.
\end{itemize}

In particular, for every $j$ the network statistics $s_{j}^*$ are independent of whether there is a robust non-link between $i$ and $j$ under this process. Also, since the Poisson intensity $\hat{M}_n(x,s|R)$ is bounded as $n$ increases, so is the (random) size of $\mathcal{N}_i$.

Lemma \ref{app_sampling_rep_lem} below shows that this alternative representation can be used to obtain probabilities for any events on $\mathbf{E}_{\mathbf{i}}$ given the sigma-field $\mathcal{F}$. Specifically, we let  $\pi_n(A_{\mathbf{E}_{\mathbf{i}}}|\mathbf{x}_{\mathbf{i}})\equiv\pi_n(A_{\mathbf{E}_{\mathbf{i}}}|\mathbf{x}_{\mathbf{i}},\mathcal{F})$ denote the conditional probability that an event $A_{\mathbf{E}_{\mathbf{i}}}:=A(\mathbf{L}_{\mathbf{E}_{\mathbf{i}}})$ on $\mathbf{E}_{\mathbf{i}}$ is supported by a PSN given $\mathcal{F}$, and $\mathbf{x_i}$, and $\tilde{\pi}_n(A_{\mathbf{E}_{\mathbf{i}}}|\mathbf{x_i})\equiv\tilde{\pi}_n(A_{\mathbf{E}_{\mathbf{i}}}|\mathbf{x_i},\mathcal{F})$ its analog under the alternative sampling scheme. In particular, the approximation is valid conditional on any public signals that are used to select among multiple pairwise stable networks. For notational ease we generally suppress explicit conditioning on $\mathcal{F}$ wherever possible.

In the following, we first consider the event whether a particular configuration $\mathbf{\mathbf{L}_{i}}$ for the subnetwork among $i_1,\dots,i_D$ is supported by a pairwise stable network on all of $\{1,\dots,n\}$, and we call such a subnetwork event \emph{elementary}. We then let $\mathcal{A}_{\mathbf{E_i}}$ denote the set of events on $\mathbf{L_{E_i}},\mathbf{X}$ that are invariant to permutations of edges $(ij)\in\mathbf{E_i}$ for $j\notin\{i_1,\dots,i_D\}$, and that can be represented as the union of mutually exclusive elementary events regarding outcomes on the subnetwork $\mathbf{\mathbf{L}_{i}}$.

\begin{lem}\label{app_sampling_rep_lem} Suppose that Assumptions \ref{finite_supp_ass}-\ref{solution_ass} hold, and that there are no edge-specific interaction effects. Then for any $A_{\mathbf{E}_{\mathbf{i}}}\in\mathcal{A}_{\mathbf{E_i}}$
\[\pi_n(A_{\mathbf{E}_{\mathbf{i}}}|\mathbf{x}_{\mathbf{i}},\mathcal{F}) = \tilde{\pi}_n(A_{\mathbf{E}_{\mathbf{i}}}|\mathbf{x}_{\mathbf{i}},\mathcal{F})\left(1 + O(n^{-1})\right)\]
almost surely. More generally the analogous conclusion holds for any finite collection of events $A_{\mathbf{E}_{\mathbf{i}_1}},\dots,A_{\mathbf{E}_{\mathbf{i}_q}}$ and $D$-tuples $\mathbf{i}_1,\dots,\mathbf{i}_q$ with $q<\infty$ fixed.
\end{lem}

See the appendix for a proof. In interpreting this result one should note that $\hat{M}_n(x,s|R)$ is generally not $\mathcal{F}$-measurable - it may be possible to obtain the analogous conclusion for a finer conditioning set, but such a strengthening of the result will not be necessary for our analysis below.

The restriction to $\mathcal{A}_{\mathbf{E}_{\mathbf{i}}}$ allows for subnetwork events depending on the subnetwork among $i_1,\dots,i_D$ as well as the node-specific network attributes $s_{i_1}^*,\dots,s_{i_D}^*$ in an arbitrary fashion, however events in that class will generally not capture interdependencies between the network neighborhoods of the individual nodes $i_1,\dots,i_D$. We posit that this is a natural restriction on what network moments the researcher may be interested in for the model without edge-specific interaction effects. This condition on $A_{\mathbf{E_i}}$ yields a major simplification relative to the general case in which potential values for each node in $\mathcal{N}_{i_d}$ have to be evaluated at the relevant overlap with respect to $\mathbf{E_i}:=\mathbf{E}_{i_1}\cup\dots\mathbf{E}_{i_D}$ rather than $\mathbf{E}_{i_d}$ alone. In the presence of edge-specific effects, these interactions can no longer be ignored, and we discuss this more general case in the appendix.

The condition that the event $A_{\mathbf{E}_{\mathbf{i}}}$ not include multiple subnetwork configurations that may be simultaneously supported by a pairwise stable network implies that such an event can be represented in terms of the marginal distributions (rather than the joint distribution) for potential values $s_j^*(R)$ for each value $R$ of the relevant overlap. Dropping this requirement does not fundamentally change our theoretical analysis, but will pose additional practical problems for implementation: As shown by \cite{Men16} the marginal distributions for $s_j^*(R)$ can be estimated directly from the data, however the joint distribution across different values of $R$ would have to be inferred from theoretical fixed point conditions for our model. Alternatively, the Bonferroni inequality gives a conservative bound that can be evaluated from the marginal distributions alone.

\section{Aggregate State Variables}

\label{sec:agg_stat_vars}


The previous section gave a simplified representation of outcomes in pairwise stable networks in terms of local network neighborhoods and potential values for network statistics. In a pairwise stable network, the distribution of potential values for network statistics and node ``availability" in a network neighborhood are endogenously determined. In this section we show that this distribution can (to an asymptotic approximation) be characterized in terms of aggregate state variables, (a) the reference distribution $M(x,s|R)$ and (b) the inclusive value function $\Eta(x;s)$. Furthermore we show that these aggregate states are in turn determined by a fixed-point (equilibrium) condition across the entire network, which completes our asymptotic characterization of the network.

\subsection{Extreme-Value Approximations}

We first consider the limit of the conditional probability that certain link proposals are accepted. As \cite{Men16} showed, in sparse networks these limits are governed by extreme-value theory, and this section strengthens his result, deriving the convergence rate for these approximations as well as other extensions needed for the central limit theory in this paper.

For the following results, we focus on events regarding the link proposals by $i$, that is concerning edge sets of the form $\mathbf{E}_{i}:=\left[\begin{array}{ccc}i&\dots&i\\1&\dots&n\end{array}\right]$. The extension to neighborhoods of polyads $i_1,\dots,i_D$ will be straightforward in light of Lemma \ref{app_independence_lem}. Adapting the notation in Lemma \ref{psn_potential_val_lem}, we let $\mathbf{s}_{-i}:=(s_1,\dots,s_{i-1},s_{i+1},\dots,s_{n})'$ and
\[\mathcal{S}_{-i}^*(\mathbf{D_{E_i}}):=\cart_{j\neq i}\mathcal{S}_j^*\left(\mathbf{E_i},\mathbf{D_{E_i}},\left[ij,ji\right],[1,D_{ji}^*]\right)\]
be the potential values for a given set of proposals $\mathbf{D_{E_i}}$. We also use the notation $\mathbf{W}\mathbf{E}_{i}$ for the set of edges directed at $i$, where $\mathbf{W}$ is the matrix $\left[\begin{array}{cc}0&1\\1&0\end{array}\right]$.
Given $\mathbf{D_{E_i}}$ and $\mathbf{D}_{\mathbf{W}\mathbf{E}_{i}}$, we then let $\mathbf{L_{E_i}}=\mathbf{D_{E_i}}\odot\mathbf{D}_{\mathbf{W}\mathbf{E}_{i}}$ and $j_1,\dots,j_r\in\{1,\dots,n\}$ be the indices of nodes such that $L_{ij}=1$. To simplify notation, we also let $U^*_{ij}:=U^*(x_i,x_j;s_i,s_j,t_{ij})$ for $j=1,\dots,n$, where $s_i:=S(\mathbf{L}(\mathbf{D}^*,\mathbf{E}_1,\mathbf{D}_{\mathbf{E}_1});i)$.

Since links are determined simultaneously, it is necessary to consider joint probabilities of the form
\[\Phi(i,j_1,\dots,j_r) = \mathbb{P}\left(\left.\mathbf{D_{E_i}}\in\mathcal{D}_{\mathbf{E_i}}^*(\mathbf{s}_{-i})\right|\mathbf{X},\mathbf{s}_{-i}\in\mathcal{S}_{-i}^*(\mathbf{D_{E_i}}),
\mathbf{D}_{\mathbf{W}\mathbf{E}_{i}}\in\mathcal{D}_{\mathbf{W}\mathbf{E}_{i}}^*(\mathbf{D_{E_i}})\right)\]
Under the sparse network sequence in Assumption \ref{asy_rate_ass}, \cite{Men16} showed probabilities of that form converge to their extreme-value analogs, with joint probabilities across multiple links converging to products of edge-wise Logit-type probabilities. However for a derivation of the asymptotic distribution, we also have to ensure that the error from this extreme-value approximation does not contribute to first order as we take limits. The following lemma strengthens Lemma 6.2 in \cite{Men16}, showing that the additional tail condition on the c.d.f. of taste shifters in \ref{set_unobs_ass} (ii) suffices to guarantee convergence at a sufficiently fast rate.

\begin{lem}\label{ev_conv_rate_lem} Suppose that Assumptions \ref{surplus_bd_ass}-\ref{asy_rate_ass} hold. Then  
\[\left|n^{r/2}\Phi(i,j_1,\dots,j_r)- \frac{r!\prod_{s=1}^{r}\exp\{U^*_{ij_s}\}}
{\left(1+n^{-1/2}\sum_{j\notin\{j_1,\dots,j_r\}}D_{ji}^*(s)\exp\{U^*_{ij}\}\right)^{r+1}}\right|=o(n^{-1/2})\]
\end{lem}

See the appendix for a proof. The implied approximation to $\Phi(i,j_1,\dots,j_r)$ has several properties that help simplify the problem. For one, the joint probability over link acceptances to nodes $j_1,\dots,j_r$ is approximated by the product of the marginal probabilities for each of these edges and a term capturing the probability that all remaining available edges will not be chosen. Furthermore, each of these (limiting) marginal probabilities is of the Logit form, so that the set of available link opportunities enters each of these marginal probabilities with the inclusive value for the alternatives that are not accepted, $n^{-1/2}\sum_{j\notin\{j_1,\dots,j_r\}}D_{ji}^*(s)\exp\{U^*_{ij}\}$. Under Assumption \ref{asy_rate_ass} the number of available alternatives grows at a $n^{1/2}$ rates so that this statistic is approximated by the analogous average over all link opportunities, including those that are accepted.

For the sampling representation in Lemma \ref{app_sampling_rep_lem} we also need conditional probabilities of the form
\[\bar{\Phi}(i,j_1,\dots,j_r) = \mathbb{P}\left(\left.\mathbf{D_{E_i}}\in\mathcal{D}_{\mathbf{E_i}}^*(\mathbf{s}_{-i})\right|\mathbf{X},\mathbf{s}_{-i}\in\mathcal{S}_{-i}^*(\mathbf{D_{E_i}}),
\mathbf{D}_{\mathbf{W}\mathbf{E}_{i}}\in\mathcal{D}_{\mathbf{W}\mathbf{E}_{i}}^*(\mathbf{D_{E_i}}),D_{ij_10},\dots D_{ij_r0}=1\right)\]
where we defined $D_{ij0}:=\dum\left\{\bar{U}+\varepsilon_{ij}\geq MC_i\right\}$. From Lemma \ref{app_logit_ccp_lem} we can immediately conclude that
\begin{equation}\label{ubar_shift}\bar{\Phi}(i,j_1,\dots,j_r) = \exp\left\{\sum_{q=1}^r\left(U_{ij_q}^*-\bar{U}\right)\right\}\end{equation}
Hence the Logit structure in the limit gives us a simple adjustment factor to evaluate the probability of a combination of proposals conditional on a set of nodes being in the network neighborhood $\mathcal{N}_i$. The analogous conclusion holds for conditioning on $\bar{U}+\varepsilon_{ik_q}\geq MC_i$, $q=1,\dots r'$ for a subset of nodes $k_1,\dots,k_{r'}\in\{j_1,\dots,j_r\}$.

\subsection{Inclusive Values}

As discussed before, the asymptotic representation in Lemma \ref{ev_conv_rate_lem} implies that in the limit, conditional acceptance probabilities depend only on the systematic utilities of the alternatives chosen, and the inclusive value of the set of available link opportunities, which we define as
\[I_i^*:=n^{-1/2}\sum_{j=1}^nD_{ji}^*\exp\{U_{ij}^*\}\]
Most importantly, the composition and size of the set of link opportunities affects the conditional choice probabilities only through the inclusive value, so to an asymptotic approximation the inclusive value serves as a scalar parameter summarizing the systematic components of payoffs for the available options, similar to the case of single-agent discrete choice (\cite{Luc59}, \cite{McF74}) or matchign models (\cite{Dag94},\cite{Men13}).

\cite{Men16} establishes a conditional law of large numbers for the inclusive values which are sample averages over the characteristics of agents $j$ with $D_{ji}^*=1$, where the size of that set, $\sum_{j\neq i}D_{ji0}$ grows at a rate $\sqrt{n}$ for any PSN. The following was proven as Lemma 6.3 in \cite{Men16} and is re-stated here for completeness. 

\begin{lem}\label{fin_inclusive_conv} Suppose Assumptions \ref{finite_supp_ass}-\ref{solution_ass} hold. Then, (a) there exists a function $\hat{\Eta}_n^*(x,s)$ such that for any pairwise stable network, the resulting inclusive values satisfy
\[I_{i}^*-\hat{\Eta}_{n}^*(x_i,s_i)=o_p(1)\]
for each $i$ drawn from a uniform distribution over $\{1,\dots,n\}$. Furthermore, (b), if the weight functions $\omega(x,x',s,s')\geq0$ are bounded and form a Glivenko-Cantelli class in $(x,s)$, then
\[\sup_{x\in\mathcal{X}\\s\in\mathcal{S}}\frac1n\sum_{j=1}^{n}\omega(x,x_j,s,s_j)(I_j^*-\hat{\Eta}_n^*(x_j,s_j))=o_p(1)\]
\end{lem}


As shown in \cite{Men16}, this implies an (approximate) fixed-point condition for the inclusive value function $\Eta(x;s)$, where we define the mapping
\begin{eqnarray}
\nonumber\hat{\Psi}_{2n}[M,\Eta](x;s)&:=&\frac1n\sum_{i=1}^n\int\frac{s_{1i}\exp\{U^*(x,x_i;s,s_i,t_0) + U^*(x_i,x;s_i,s,t_0)-2\bar{U}\}}{1+\Eta(x_i;s_j)}M(x_i,s|R)dx_ids_i\\
\label{def_Psi2_hat}&=:&\frac1n\sum_{i=1}^n\tilde{\psi}_{2ni}[M,\Eta](x;s)
\end{eqnarray}

Letting $\hat{M}_{n}^*(x,s|R)$ denote the empirical distribution of endogenous network characteristics defined in (\ref{emp_ref_dist}), we then have the following:

\begin{lem}\label{fp_representation_lem} Suppose that Assumptions \ref{finite_supp_ass}-\ref{solution_ass} hold. Then the inclusive value function $\hat{\Eta}_n^*(x,s)$ resulting from a PSN has to satisfy the approximate fixed-point condition
\begin{equation}\label{Gamma_hat_fp}
\hat{\Eta}_n^*(x;s) = \hat{\Psi}_{2n}[\hat{M}_{n}^*,\hat{\Eta}_n^*](x;s)+o_p(1)
\end{equation}
where the remainder converges in probability uniformly in the arguments $x,s$.
\end{lem}

This was proven as Lemma 6.4 in \cite{Men16}. We do not derive the stochastic order of the approximation errors in Lemmas \ref{fin_inclusive_conv} and \ref{fp_representation_lem} at this point. But as part of the proofs for Lemma \ref{asy_bias_lem} and Theorem \ref{asy_Gauss_thm}, we find that under the conditions of our main results, $I_{i}^*-\hat{\Eta}_{n}^*(x_i,s_i)= O_o(n^{-1/4}$ and that the remainder term in (\ref{Gamma_hat_fp}) is of the order $O_p(n^{-1/2})$. Most importantly, both stochasticssL terms generally contribute to the asymptotic distribution of the network moment.

\subsection{Reference Distribution}


We next show that the representation result in Lemma \ref{app_sampling_rep_lem} implies that $\hat{M}_n(x,s|R)$ is sufficient to characterize the distribution of potential values for the nodes forming the links in the edge set $\mathbf{E}_k$.

\subsubsection*{Equilibrium Condition in Finite Network}

We first give the distribution of the r.h.s. variables in (\ref{emp_ref_dist}) to establish stochastic restrictions on the reference distribution. We define
\[\mathcal{L}(s,R_0;k,i):=\left\{\mathbf{L_{E_k}}: S(\mathbf{L_{E_k}},\mathbf{X};i)=s,R(\mathbf{E_k},\mathbf{L_{E_k}},\mathbf{X};i)=\mathbf{R_0}\right\}.\]
and similarly,
\[\mathcal{L}(R_0;k,i):=\left\{\mathbf{L_{E_k}}:R(\mathbf{E_k},\mathbf{L_{E_k}},\mathbf{X};i)=\mathbf{R_0}\right\}.\]
If we then let $\mathcal{L}_{k,i}^*$ be the set of subnetworks $(\mathbf{D_{E_k}},\mathbf{D_{E_{-k}}})$ that are supported by a pairwise stable network, we can use
\[A_{k,i}(s,R_0,x)=\left\{\mathcal{L}_{k,i}^*\cap
\mathcal{L}(s,R_0;k,i)\neq\emptyset,x_i= x,D_{ik0}=D_{ki0}=1\right\}\] to denote the event that $s_i^*=s$ and $R(\mathbf{E_k},\mathbf{L_{E_k}},\mathbf{X};i)=R_0$ are supported by a pairwise stable subnetwork on $\mathbf{E}_k$, and $x_i=x$. Similarly,
\[A_{k,i}(R_0)=\left\{\mathcal{L}_{k,i}^*\cap
\mathcal{L}(R_0;k,i)\neq\emptyset\right\}\]
In that notation,
\[\psi^*_{1ni}(s,R_0,x):=\sum_{k\neq i}\dum\left\{A_{k,i}(s,R_0,x)\right\}\]
and $\psi^*_{1ni}(R_0):=\sum_{k\neq i}\dum\left\{A_{k,i}(R_0)\right\}$
so that we can write
\begin{equation}\label{M_hat_upper_bound}\hat{M}_n(s,x|R_0)\leq \hat{\Psi}_{1n}^*(x,s|R_0)
\end{equation}
for each $(s,R_0,x)$, where
\[\hat{\Psi}_{1n}^*(x,s|R_0):=\frac{\frac1{n}\sum_{i=1}^n\psi_{1ni}^*(s,R_0,x)}{\frac1{n^2}\sum_{i=1}^n\psi_{1ni}^*(R_0)}\]

\subsubsection*{Equilibrium Condition under Sampling Representation}

We next define an analog of $\hat{\Psi}_{1n}^*$ under the sampling representation in Lemma \ref{app_sampling_rep_lem}, where we let $\tilde{\mathcal{N}}_i$ be the random network neighborhood of node $i$ given a reference distribution $M$ and let $\mathbf{E}_{i,\{j_1,\dots,j_r\}}$ denote the set of edges $(ij_1),\dots,(i,j_r)$. For notational simplicity, we identify $\tilde{\mathcal{N}}_i$ with the attribute values $(x_{j_1},s_{j_1},\dots,x_{j_r},s_{j_r})$ where we denote a typical draw $N_{i,\{j_1,\dots,j_r\}}$. We then denote
\begin{eqnarray}\nonumber\mathcal{L}(s,R_0;k,i,N_{i,\{j_1,\dots,j_r\}})&:=&\left\{\mathbf{L_{E_{i,\{j_1,\dots,j_r\}}}}: S(\mathbf{L_{E_{i,\{j_1,\dots,j_r\}}}},\mathbf{X};i)=s,\right.\\
\nonumber&&\left.R(\mathbf{L_{E_{i,\{j_1,\dots,j_r\}}}},\mathbf{X};i)
=\mathbf{R_0}\right\}
\end{eqnarray}
and let $\tilde{\mathcal{L}}_{k,i,\{j_1,\dots,j_r\}}^*$ be the set of networks on $E_{i,\{j_1,\dots,j_r\}}$ such that the links $L_{kj}$ are pairwise stable for $j\in\{j_1,\dots,j_r\}$ given potential values in $\mathcal{N}_k$. We also let
\[\tilde{A}_{k,i}(s,R_0,x)=\left\{\tilde{\mathcal{L}}_{k,i,\{j_1,\dots,j_r\}}^*\cap
\mathcal{L}(s,R_0;k,i,\{j_1,\dots,j_r\})\neq\emptyset,x_i= x,\tilde{D}_{ik0}=1\right\}\]
We can then define an analog to $\psi_{1ni}^*$ in terms of the sampling representation,
\begin{equation}\label{def_Psi1_hat}\tilde{\psi}_{1ni}[M](s,R_0,x):=\sum_{k\neq i}\dum\left\{\tilde{A}_{k,i}(s,R_0,x)\right\}.\end{equation}
and $\tilde{\psi}_{1ni}[M](R_0)$ in an analogous fashion. Note that the first argument expresses dependence of $\tilde{\psi}_{1ni}$ on the reference distribution $M$ from which the network neighborhood $\tilde{\mathcal{N}}_i$ was drawn.

Lemma \ref{app_ref_fp_lem} below establishes that (\ref{M_hat_upper_bound}) gives rise to an approximate fixed point condition
\[\hat{M}_n(s,R_0,x)\leq \bar{\Psi}_{1n}[\hat{M}_n,\hat{\Eta}_n](x,s|R_0)+o_p(1)\]
for a mapping of the form
\begin{eqnarray}
\nonumber \bar{\Psi}_{1n}[M,\Eta](x,s|R_0)&:=&\frac{\frac1n\sum_{i=1}^n\bar{\psi}_{1ni}[M,\Eta](s,R_0,x)}
{\frac1n\sum_{i=1}^n\bar{\psi}_{1ni}[M,\Eta](R_0)}
\end{eqnarray}
where the functions $\bar{\psi}_{1ni}$ derive from expectations of $\tilde{\psi}_{1ni}$ according to asymptotic probabilities of subnetwork events. Specifically, for a given realization of the network neighborhood $N_{i,\{j_1,\dots,j_r\}}$ we define
\begin{eqnarray}\nonumber\bar{\psi}_{1ni,j_1\dots j_r}[M,\Eta](\mathbf{L_{E_{i,\{j_1,\dots,j_r\}}}},N_{i,\{j_1,\dots,j_r\}},x):=\frac{\exp\{\bar{U}\}}{1+\Eta(x_i,s_i)}
\hspace{4.5cm}\textnormal{ }\\
\nonumber\textnormal{ }\hspace{5.5cm}\times\left[\prod_{q=1}^r
\frac{\exp\left\{D_{ij_q}U^*_{ij_q} + D_{j_qi}U^*_{j_qi} - 2\bar{U}\right\}}{(1+\Eta(x_i,s_i))^{D_{ij_q}}}M(s_{j_q},x|R_{j_q})\right]
\end{eqnarray}
which can be aggregated at the level of each node as
\begin{eqnarray}
\nonumber \bar{\psi}_{1ni}[M,\Eta](s,R_0,x) &:=&\sum_{r\geq0}\sum_{N_{i,\{j_1,\dots,j_r\}}}\sum_{\mathbf{L}\in\mathcal{L}_{i,j_1,\dots,j_r}}
\bar{\psi}_{1ni,j_1\dots j_r}[M,\Eta](\mathbf{L},N_{i,\{j_1,\dots,j_r\}},x)
\end{eqnarray}
where $\mathcal{L}_{i,j_1,\dots,j_r}:=\mathcal{L}(s,R_0;i,N_{i,\{j_1,\dots,j_r\}})$ and
$\bar{\psi}_{1ni}[M,\Eta](R_0)$ is defined analogously.

Sharp bounds for the reference distribution have to incorporate additional restrictions between various values of $s,R_0$. For a set of outcomes $\{(s_1,R_1),\dots,(s_K,R_K)\}$, let the set
\begin{eqnarray}\nonumber\mathcal{L}(\{(s_1,R_1),\dots,(s_K,R_K)\};i,\mathcal{N}_i)
&:=&\left\{\mathbf{L_{E_i}}:
L_{ij}=0\textnormal{ for all }j\notin\mathcal{N}_i,S(\mathbf{L_{E_i}}\mathbf{X};i)=s_k,\textnormal{ and }\frac{}{}\right.\\
\nonumber&&\left.\frac{}{}R(\mathbf{E_i},\mathbf{L_{E_i}},\mathbf{X};i)=\mathbf{R_k}\textnormal{ for some }k\leq K\right\}
\end{eqnarray}
The set of distributions that can then be characterized by a capacity
\begin{eqnarray}\nonumber\bar{\Psi}_{1n}[\hat{M}_n,\hat{\Eta}_n^*]\left(x,s_1,\dots,s_K\left|\bigcup_{k=1}^KR_k\right.\right)
\nonumber&=&\frac{\frac1n\sum_{i=1}^n\bar{\psi}_{1ni}[\hat{M}_n,\hat{\Eta}_n](\{(s_1,R_1),\dots,(s_K,R_K)\},x)}
{\frac1n\sum_{i=1}^n\bar{\psi}_{1ni}[\hat{M}_n,\hat{\Eta}_n](R_1,\dots,R_K)}
\end{eqnarray}
where
\begin{eqnarray}
\nonumber \bar{\psi}_{1ni}[\hat{M}_n,\hat{\Eta}_n](\{(s_1,R_1),\dots,(s_K,R_K)\},x)\hspace*{9cm}\textnormal{ }\\
\nonumber\textnormal{ }\hspace*{1cm}=
\sum_{r\geq0}\sum_{N_{i,\{j_1,\dots,j_r\}}}\sum_{\mathbf{L}\in\mathcal{L}_{K,i,j_1,\dots,j_r}}
\bar{\psi}_{1ni,j_1\dots j_r}[M,\Eta](\mathbf{L},N_{i,\{j_1,\dots,j_r\}},x)
\end{eqnarray}
where $\mathcal{L}_{K,i,j_1,\dots,j_r}:=\mathcal{L}((s_1,R_1),\dots,(s_K,R_K);i,N_{i,\{j_1,\dots,j_r\}})$. Taken together, these arguments imply that any reference distribution that results from a pairwise stable network must satisfy an approximate fixed point condition which we summarize in the following Lemma:

\begin{lem}\label{app_ref_fp_lem} Suppose that Assumptions \ref{finite_supp_ass}-\ref{solution_ass} hold and that $n\mathbb{P}(R_{ij}=R_0|D_{ij0}=D_{ji0}=1)\rightarrow\infty$ for each value of $R_0$. Then the reference distribution  $\hat{M}_n^*(x,s,R)$ resulting from a PSN has to satisfy the approximate fixed-point condition
\begin{equation}\label{M_hat_fp}
\hat{M}_n^*(x,s|R) = \bar{\Psi}_{1n}[\hat{M}_{n}^*,\hat{\Eta}_n^*](x,s|R)+o_P(1)
\end{equation}
for any $s,R_0$, and
\[\sum_{\{k=1,\dots,K\}}\hat{M}_n(s_k,R_k,x)\leq \bar{\Psi}_{1n}[\hat{M}_n]\left(x,s_1,\dots,s_K\left|\bigcup_{k=1}^KR_k\right.\right) + o_P(1)\]
where the remainder converges in probability uniformly in the arguments $x,s$.
\end{lem}
See the appendix for a proof.


%

\subsection{Limit of Fixed-Point Mapping}


Taken together, Lemmas \ref{fp_representation_lem} and \ref{app_ref_fp_lem} give a joint fixed point condition for the reference distribution $\hat{M}_n^*(x,s|R_0)$ and the inclusive value function $\hat{\Eta}_n^*(x,s)$, where we stack the mappings $\hat{\Psi}_n:=[\hat{\Psi}_{1n}',\hat{\Psi}_{2n}']'$, to write
\begin{eqnarray}
\nonumber \hat{M}_n^*&\leq&\hat{\Psi}_{1n}[\hat{M}_n^*,\hat{\Eta}_n^*] + o_p(1)\\
\label{agg_fp_condition} \hat{\Eta}_n^*&=&\hat{\Psi}_{2n}[\hat{M}_n^*,\hat{\Eta}_n^*] + o_p(1)
\end{eqnarray}
Under regularity conditions, we next show that the mapping $\hat{\Psi}_n[M,\Eta]$ and its fixed points converge to a proper limit $\Psi_0[M,\Eta]:=[\Psi_{10}[M,\Eta]',\Psi_{20}[M,\Eta]']'$, and its fixed points to the solutions of the limiting problem
\begin{eqnarray}
\nonumber M_0^*&\leq&\Psi_{10}[M_0^*,\Eta_0^*]\\
\label{agg_lim_fp_condition}\Eta_0^*&=&\Psi_{20}[M_0^*,\Eta_0^*]
\end{eqnarray}

In order to define stochastic convergence when the limit may remain stochastic (e.g. when pairwise stable networks are selected at random), we furthermore need to ensure that for each limiting point $\eta_0\in\Psi_0[\eta_0]$ there exists a sequence of values for the aggregate state variable $\hat{\eta}_n\in\hat{\Psi}_n[\hat{\eta}_n]$ such that $\hat{\eta}_n\rightarrow\eta_0$. To this end, we rely on the notion of regularity for the solutions of the fixed-point problem which correspond to standard local stability conditions in optimization theory (see e.g. chapter 9 in \cite{Lue69}, or chapter 3 in \cite{AFr90}): The \emph{contingent derivative} of $\Psi_0$ at $(\eta',\psi')'\in\textnormal{gph }\Phi_0\subset\mathcal{H}^2$ is defined as the set-valued mapping $D\Psi_0(\eta,\psi):\mathcal{H}\rightrightarrows\mathcal{H}$ such that for any $u\in\mathcal{H}$
\[v\in D\Psi_0(\eta,\psi)(u)\Leftrightarrow\liminf_{h\downarrow0,\\ u'\rightarrow u} d\left(v,\frac{\Psi_0(\eta+hu')-\psi}{h}\right)\]
where $d(a,B)$ is taken to be the distance of a point $a$ to a set $B$.\footnote{See Definition 5.1.1 and Proposition 5.1.4 in \cite{AFr90}} Note that if the correspondence $\Psi_0$ is singleton-valued and differentiable, the contingent derivative is also singleton-valued and equal to the derivative of the function $\Psi_0(\eta)$. The contingent derivative of $\Psi_0$ is \emph{surjective} at $\eta_0$ if the range of $D\Psi_0(\eta_0,\psi_0)$ is equal to $\mathcal{H}$. If the fixed point $\eta_0\in\Psi_0[\eta_0]$ is such that the contingent derivative  of $\Psi_0$ at $\eta_0$ is surjective, then we say that it is a \emph{regular fixed point} of the mapping.

We can now summarize the main assumptions on the fixed-point mappings $\hat{\Psi}_n$ and $\Psi_0$ for the aggregate state variables in the finite network and the limiting economy, respectively:


\begin{ass}\label{fp_map_ass} 
(i) The mapping $\Psi_0$ is compact and upper hemi-continuous in $\Eta,M$ for all $x\in\mathcal{X}$ and $S\subset\mathcal{R}$, and (ii) the core of $\Psi_0[\Eta,M]$ is nonempty, where the boundary points of the core of the capacity $\Psi_0[\Eta,M]$ is in some compact subset $\mathcal{U}\subset\Delta(\mathcal{X}\times\mathcal{R})$ for all values of $\Eta,M$. (iii) $\sup_{x,Z\subset\mathcal{R}}\left|\hat{\Psi}_n[M,\Eta](Z)-\Psi_0[M,\Eta](Z)\right|\rightarrow0$ uniformly in $\Eta\in\mathcal{G}$ and distributions $M\in\mathcal{U}$. (iv) Each point $(M_0,\Eta_0)$ satisfying $(M_0,\Eta_0)\in\Psi_0[M_0,\Eta_0]$ is a regular fixed point of $\Psi_0$. (v) The fixed-point mapping $\Psi_0[M,\Eta]$ is singleton-valued and continuously differentiable at each $\eta$.
\end{ass}

These high-level assumptions on the equilibrium mapping $\Psi_0$ have to be verified on a case by case basis. Existence of a fixed point for the limiting model as well as the law of large numbers require parts (i)-(iii) of this assumption, parts (iv) and (v) are only used to derive the asymptotic distribution.

Uniform convergence of $\hat{\Psi}_n$ with respect to $M$ in part (iii) is only stated only as a high-level condition in order to keep the result as general as possible. \cite{Men16} discussed primitive conditions and also showed that for some cases of applied interest - e.g. interactions through the degree distribution - the mapping $\Psi_0$ does not depend on the sampling distribution of types, in which case uniform convergence as in part (iii) trivially holds.


These high-level conditions are sufficient to ensure that the limiting model for the network formation problem is indeed well-defined. The following result was proven as Theorem 4.1 in \cite{Men16}:

\begin{thm}\textbf{(Fixed Point Existence)}\label{fp_existence_thm} Suppose that Assumptions \ref{surplus_bd_ass} and \ref{fp_map_ass} (i)-(ii) hold. Then the mapping $(\Eta,M)\rightrightarrows(\Psi_0,\Omega_0)[\Eta,M]$ has a fixed point.
\end{thm}

As shown in \cite{Men16}, the reference distribution can be estimated from observable network data under certain conditions, however the inclusive value function $\Eta^*(x;s)$ cannot. That said, we can give conditions that ensure that the inclusive value function $\Eta^*$ is uniquely defined given a reference distribution $M(x,s|R)$:

\begin{prp}\label{Psi_0_contraction_prp} Suppose that Assumptions \ref{surplus_bd_ass}-\ref{asy_rate_ass} hold. Then (i) for any given reference distribution $M^*(s|x)$ for which the network degree $s_{1i}$ satisfies $\mathbb{E}[s_{1i}|x_i]< B_s<\infty$ almost surely, the mapping $\log \Eta\mapsto \log \Psi_{20}[\Eta]$ is a contraction mapping with contraction constant $\lambda< \frac{B_s\exp\{2\bar{U}\}}{1+B_s\exp\{2\bar{U}\}}$. Moreover, (ii) the fixed points in (\ref{agg_lim_fp_condition}) are continuous functions that have bounded partial derivatives at least up to order $p$.
\end{prp}
The formal argument for this result closely parallels the proof of Theorem 3.1 in \cite{Men13} with contraction constant equal to $\frac{B_s\exp\{2\bar{U}\}}{1+B_s\exp\{2\bar{U}\}}$, a separate proof is therefore omitted.

%

Below we rely on these uniqueness results for conditional inference given $\hat{M}_{n}(x,s|R)$ to rule out multiplicity for unobserved aggregate state variables.

To summarize, this section establishes an approximate representation of aggregate states of pairwise stable networks that govern the process generating the network neighborhoods in the representation in Lemma \ref{app_sampling_rep_lem}. Those states are (a) the inclusive value function $\Eta^*$ characterizing the set of link opportunities available to the ``typical" node with certain attributes, and (b) the reference distribution $M^*$ of potential values for the (endogenous) payoff-relevant network statistics. We give a fixed-point characterization for these states, where the finite-network fixed-point mapping can be computed from observable data, and its solutions are shown to converge to fixed points of the limiting map. This structure will serve as the basis for estimation and inference, using the distribution theory given in the next section.

\section{Asymptotic Results}

\label{sec:struct_limit}

We can now state the main formal results of this paper, a law of large numbers and a central limit theorem for the network moments. Our approach is based on our finding (most importantly the conclusion of Lemma \ref{app_sampling_rep_lem}) that the behavior of network moments can be approximated by a representation in which the contributions from a finite empirical population (``drawing without replacement") are replaced by independent draws from a common distribution (``drawing with replacement"). Specifically, we approximate the array $\left(m_{i_1\dots i_D}\right)_{i_1\dots i_D}$ with the array $\left(\tilde{m}_{i_1\dots i_D}\right)_{i_1\dots i_D}$ resulting from drawing network statistics $s_j^*$ from the distribution $\hat{M}_n(s,R_0,x)$. Furthermore, $\hat{M}_n(s,R_0,x)$ is itself random and determined endogenously by the same network formation process. We therefore characterize the distribution of $\hat{m}_n$ and $\hat{M}_n(s,R_0,x)$ jointly by stacking the network moment and the fixed point mapping, which is an average across the $n$ nodes.

To that end, we let
\[\eta:=[M',\Eta']\]
denote a vector that consists of the reference distribution $M$ and inclusive value function $\Eta$. Under Assumption \ref{finite_supp_ass}, $\eta$ can be taken to be a finite-dimensional vector. By Lemmas \ref{fp_representation_lem} and \ref{app_ref_fp_lem} we can then write the moments of $\tilde{m}_{i_1\dots i_D}$ in more compact notation as a function of $\eta$,
\begin{eqnarray}
\nonumber m_0(\theta;\eta)&:=& \mathbb{E}_{\eta}\left[\tilde{m}_{i_1\dots i_D}(\theta)\right]
\end{eqnarray}
where $\mathbf{E}_{\eta}[\cdot]$ denotes the expectation for network outcomes with random network neighborhoods drawn according to a distribution governed by the aggregate state $\eta$. In particular, $m_0(\theta;\eta)$ may in general remain random in the limit if $\eta$ is stochastic. In the infinite-population model, that aggregate state satisfies the fixed-point condition
\begin{eqnarray}
\nonumber \eta_0&\in&\Psi_0(\eta_0)
\end{eqnarray}
where $\Psi_0(\eta)$ was given as a function of the expectations $\mathbb{E}\left[\tilde{\psi}_i(\eta)\right]$ in Section \ref{sec:agg_stat_vars}.

Similarly, we can write the sampling approximation to the empirical moment as
\begin{eqnarray}
\nonumber \tilde{m}_n(\theta;\hat{\eta}_n)&:=&\left[\binom{n}{D}p_n\right]^{-1}\sum_{i_1<\dots<i_D}\tilde{m}_{i_1\dots i_D}(\mathbf{L},\mathbf{X};\theta)
\end{eqnarray}
where $\hat{\eta}_n$ denotes the value of the relevant aggregate state variables for the finite network which determine the sampling representation according to which $\tilde{m}_{i_1\dots i_D}(\cdot)$ is drawn. Furthermore, as shown in Section \ref{sec:agg_stat_vars}, $\hat{\eta}_n$ must satsify the fixed point condition
\begin{eqnarray}
\nonumber \hat{\eta}_n&\in&\hat{\Psi}_n(\hat{\eta}_n)
\end{eqnarray}
where $\tilde{\Psi}_n(\eta)$ was defined as a function of averages $\frac1n\sum_{i=1}^n\tilde{\psi}_i(\eta)$.

Hence, the asymptotic distribution of $\tilde{m}_n(\theta)$ can be derived in terms of properties of $\left(\hat{m}_n(\theta;\eta)-m_0(\theta;\eta),\hat{\Psi}_n(\eta)-\Psi_0(\eta)\right)$ as a function of $\eta$. Specifically, we can expand
\begin{eqnarray}
\nonumber \hat{m}_n(\theta) &=& \tilde{m}_n(\theta;\hat{\eta}_n) + o_P(n^{-1})\\
\nonumber&=&m_0(\theta;\eta_0) + \left[m_0(\theta;\hat{\eta}_n) - m_0(\theta;\eta_0)\right] + \left[\tilde{m}_n(\theta;\hat{\eta}_n)
-m_0(\theta;\hat{\eta}_n)\right] + o_P(n^{-1})
\end{eqnarray}
Similarly, for the fixed point equation
\begin{eqnarray}
\nonumber \hat{\eta}_n&\in&\tilde{\Psi}_n(\hat{\eta}_n)\\
\nonumber&=&\Psi_0(\eta_0) + \left[\Psi_0(\hat{\eta}_n) - \Psi_0(\eta_0)\right] + \left[\tilde{\Psi}_n(\hat{\eta}_n)
-\Psi_0(\hat{\eta}_n)\right]
\end{eqnarray}

In particular, we can show that given the set of fixed points $\mathcal{H}_0:=\left\{\eta_0:\eta_0\in\Psi_0[\eta_0]\right\}$, the distance $d(\hat{\eta}_n,\mathcal{H}_0)$ converges to zero almost surely. Under the additional regularity conditions on the moment functions we can therefore obtain a law of large numbers for the network moments, which is proven in the appendix.

\begin{thm}\label{primitive_lln_thm} Suppose that Assumptions \ref{moment_ass}-\ref{asy_rate_ass} and \ref{solution_ass} (i) hold. In addition assume that the fixed-point mapping $\Psi_0$ satisfies Assumption \ref{fp_map_ass} (i)-(iii). Then we can find a $\mathcal{F}$-measurable sequence $\eta_n$ such that $\eta_n\in\Psi_0(\eta_n)$ and
\[\hat{m}_n(\theta) - m_0(\theta;\eta_n)\stackrel{p}{\rightarrow}0\]
\end{thm}


Given this law of large numbers with respect to the set of possible limits for the network formation model, we next provide an asymptotic distribution theory under somewhat more stringent conditions. The central limit theorem for network moments assumes stronger conditions on the fixed-point mapping, \ref{fp_map_ass} (iv)-(v) which were not needed for the law of large numbers. In particular, the CLT will require $\Psi_0[\eta]$ to be singleton valued, which may either hold directly given the nature of strategic interaction effects, or under additional constraints on the mechanism for selecting among multiple pairwise stable networks.


Since the limiting model need not have a unique solution and the observed network may be generated according to a stochastic selection rule, we formulate an asymptotic theory that is conditional on the selected equilibrium. Specifically, we assume that agents to coordinate on a pairwise stable network via a public signal with a selection mechanism satisfying Assumption \ref{solution_ass} (ii)-(iii). In particular the remaining results will assume that selection among multiple stable networks is based on a signal that is independent of the payoff-relevant variables $x_i,MC_i,\varepsilon_{ij}$.

We first give a result on the rate of the asymptotic bias. Let $\bar{m}_n(\theta;\eta):=\mathbb{E}_n[m_{i_1\dots i_D}(\theta;\eta)]$ and $\bar{\Psi}_n(\theta;\eta):=\mathbb{E}_n[m_{i_1\dots i_D}(\theta;\eta)]$, where $\mathbb{E}_n[\cdot]$ denotes the expectation under the respective distributions of $\tilde{m}_{i_1\dots i_D}$ and $\tilde{\psi}_{i_1}$ for a network of size $n$. We then define
\begin{eqnarray}
\nonumber B_{m,n}(\theta;\eta)&:=&\sqrt{n}(\bar{m}_n(\theta;\eta) - m_0(\theta;\eta))\\
\nonumber B_{\Psi,n}(\theta;\eta)&:=&\sqrt{n}(\bar{\Psi}_n(\theta;\eta) - \Psi_0(\theta;\eta))
\end{eqnarray}
and let $B_n:=[B_{m,n}',B_{\Psi,n}']'$.

\begin{lem}\label{asy_bias_lem} Assume the network formation model in Assumptions \ref{moment_ass}-\ref{solution_ass}, and \ref{fp_map_ass}. Furthermore suppose that the network event $\mathcal{A}_{i_1\dots i_D}$ implies that $L_{i_1i_s}=1$ for at least one $s\in\{2,\dots,D\}$. Then $B(\theta;\eta):=\lim_nB_n(\theta;\eta)$ exists and is finite almost surely.
\end{lem}

This bias arises from the approximation of conditional link acceptance probabilities with the inclusive value function and effectively consists of two separate contributions which can be addressed separately. For one, the probabilities given in Lemma \ref{app_logit_ccp_lem} are in terms of the inclusive value of the link opportunity set \emph{excluding} the contribution of the link proposals that were accepted. This source of bias can immediately be removed by evaluating conditional link probabilities at the suitably adjusted inclusive value, resulting in a correction of the order $n^{-1/2}$. The second bias contribution due to the nonlinearity in conditional link acceptance probabilities could in principle be estimated analytically, using the asymptotic moments of $n^{1/4}(I_{i}^*-\Eta^*(x_i,s_i))$, or via resampling. However establishing the asymptotic validity of either procedure is beyond the scope of this paper, and we will be left for future research.

In order to establish a central limit theorem, we approximate the exact, finite-network distribution of the network moment by that of an exchangeable array. The finite-agent network is generally not jointly exchangeable under pairwise stability, instead links are formed simultaneously given realized payoffs for finitely many nodes. Nevertheless we can construct a sequence of jointly exchangeable arrays that approximate the finite-agent moment sufficiently closely. We show how to define that approximating sequence on a common probability space, allowing to establish convergence in distribution that is mixing with respect to the selected fixed point $\eta_0$.

Then, by an Aldous-Hoover representation any jointly exchangeable array can be represented as a function
\[\tilde{Y}_{i_1\dots i_D} = h\left(\alpha^{(0)},\alpha_{i_1}^{(1)},\dots,\alpha_{i_D}^{(1)},\alpha_{\{i_1,i_2\}}^{(2)},\dots,\alpha_{\{i_1,\dots, i_D\}}^{(D)}\right)\]
of i.i.d. uniform random variables $\alpha^{(0)},\alpha_{i_1}^{(1)},\dots,\alpha_{i_D}^{(1)},\alpha_{\{i_1,i_2\}}^{(2)},\dots,\alpha_{\{i_1,\dots, i_D\}}^{(D)}$ indexed by the subsets of $\{i_1,\dots,i_D\}$. We therefore can analyze the $D$-adic mean of $\tilde{Y}_{i_1\dots i_D}$ in terms of its Hoeffding decomposition, i.e. in terms of a conditional expectations projection on nested subsets of the Aldous-Hoover factors $\alpha^{(0)},\alpha_{i_1}^{(1)},\dots,\alpha_{i_D}^{(1)},\alpha_{\{i_1,i_2\}}^{(2)},\dots,\alpha_{\{i_1,\dots, i_D\}}^{(D)}$. Asymptotic normality is then established by showing that the leading terms of that Hoeffding decomposition correspond to projections on single Aldous-Hoover factors which dominate the respective contributions of the terms depending on two or more factors at a time.

Given these constructions, we can state our main result as follows:

\begin{thm}\label{asy_Gauss_thm} Suppose Assumptions \ref{moment_ass}-\ref{solution_ass} and \ref{fp_map_ass} hold. Furthermore we assume that the network event $\mathcal{A}_{i_1\dots i_D}$ is such that $L_{i_1i_s}=1$ for at least one $s\in\{2,\dots,D\}$ and $\binom{n}{D}p_n\rightarrow\infty$. We then have
\[\sqrt{n}\left(M'V M\right)^{-1/2}\left(\hat{m}_n(\theta) - \mathbb{E}\left[\left.\tilde{m}_{i_1\dots i_D}(\theta) \right|\mathcal{F}\right]\right)-M'B\stackrel{d}{\rightarrow}N(0,\mathbf{I}_q)\]
mixing in $\mathcal{F}$, where $\eta_0$ is $\mathcal{F}$-measurable,
\[M:=\left[\mathbf{I},\nabla_{\eta}m_0(\theta;\eta_0)(\mathbf{I}-\nabla_{\eta}\Psi_0(\eta_0))^{-1}\right]'\]
and $V$ is the conditional asymptotic variance matrix of $[\hat{m}_n(\theta,\eta_0),\hat{\Psi}_n(\eta_0)]$ given $\eta_0$.
\end{thm}

See the appendix for a proof. Note that Assumption \ref{fp_map_ass} (v) requires that $\Psi_0[\eta]$ is singleton-valued, so the derivative of the fixed-point mapping is well-defined.

Since this result is obtained from a statistical theory that treats nodes in the (large) network as random, we can interpret the central limit theorem as a first-order approximation to the sampling error in network moments relative to their large-population counterparts. Since the main result in \cite{Men16} allows to characterize the mapping from payoff parameters to those limiting population moments, Theorem \ref{asy_Gauss_thm} can therefore serve as the basis for an inferential theory relying on these asymptotic representations. In particular with appropriate corrections for the asymptotic bias in Lemma \ref{asy_bias_lem}, this result can be used to construct hypothesis tests or confidence intervals for structural parameters of the network formation model in an otherwise conventional manner. When the limiting distribution is not unique, inference will be conditional on $\eta_0$ in a completely analogous fashion to the approach in \cite{Men12}. In future versions of this paper, we will also provide formulae to correct for the asymptotic bias and variance for $\hat{m}_n(\theta)$.


\section{Conclusion}

This paper develops an asymptotic distribution theory for network formation models which allow for strategic interaction effects. Our results are based on large-network asymptotics, i.e. approximations that become more accurate as the number of agents (nodes) in the network increases. Tractable approximations to probabilities of network events (most importantly conditional link frequencies) for our framework were first derived by \cite{Men16}. Our analysis in the present paper gives a law of large numbers and a (conditional) central limit theorem for a class of network moments which make it possible to develop estimation and inference procedures based on those approximations. For certain cases of interest these approximations are available in closed form and can therefore be implemented without relying on simulation methods.

The limiting model is characterized in terms of stable subnetworks on random network neighborhoods that are generated from a common distribution of nodes and their (endogenous) network attributes. This approximation has some interesting qualitative features. For one, any ``long-range" interdependence between link formation decisions is entirely captured by the aggregate state variables $\Eta^*$ and $M^*$, whereas ``short-range" interdependence can be characterized through the edge-level response to a given realization of potential values of payoff-relevant network variables. This dichotomy - eliminating any additional ``medium-range," indirect effects in the limit - is a result of symmetry (exchangeability) of a form that is implicitly assumed for typical economic models of networks. Non-uniqueness of the pairwise stable network in the finite economy may still manifest itself in the limiting distribution, both as ``global" multiplicity corresponding to different solutions for the equilibrium conditions for aggregate state variables, as well as ``local" multiplicity of stable outcomes in a network neighborhood.


\bibliographystyle{econometrica}
\bibliography{mybibnew}
\footnotesize
\appendix

\section{Edge-Specific Interaction Effects}

\label{sec:edge_spec_samp_app}

This appendix generalizes the sampling representation from Lemma \ref{app_sampling_rep_lem} to the case of edge-specific interaction effects when strategic effects may be jointly determined jointly by subsets of the network neighborhood of $\{i_1,\dots,i_D\}$ that are larger than in the case of node-specific interaction effects.

To define the expanded network neighborhoods, we let
\[\mathcal{T}_{q}:=\left\{T(\mathbf{L}_q,\mathbf{X};i,j):\mathbf{L}_q\in\{0,1\}^{q(q-1)/2}\right\}\] denote the set of values for the statistic $T_{ij}$ that can be generated on a network of $q$ nodes including $i,j$. We also write
\[\bar{U}_{q}:=\sup_{t_{12}\in\mathcal{T}_{q}}\sup_{x_1,x_2,s_1,s_2}U(x_1,x_2;s_1,s_2,t_{12})\]
for the largest value of the systematic part of marginal utility that can be attained on $\mathcal{T}_{q}$. We then let
\[\bar{v}(q):=\min\left\{v:\sup_{x_1,x_2,s_1,s_2}U(x_1,x_2;s_1,s_2,T(\mathbf{L}_q,\mathbf{X};i,j))=\bar{U}_q,\|\mathbf{L}_q\|=v \right\}\]
be the smallest numbers of edges on a network of size $q$ that attains $\bar{U}_q$, and $\bar{r}$ denote the smallest integer that attains the upper bound for $\bar{U}_q$ with respect to $q$,
\[\bar{r}:=\min\left\{r\in\mathbb{Z}:\sup_{v\in\mathbb{Z}}\bar{U}_{rv}\geq \sup_{q,v\in\mathbb{Z}}\bar{U}_{qv}\right\}\]
While the approach outlined here will only be practical when $\bar{r}$ is fairly small, definitions and notation will cover the general case of any finite value.


For any integers $r,p$ we define the $p,v$-neighborhood of nodes $i_1,\dots,i_r$ as
\[\mathcal{N}_{i_1\dots i_r;v}^{(p)}:=\left\{k_1,\dots,k_p:\sum_{s=1}^r\sum_{q=1}^pD_{i_sk_q;v0}D_{k_qi_s;r0}\geq \bar{v}(r+p)-\frac{r(r-1)}2\right\}\]
where
\[D_{ki;r0}:=\dum\left\{\bar{U}_{r}+\varepsilon_{ki}\geq MC_k\right\}\]
That is, $\mathcal{N}_{i_1\dots i_r;v}^{(p)}$ is the set of $p$ distinct nodes $k$ that are contained in a $p$-tuple $(k_1,\dots,k_p)$ such that at least $\bar{v}(r+p)$ links among $(i_1,\dots,i_r,k_1,\dots,k_p)$ can be pairwise stable under some configuration of the subnetwork among these nodes, assuming that $i_1,\dots,i_r$ are fully connected. For such a $p$-tuple $(k_1,\dots,k_p)$, we also denote the relevant overlap with the $r$-tuple $(i_1,\dots,i_r)$ with
\[R_{i_1\dots i_r;k_1\dots k_p}:=R\left(\mathbf{E}_{i_1\dots i_r},\mathbf{L}_{\mathbf{E}_{i_1\dots i_r}}(\mathbf{D}_0,\mathbf{D_{\mathbf{E}_{i_1\dots i_r}}});k_1,\dots,k_p\right)\]

As in the baseline case, the probability of $\mathbf{A_{E_i}}$ depends on the global network structure through the empirical (Poisson) rates at which groups of nodes $\mathbf{k}:=(k_1,\dots,k_p)$ with attributes $x_{k_1},\dots,x_{k_p}$ and potential values for network attributes $s_{k_1}^*(R_{i_1\dots i_r;k_1\dots k_p}),\dots,s_{k_p}^*(R_{i_1\dots i_r;k_1\dots k_p})$ are jointly selected into the network neighborhood of the tuple $\mathbf{i}:=(i_1,\dots,i_r)$. To characterize these Poisson rates for different values of $r,p$, we extend our definition of the reference distribution as follows
\begin{equation}
\hat{M}_{rp,n}(x_1,\dots,x_p,s_1,\dots,s_p|R):=
\frac{\binom{n}{p}^{-1}p_n^{-1}\sum_{\mathbf{k}}\sum_{\mathbf{k}\cap\mathbf{i}=\emptyset}\dum\{A_{\mathbf{i}\mathbf{k}}(x_1,\dots,x_p,s_1,\dots,s_p,R,\bar{v})\}}
{\binom{n}{p}^{-1}\sum_{\mathbf{k}}\sum_{\mathbf{k}\cap\mathbf{i}=\emptyset}\dum\left\{A_{\mathbf{ik}}(R)\right\}}
\end{equation}
where
\begin{eqnarray}\nonumber A_{\mathbf{i}\mathbf{k}}(x_1,\dots,x_p,s_1,\dots,s_p,R,\bar{v})&:=&\left\{x_{k_1}=x_1,\dots,x_{k_p}=x_p,
s_{k_1}^*=s_1,\dots,s_{k_p}^*=s_p,
R_{i_1\dots i_r;k_1\dots k_p}=R,\frac{}{}\right.\\
\nonumber&&\left.\frac{}{}\sum_{s=1}^r\sum_{q=1}^pD_{k_qi_s0}=\bar{v}(r+p)-\frac{r(r-1)}2\right\},\\
\nonumber\mathbf{A}_{\mathbf{i}\mathbf{k}}(R)&:=&\left\{R_{i_1\dots i_r;k_1\dots k_p}=R\right\},
\end{eqnarray}
and $p_n$ is a normalizing sequence that depends on $\bar{v}(r+p)-\frac{r(r-1)}2$ and the rate at which $\bar{U}_{r}$ grows in $n$. To approximate the probability of an event $\mathbf{A_{E_i}}$ we then generate a network neighborhood $\tilde{\mathcal{N}}_{\mathbf{i}}$ of $\mathbf{i}$ at random by drawing sets of nodes that are individually (or jointly) acceptable to some subset of nodes in $\{i_1,\dots,i_D\}$ under some configuration of that subnetwork. Specifically,

\begin{itemize}

\item For each subset $\{i_1,\dots,i_r\}\subset \{i_1,\dots,i_D\}$ of size $r=1,\dots,\bar{r}-1$ and $p=1,\dots,\bar{r}-r$ we then generate the degree-$p$ neighborhood of $\{i_1,\dots,i_r\}$, $\tilde{\mathcal{N}}_{i_1\dots i_r}^{(p)}$ independently where potential values in $\tilde{\mathcal{N}}_{i_1\dots i_r}^{(p)}$ are with respect to the edge set $\bigcup_{s=1}^r\mathbf{E}_{i_s}$. Specifically, the set $\tilde{\mathcal{N}}_{i_1\dots i_r}^{(p)}$ consists of $p$-tuples with attributes and potential values \[\left((x_j,s_j^*(R),t_{ij}^*(R)):i\in\{i_1,\dots,i_r\},j\in\{j_1,\dots,j_p\}\in\tilde{\mathcal{N}}_{i_1\dots i_r}^{(p)}\right)\]
    that are the realizations of a point process with Poisson intensity $\hat{M}_{rp,n}(x_1,\dots,x_p,s_1,\dots,s_p|R)$.
\item For each node pair $(i,j)$ with $i\in\{i_1,\dots,i_D\}$ and  $j\in\left(\{i_1,\dots,i_D\}\cup\left(\bigcup_{r,p}\bigcup_{i\in\{i_1,\dots,i_r\}}\mathcal{N}_\{i_1\dots\i_r\}^{(p)}\right)\right)\backslash\{i\}$  we also generate i.i.d. draws $\widetilde{MC}_i$ and $\tilde{\varepsilon}_{ij}$ from their respective marginal distributions.
\item Give these network neighborhoods, we let the indicator
    \[L_{ik;i_1\dots i_r,k_1\dots k_p}^{(p)}=\dum\left\{U^*(x_{i},x_{k};s_{i},s_{k}^*(R_{i_1\dots i_r;k_1\dots k_p}),t_{ik}^*(R_{i_1\dots i_r;k_1\dots k_p}))+\tilde{\varepsilon}_{ik}\geq \widetilde{MC}_{i}\right\}D_{ki}(s_{i},R_{i_1\dots i_r;k_1\dots k_p})\]
    denote the event that the link $ik$ is supported on the neighborhood $\tilde{\mathcal{N}}_{i_1\dots i_r}^{(p)}$.
\item The subnetwork $\mathbf{L_{E_{i_1}\cup\dots E_{i_{D+1}}}}$ on $E_{i_1}\cup\dots E_{i_{D+1}}$ is then supported if for each $d=1,\dots,D$ and $k\in\bigcup_{r,p}\bigcup_{i\in\{i_1\dots,i_r\}}\tilde{\mathcal{N}}_{i_1\dots i_r}^{(p)}$
    \[L_{i_dk}=\max_{r\leq\bar{r}}\max_{i\in\{i_1\dots i_r\}}\max_{k\in\{k_1\dots,k_p\}}L_{ik;i_1\dots i_r,k_1\dots k_p}^{(p)},\]
    and $L_{i_dk}=0$ for all $k\notin\bigcup_{r,p}\bigcup_{i\in\{i_1\dots,i_r\}}\tilde{\mathcal{N}}_{i_1\dots i_r}^{(p)}$.
\end{itemize}

Under Assumption \ref{asy_rate_ass}, the size of each neighborhood $|\tilde{\mathcal{N}}_{i_1\dots i_r}^{(p)}|$ is random, but stochastically bounded as $n$ increases. By an argument analogous to that in the proof of Lemma \ref{app_sampling_rep_lem}, the probability of ties across $\{i_1,\dots,i_D\}$ and the neighborhoods $\mathcal{N}_{i_1\dots i_r}^{(p)}$ for $r=1,\dots,\bar{r}$ and $p=1,\dots,\bar{r}-1$ goes to zero at a $\frac1n$ rate. In particular the probability that the event $\mathbf{A_{E_i}}$ is supported by a pairwise stable network on the full set of nodes $\{1,\dots,n\}$ is also approximated up to the order $\frac1n$ by the probability of a pairwise stable subnetwork on $\{i_1,\dots,i_D\}$ given the network neighborhood $\tilde{\mathcal{N}}_{\mathbf{i}}$.

While this extension substantially complicates notation, we assert - without rigorous proof - that the main arguments for the baseline case continue to go through with minor modifications.

\section{Proofs for Section \ref{sec:potential_values_sec}}

\subsection{Proof of Lemma \ref{app_independence_lem}}

By construction, the potential values given $\mathbf{D_{E_i}}$ are fully determined by the ``complementary" edge set
\[\mathbf{E_i}^c:=
\left[\begin{array}{ccc}(i_{D+1},\dots,i_n)&\otimes&\iota_n\\ \iota_{n-D}&\otimes&(1,\dots,n)\end{array}\right]\]
For a given proposal subnetwork $\mathbf{D_{E_i}}$, the pairwise stability conditions for proposals on $\mathbf{E_i}^c$ are fully determined by $x_1,\dots,x_n$ together with $MC_{i_{D+1}},\dots, MC_{i_n}$ as well as $\varepsilon_{i_{D+1}1},\dots,\varepsilon_{i_{n}n}$. On the other hand, the proposals on $\mathbf{E_i}$ are determined by $x_1,\dots,x_n$ together with $MC_{i_1},\dots, MC_{i_D}$ as well as $\varepsilon_{i_{1}1},\dots,\varepsilon_{i_{D}n}$. In particular, there is no overlap between the taste shocks determining pairwise stability of $\mathbf{D_{E_i}}$ and $\mathbf{D_{E_i^c}}$, respectively. Since the taste shocks $MC_1,\dots,MC_n$ as well as $\varepsilon_{12},\dots,\varepsilon_{(n-1)n}$ are i.i.d. draws from their respective distributions by Assumption \ref{set_unobs_ass}, this establishes the main conclusion of the lemma. The second conclusion follows immediately from the observation that the taste shocks $\varepsilon_{k_1l_1},MC_{k_1},\dots,\varepsilon_{k_ql_q},MC_{k_q}$ are independent of those determining $\mathbf{D_{E_i}}$.\qed

\subsection{Proof of Lemma \ref{app_invariance_lem}}

Consider first a permutation $\tau$ such that for $j,k$, $\tau(j)=k$ and $\tau(k)=j$, but $\tau(i)=i$ for any $i\neq j,k$. To establish invariance, we now compare the distribution of potential values $\mathcal{S}_{j}^*(\mathbf{E}_1,\mathbf{D}_{\mathbf{E}_1},\mathbf{E}_2,\mathbf{D}_{\mathbf{E}_2})$ to that of their permuted analogs $\mathcal{S}_{\tau(j)}^*(\mathbf{E}_1,\mathbf{D}_{\mathbf{E}_1},\mathbf{E}_2^{\tau},\mathbf{D}_{\mathbf{E}_2})$ under $\tau$.

We establish the conclusion by induction. For the start of induction, we can immediately see that the conclusion of the Lemma holds for $r=0$, i.e. $\{\mathbf{E}_1\}=\emptyset$ since node attributes and taste shocks are identically distributed, and payoffs are therefore jointly exchangeable. Therefore we only need to establish the inductive step where an edge $ij$ is added to $\mathbf{E}_1$.

Specifically, suppose that invariance with respect to the permutation $\tau$ holds for $r-1$ with the edge set $\mathbf{E_{1,r-1}}$ and any values of $\mathbf{D}_{\mathbf{E}_{1,r-1}}\in\{0,1\}^{r-1}$. We now add the directed edge $i_r$ and $j_r$ to that set,
\[\mathbf{E_{1,r}}:=\left(\begin{array}{cccc}i_1&\cdots&i_{r-1}&i_r\\j_1&\cdots&j_{r-1}&j_r\end{array}\right),\]
and let $\mathbf{D}_{\mathbf{E}_{1,r}}:=(D_{i_1j_1},\dots,D_{i_rj_r})\in\{0,1\}^r$. Given the inductive hypothesis, it now suffices to show that for any fixed values $\mathbf{D}_{\mathbf{E}_{1,r-1}}\in\{0,1\}^{r-1}$, the distribution of potential values is invariant to the permutation $\tau$ for either value of $D_{i_rj_r}\in\{0,1\}$.

To this end, we need to distinguish whether or not there may be indirect ``interference" effects from a change to $D_{i_rj_r}$ on the potential values $s_j^*,s_k^*$ that are supported by a pairwise stable network given $\mathbf{D}_{\mathbf{E}_{1,r-1}}\in\{0,1\}^{r-1}$. Following \cite{Leu16}, we say that a link $ij$ is \emph{not robust} if there exist values of $s_1,s_2,t_{12}$ and $s_1',s_2',t_{12}'$ such that $U^*(x_i,x_j;s_1,s_2,t_{12})+\varepsilon_{ij}\geq MC_i$ and $U^*(x_j,x_i;s_2,s_1,t_{21})+\varepsilon_{ji}\geq MC_j$, but also either
$U^*(x_i,x_j;s_1',s_2',t_{12}')+\varepsilon_{ij}< MC_i$ or $U^*(x_j,x_i;s_2',s_1',t_{21}')+\varepsilon_{ji}< MC_j$. That is, a link is not robust given realized payoffs if there exist one configuration of values of $s_i,s_j,t_{ij}$ such that $L_{ij}=1$ is pairwise stable, and another such that it is not.

To make this argument precise, we now characterize events regarding whether a switch of $D_{i_rj_r}$ triggers a chain of non-robust link formation decisions given the network $L^*$ that was pairwise stable given the restrictions on the edge set $\mathbf{E}_{1,r-1}$. For any $l\notin\{i_r,j_r\}$, let $A_{kl}$ denote an indicator for the event that there exists a proposal network $\mathbf{D}^*\in \mathcal{D}^*(\mathbf{E_{1,r-1}})$ given the link proposals $\mathbf{D}_{\mathbf{E_{1,r-1}}}$ such that for the resulting network $\mathbf{L}^*:=\mathbf{L}(\mathbf{D}^*)$,
\[\dum\{U_{kl}(\mathbf{L}^*+\{i_rj_r\})\geq MC_k,U_{lk}(\mathbf{L}^*+\{i_rj_r\})-MC_l\}\neq
\dum\{U_{kl}(\mathbf{L}^*-\{i_rj_r\})\geq MC_k,U_{lk}(\mathbf{L}^*-\{i_rj_r\})-MC_l\}\]
In words, $A_{kl}$ is an indicator for whether a change to the link $L_{i_rj_r}$ changes whether the link $jl$ is pairwise stable.

We then let $B_{j}(q;F)$ be an indicator for the event that there exists a proposal network $D^*\in \mathcal{D}^*(\mathbf{E_{1,r-1}},\mathbf{D_{E_{1,r-1}}})$ such that $D_{j_ri_r}^*=1$, $\sum_{k,l} A_{jl}=q$ and the conditional empirical distribution of $x_l,s^*_l,t^*_{jl}$ given $A_{jl}=1$ is equal to $F$. By Assumptions \ref{finite_supp_ass} (ii) and \ref{asy_rate_ass} (ii), $q$ is stochastically bounded. We then define $\mathcal{B}_{j,i_rj_r}$ as the sigma-field generated by $\{B_{j}(q;F):q=0,1,\dots,n-1,\hspace{0.3cm} F\textnormal{ is a c.d.f.}\}$. We define $\mathcal{B}_{k,i_rj_r}$ analogously.

Now conditional on $j$ and $k$ having the same relevant overlap with $\mathbf{E_{1,r}}$ for some permutation $\tau$, \[R(\mathbf{E_1}^{\tau},\mathbf{L}_{\mathbf{E}_1}(\mathbf{D}_0,\mathbf{D_{E_1}}),\mathbf{X}^{\tau};j)
=R(\mathbf{E_1},\mathbf{L}_{\mathbf{E}_1}(\mathbf{D}_0,\mathbf{D_{E_1}}),\mathbf{X};k),\] we have by the inductive hypothesis that for each $q$ and $F$, $B_{j}(q;F)$ and $B_{k}(q;F)$ have the same probability. We then distinguish all possible cases whether or not a change to $D_{i_rj_r}$ triggers a chain of adjustments through $j$ or $k$, or neither, corresponding to the events $\mathcal{B}_{j,i_rj_r}(q_j,F_j)$ and $\mathcal{B}_{k,i_rj_r}(q_k,F_k)$ for all combinations of $(q_j,F_j)$ and $q_k,F_k$. If we can show that conditional on a partition in terms of these events, the distribution of network statistics is the same for nodes $j$ and $k$, then conditional invariance given $x_i,x_j$ and equal relevant supports follows from the law of total probability.

Specifically we first consider the (potentially overlapping) events
\[C(q_1,q_2,F_1,F_2):=\{B_j(q_1,F_1)\cap B_k(q_2,F_2)\}\]
as the arguments $q_1,q_2$ and $F_1,F_2$ vary freely. We first consider the case $q_2=q_1$ and $F_2=F_1$: by definition of the event $C(q_1,q_1,F_1,F_1)$, a switch in $D_{i_1j_1}$ triggers $q_1$ chains of adjustments starting at node $j$. Any such chain may reach node $k$ after a change to $l$, and depending only on the signs of $U^*(x_l,x_k;s_l,s_k,t_{lk})+\varepsilon_{lk}-MC_l$ and $U^*(x_k,x_l;s_k,s_l,t_{lk})+\varepsilon_{kl}-MC_k$. Similarly, the chain reaches $j$ after that same change depending only on the signs of
$U^*(x_l,x_j;s_l,s_j,t_{lj})+\varepsilon_{lj}-MC_l$ and $U^*(x_j,x_l;s_j,s_l,t_{lj})+\varepsilon_{jl}-MC_j$. By inspection, conditional on $\mathcal{B}_{\mathbf{E_1},j}$ and $\mathcal{B}_{\mathbf{E_1}^{\tau},k}$ the distributions of $(\varepsilon_{lk},\varepsilon_{kl},MC_k,x_k)$ and $(\varepsilon_{lj},\varepsilon_{jl},MC_j,x_j)$ are the same, so that for $s_j=s_k$, $t_{lk}=t_{lj}$, and $t_{jl}=t_{kl}$, the probability that the chain proceeds to $j$ is equal to that of the chain proceeding to $k$. Finally, conditional on  $R(\mathbf{E_1}^{\tau},\mathbf{L}_{\mathbf{E}_1}(\mathbf{D}_0,\mathbf{D_{E_1}}),\mathbf{X}^{\tau};k)
=R(\mathbf{E_1},\mathbf{L}_{\mathbf{E}_1}(\mathbf{D}_0,\mathbf{D_{E_1}}),\mathbf{X};j)$, the resulting changes on the sets $\mathcal{S}_{j}^*(\mathbf{E}_1,\mathbf{D}_{\mathbf{E}_1},\mathbf{E}_2,\mathbf{D}_{\mathbf{E}_2})$ and $\mathcal{S}_{\tau(j)}^*(\mathbf{E}_1,\mathbf{D}_{\mathbf{E}_1},\mathbf{E}_2^{\tau},\mathbf{D}_{\mathbf{E}_2})$ are the same.

Hence, given $R(\mathbf{E_1}^{\tau},\mathbf{L}_{\mathbf{E}_1}(\mathbf{D}_0,\mathbf{D_{E_1}}),\mathbf{X}^{\tau};k)
=R(\mathbf{E}_1,\mathbf{L}_{\mathbf{E}_1}(\mathbf{D}_0,\mathbf{D_{E_1}}),\mathbf{X};j)$, the conditional probability that $\mathcal{S}_{j}^*(\mathbf{E}_1,\mathbf{D}_{\mathbf{E}_1},\mathbf{E}_2,\mathbf{D}_{\mathbf{E}_2})$ is supported given $C(q_1,q_1;F_1,F_1)$ is equal to that of $\mathcal{S}_{k}^*(\mathbf{E}_1,\mathbf{D}_{\mathbf{E}_1},\mathbf{E}_2^{\tau},\mathbf{D}_{\mathbf{E}_2})$ being supported given $C(q_1,q_1;F_1,F_1)$.

For the case $(q_1,F_1)\neq (q_2,F_2)$, we can follow an analogous line of argument to conclude that conditional on $R(\mathbf{E_1}^{\tau},\mathbf{L}_{\mathbf{E}_1}(\mathbf{D}_0,\mathbf{D_{E_1}}),\mathbf{X}^{\tau};k)
=R(\mathbf{E}_1,\mathbf{L}_{\mathbf{E}_1}(\mathbf{D}_0,\mathbf{D_{E_1}}),\mathbf{X};j)$, the conditional probability that $\mathcal{S}_{j}^*(\mathbf{E}_1,\mathbf{D}_{\mathbf{E}_1},\mathbf{E}_2,\mathbf{D}_{\mathbf{E}_2})$ is supported given $C(q_1,q_2;F_1,F_2)$ is equal to that of $\mathcal{S}_{k}^*(\mathbf{E}_1,\mathbf{D}_{\mathbf{E}_1},\mathbf{E}_2^{\tau},\mathbf{D}_{\mathbf{E}_2})$ being supported given $C(q_2,q_1;F_2,F_1)$. Since the events $C(q_2,q_1;F_2,F_1)$ and $C(q_2,q_1;F_2,F_1)$ have equal probability, it follows from the law of total probability that the conditional distributions are the same given the union $C(q_1,q_2;F_1,F_2)\cup C(q_2,q_1;F_2,F_1)$. A similar line of reasoning gives us the analogous conclusion conditional on any intersections of ``symmetrized" events $C(q_1,q_2;F_1,F_2)\cup C(q_2,q_1;F_2,F_1),\dots,C(q_R,q_{R+1};F_R,F_{R+1})\cup C(q_{R+1},q_R;F_{R+1},F_R)$, allowing us to construct a partition of the event $R(\mathbf{E_1}^{\tau},\mathbf{L}_{\mathbf{E}_1}(\mathbf{D}_0,\mathbf{D_{E_1}}),\mathbf{X}^{\tau};k)
=R(\mathbf{E}_1,\mathbf{L}_{\mathbf{E}_1}(\mathbf{D}_0,\mathbf{D_{E_1}}),\mathbf{X};j)$ such that invariance with respect to $\tau$ holds conditional on each element in that partition. By the law of total probability, this completes the inductive step from $r-1$ to $r$.

This establishes the conclusion of the Lemma for a binary permutation $\tau$ such that for $j,k$, $\tau(j)=k$ and $\tau(k)=j$, but $\tau(i)=i$ for any $i\neq j,k$. Since an arbitrary permutation can be generated by a sequence of pairwise swaps of indices, this is sufficient to establish the first conclusion of the Lemma.

To establish the second part of the claim we can adapt the same argument to the level of an edge, requiring only notational adjustments. Since the argument is otherwise completely analogous to the case of node-level statistics, we do not give it here explicitly.\qed

For future reference we now also state without proof an immediate generalization of the result for the leading case to the joint distribution for multiple nodes.

\begin{lem}Suppose Assumptions \ref{finite_supp_ass}-\ref{asy_rate_ass} hold. Then for any $i_1,\dots,i_D=1,\dots,n$, $\mathbf{E}$, and $\mathbf{D}_{\mathbf{E}}$ we have
\begin{eqnarray}
\nonumber \left(\mathcal{S}_{i_1}^*(\mathbf{E}_1,\mathbf{D}_{\mathbf{E}_1},\mathbf{E}_2,\mathbf{D}_{\mathbf{E}_2}),\right.&\dots,&\left. \mathcal{S}_{i_D}^*(\mathbf{E}_1,\mathbf{D}_{\mathbf{E}_1},\mathbf{E}_2,\mathbf{D}_{\mathbf{E}_2})\right)\\
\nonumber&\stackrel{d}{=}&\left(\mathcal{S}_{\tau(i_1)}^*(\mathbf{E}_1,\mathbf{D}_{\mathbf{E}_1},\mathbf{E}_2^{\tau},\mathbf{D}_{\mathbf{E}_2}),
\dots, \mathcal{S}_{\tau(i_D)}^*(\mathbf{E}_1,\mathbf{D}_{\mathbf{E}_1},\mathbf{E}_2^{\tau},\mathbf{D}_{\mathbf{E}_2})\right)
\end{eqnarray}
conditional on
\begin{eqnarray}\nonumber R(\mathbf{E_1}^{\tau},\mathbf{L}_{\mathbf{E}_1}(\mathbf{D}_0,\mathbf{D_{E_1}}),\mathbf{X}^{\tau};\tau(i_1))
&=&R(\mathbf{E}_1,\mathbf{L}_{\mathbf{E}_1}(\mathbf{D}_0,\mathbf{D_{E_1}}),\mathbf{X};i_1)\\
\nonumber&\vdots&\\
\nonumber R(\mathbf{E_1}^{\tau},\mathbf{L}_{\mathbf{E}_1}(\mathbf{D}_0,\mathbf{D_{E_1}}),\mathbf{X}^{\tau};\tau(i_D))
&=&R(\mathbf{E}_1,\mathbf{L}_{\mathbf{E}_1}(\mathbf{D}_0,\mathbf{D_{E_1}}),\mathbf{X};i_D),
\end{eqnarray}
$x_{\tau(i_1)}=x_{i_1},\dots,x_{\tau(i_D)}=x_{i_D}$, and $\mathcal{F}$. The analogous conclusion holds for edge-level statistics among $i_1,\dots,i_D$ as well.
\end{lem}

It can be seen from the proof of the main result that this generalization of the leading case, where we only need to account for the rates at which any chain of non-robust edges links back to either node $i_1,\dots,i_D$ relative to its image under the permutation $\tau$.

\subsection*{Proof of Lemma \ref{app_sampling_rep_lem}}

First note that a Poisson process with an intensity equal to the empirical distribution over $n$ values of the variable can equivalently be represented as a Poisson process on $\{1,\dots,n\}$ with uniform mean measure. In contrast, $\mathcal{N}_i$ is generated by drawing from $\{1,\dots,n\}\backslash\{i_1,\dots,i_D\}$ uniformly at random and without replacement  conditional on realizations $(x_j,s_j^*)$ for $j\notin\{i_1,\dots,i_D\}$. Hence the point measure associated with $\mathcal{N}_i$ has the same distribution as that associated with $\tilde{\mathcal{N}}_i$ conditional on $\tilde{\mathcal{N}}_i$ assigning zero weight to the values corresponding to $i_1,\dots,i_D$ and attaining no multiplicities greater than one for any $i\in\{1,\dots,n\}$.

Furthermore, if the neighborhoods $\mathcal{N}_{i_1},\dots,\mathcal{N}_{i_D}$ are disjoint, the relevant overlap for any  $j\in\mathcal{N}_{i_d}$ with $i_{d'}, d'\neq d$ corresponds to the empty graph. Hence in the absence of ties in $\mathcal{N}_{i_1},\dots,\mathcal{N}_{i_D}$ and $\tilde{\mathcal{N}}_{i_1},\dots,\tilde{\mathcal{N}}_{i_D}$, respectively, the relevant overlap of any node $j$ with $\{i_1,\dots,i_D\}$ can without loss of generality be taken to be $R_{i_dj}$ if $j\in\mathcal{N}_{i_d}$.

We therefore next verify that the probability of ties across $\mathcal{N}_{i_1},\dots,\mathcal{N}_{i_D}$ and $\{i_1,\dots,i_D\}$ is bounded by a sequence of the order $O(n^{-1})$: Denoting $\kappa_n:=n\mathbb{P}\left(D_{ij0}=D_{ji0}=1\right)$, Assumption \ref{asy_rate_ass} immediately implies that $\kappa_n$ is bounded. We then let $q_{i_d}:=|\tilde{\mathcal{N}}_{i_d}|$ and $Q:=\sum_{d=1}^Dq_{i_d}$, we have that conditional on $Q$, the probability that the $q$th draw for $\mathcal{N}_{i_d}$ constitutes a tie is at most $\frac{D+Q-1}{n}$, so the conditional probability of a tie given $Q$ is bounded from above by
\[\bar{\mu}_n:=1 - \left(1-\frac{D+Q-1}{n}\right)^Q= \frac{D+Q-1}{n} + o\left(n^{-1}\right)\]
for any finite $Q$.

We can then use the law of iterated expectations with respect to $Q$ together with the fact that the Poisson distribution for $Q$ has exponential tails to conclude that the sampling representation approximates probability of the network event up to an error $\frac{2D(1+\kappa_n)}{n}$. Specifically, for a truncation argument we can choose $Q_n=O(\log n)$ such that $P(Q\geq Q_n)\leq\frac{2D(1+\kappa_n)}{n}$. We can then bound
\begin{eqnarray}
\nonumber \mathbb{E}\left[\left(1-\frac{Q}n\right)^Q\right] &=& \sum_{q=0}^{\infty}\left(1-\frac{q}n\right)^q\frac{(D\kappa_n)^qe^{-D\kappa_n}}{q!}\\
\nonumber &\geq&\sum_{q=0}^{Q_n}\left(1-\frac{q}n\right)\left(1-\frac{Q_n}n\right)^{q-1}\frac{(D\kappa_n)^qe^{-D\kappa_n}}{q!}\\
\nonumber &=&\left(1-\frac{\mathbb{E}[Q]}n\right)\left(1-P(Q\geq Q_n) - o_p(n^{-1})\right)\\
\nonumber &\geq&1-\frac{2D(1+\kappa_n)}{n} + o_p(n^{-1})
\end{eqnarray}
which is of the desired order of magnitude. Note in particular that this bound does not depend on the empirical distribution of realizations for $(x_1,s_1^*),\dots,(x_n,s_n^*)$ and therefore also holds conditional on $\mathcal{F}$. Furthermore, by Assumption \ref{asy_rate_ass}, the size of $\mathcal{N}_{i_d}$ is also bounded for each $d=1,\dots,D$ so that by an analogous argument the probability of ties among $\mathcal{N}_{i_1},\dots,\mathcal{N}_{i_D}$ is also of the order $O(n^{-1})$ with the same bounding constants.

We next argue that conditional on the event $\mathbf{B}_{\mathbf{i}}$ of no ties in the sets $\mathcal{N}_{i_1},\dots,\mathcal{N}_{i_D},\{i_1,\dots,i_D\}$ and no ties in
$\tilde{\mathcal{N}}_{i_1},\dots,\tilde{\mathcal{N}}_{i_D},\{i_1,\dots,i_D\}$, the probability of the subnetwork event on $\{i_1,\dots,i_D\}$ on the $n$-agent network is the same as that generated by the sampling process. To this end, consider any node $j\in\mathcal{N}_i$ in the original network. It then follows from Lemma \ref{app_invariance_lem} that conditional on $R_{j}$, $(x_j,s_j^*)$ has the same distribution as $(x_j,s_{\tilde{j}}^*(R_{j}),t_{i\tilde{j}}^*(R_{j}))$ for any node $\tilde{j}\in\mathcal{\tilde{N}}_i$. Moreover, Lemma \ref{app_invariance_lem} also implies that conditional on $\mathcal{F}$ drawing from $(x_j,s_{\tilde{j}}^*(R_{j}),t_{i\tilde{j}}^*(R_{j}))_{j=1}^n$ with replacement results in the same conditional probabilities as drawing from the distribution $\hat{M}_n(x,s|R_j)$.

By Lemma \ref{app_independence_lem}, for all $i\in\{i_1,\dots,i_D\}$ and $j\in\mathcal{N}_{\mathbf{i}}$, $\varepsilon_{ij},MC_i$ are furthermore independent of $s_j^*(R)$ for each $R$ conditional on $\mathcal{F}$. In particular, proposals
\[D_{ji}^*(s,R_{\mathbf{i}j}):=\dum\left\{U^*(x_j,x_i;s_j^*(R_{\mathbf{i}j}))+\varepsilon_{ji}\geq MC_j\right\}\]
are fully determined by initial taste shocks, and potential values $s_{j}^*(R_{\mathbf{i}j})$ for each $s\in\mathcal{S}$ and therefore also independent of $\varepsilon_{ij},MC_i$.

Since the event $\mathbf{A_{E_i}}$ was assumed to be the union of disjoint elementary events regarding the subnetwork $\mathbf{\mathbf{L}_{i}}$, these steps establish that $\pi_n(\mathbf{A_{E_i}}|\mathbf{x}_{\mathbf{i}},\mathbf{B_i},\mathcal{F})=\tilde{\pi}_n(\mathbf{A_{E_i}}|\mathbf{x}_{\mathbf{i}},\mathbf{B_i},\mathcal{F})$ and that $\mathbb{P}(\mathbf{B_i}^c|\mathcal{F})=O(n^{-1})$, where $\mathbf{B_i}^c$ denotes the complement of $\mathbf{B}_{\mathbf{i}}$. By the law of total probability,
\[|\pi_n(\mathbf{A_{E_i}}|\mathbf{x}_{\mathbf{i}},\mathcal{F})-\tilde{\pi}_n(\mathbf{A_{E_i}}|\mathbf{x}_{\mathbf{i}},\mathcal{F})|\leq |\pi_n(\mathbf{A_{E_i}}|\mathbf{x}_{\mathbf{i}},\mathbf{B_i}^c,\mathcal{F})-\tilde{\pi}_n(\mathbf{A_{E_i}}|\mathbf{x}_{\mathbf{i}},\mathbf{B_i}^c,\mathcal{F})|
\mathbb{P}(\mathbf{B_i}^c|\mathcal{F})\]
By assumption, the event $\mathbf{A_{E_i}}$ is invariant to permutations of edges $(ij)\in\mathbf{E_i}$ for $j\notin\{i_1,\dots,i_D\}$, so that $\pi_n(\mathbf{A_{E_i}}|\mathbf{x}_{\mathbf{i}},\mathbf{B_i}^c,\mathcal{F})=\pi_n(\mathbf{A_{E_i}}|\mathbf{x}_{\mathbf{i}},\mathcal{F})$ and $\tilde{\pi}_n(\mathbf{A_{E_i}}|\mathbf{x}_{\mathbf{i}},\mathbf{B_i}^c,\mathcal{F})=\tilde{\pi}_n(\mathbf{A_{E_i}}|\mathbf{x}_{\mathbf{i}},\mathcal{F})$ so that $\frac{\pi_n(\mathbf{A_{E_i}}|\mathbf{x}_{\mathbf{i}},\mathbf{B_i}^c,\mathcal{F})}{\tilde{\pi}_n(\mathbf{A_{E_i}}|\mathbf{x}_{\mathbf{i}},\mathcal{F})}$ is also bounded. By the triangle inequality it therefore follows that
\[\frac{|\pi_n(\mathbf{A_{E_i}}|\mathbf{x}_{\mathbf{i}},\mathcal{F})-\tilde{\pi}_n(\mathbf{A_{E_i}}|\mathbf{x}_{\mathbf{i}},\mathcal{F})|}{\tilde{\pi}_n(\mathbf{A_{E_i}}|\mathbf{x}_{\mathbf{i}},\mathcal{F})}=O(n^{-1})\]
establishing the main claim. The generalization to finite collections of events is immediate\qed

\section{Proofs for Section \ref{sec:agg_stat_vars}}

Before establishing the rate of convergence to an extreme value distribution in Lemma \ref{ev_conv_rate_lem}, we first re-state the result from \cite{Men16} establishing that the limits for conditional link acceptance probabilities essentially correspond to their analogs under the assumption of independent extreme-value type-I taste shifters:

\begin{lem}\label{app_logit_ccp_lem} Suppose that Assumptions \ref{surplus_bd_ass}-\ref{asy_rate_ass} hold. Then as $n\rightarrow\infty$,
\begin{equation}\label{ccp_limit} \left|n^{r/2}\Phi(i,j_1,\dots,j_r)- \frac{r!\prod_{s=1}^{r}\exp\{U^*_{ij_s}\}}
{\left(1+n^{-1/2}\sum_{i=1}^nD_{ji}^*(s)\exp\{U^*_{ij}\}\right)^{r+1}}\right|\rightarrow 0\end{equation}
for any $r=0,1,2,\dots$.
\end{lem}
This is Lemma 6.2 in \cite{Men16}, and we include the proof below for easier reference.

\subsection*{Proof of Lemma \ref{app_logit_ccp_lem}} This result is a generalization of Lemma B.1 in \cite{Men13}. We therefore refer to the proof of that result for some of the intermediate technical steps below.

Denote the event $\mathcal{B}_i:=\left\{\mathbf{X},\mathbf{s}_{-i}\in\mathcal{S}_{-i}^*(\mathbf{D_{E_i}}),
\mathbf{D_{E_{-i}}}\in\mathcal{D}_{\mathbf{{E_{-i}}}}^*(\mathbf{D_{E_i}})\right\}$. We also let $J=\lceil n^{1/2}\rceil$ as in Assumption \ref{asy_rate_ass}, and $j_{r+1},\dots,j_n$ be an enumeration of the indices in $\{1,\dots,n\}\backslash\{j_1,\dots,j_r\}$. Then by independence of $\eta_{i1},\dots,\eta_{iN}$,
\begin{eqnarray}
\nonumber J^r\Phi(i,j_1,\dots,j_r)&=&J^rP(U_{ij_1}\geq MC_i,\dots,U_{ij_r}\geq MC_i,U_{ij_{r+1}}<MC_i,\dots,U_{ij_{J}}<MC_i|\mathcal{B}_i)\\
\nonumber&=&J^r\int\left(\prod_{q=1}^r P(U_{ij_q}\geq \sigma s|\mathcal{B}_i)\right)\left(\prod_{q=r+1}^nP(U_{ij_q}<\sigma s|\mathcal{B}_i)^{D_{j_qi}}\right)JG(s)^{J-1}g(s)ds\\
\nonumber&=&J^r\int \left(\prod_{q=1}^r(1-G(s-\sigma^{-1}U^*_{ij_q}))\right)\left(\prod_{q=r+1}^nG(s-\sigma^{-1}U^*_{ij_q})^{D_{j_qi}}\right)
JG(s)^{J-1}g(s)ds\\
\nonumber&=&\int \left(\prod_{q=1}^rJ(1-G(s-\sigma^{-1}U^*_{ij_q}))\right)J\frac{g(s)}{G(s)}\\
\label{app_ccp_eqn}&&\times\exp\left\{J\log G(s) + \frac1J\sum_{q=r+1}^nJ\log G(s-\sigma^{-1}U^*_{ij_q})D_{j_qi}\right\}ds
\end{eqnarray}

Now let $b_J:=G^{-1}\left(1-\frac1J\right)$ and $a_J=a(b_J)$, where $a(\cdot)$ is the auxiliary function in Assumption \ref{set_unobs_ass} (ii). By Assumption \ref{asy_rate_ass} (iii), $\sigma =\frac{1}{a(b_J)}$, so that a change of variables $s=a_Jt + b_J$ yields
\begin{eqnarray}\label{ccp_lem_change_vars}\nonumber J^r\Phi(i,j_1,\dots,j_r)&=&\int \left(\prod_{q=1}^rJ(1-G(b_J + a_J(t-U^*_{ij_q})))\right)J\frac{a_Jg(b_J + a_Jt)}{G(b_J + a_Jt)}\\
\nonumber &&\times\exp\left\{J\log G(b_J + a_Jt) + \frac1J\sum_{q=r+1}^nJ\log G(b_J + a_J(t-U^*_{ij_q}))D_{ji}\right\}dt\end{eqnarray}

By Assumption \ref{set_unobs_ass} (ii), $J(1-G(b_J+a_Jt))\rightarrow e^{-t}$ and
\[Ja_Jg(b_J + a_Jt)=Ja(b_J)g(b_J+a(b_J)t)=a(b_J)\frac{1-G(b_J+a_Jt)}{a(b_J+a_Jt)(1-G(b_J))}\rightarrow e^{-t}\]
where the last step uses Lemma 1.3 in \cite{Res87}. Also, following steps analogous to the proof of Lemma B.1 in \cite{Men13}, we can take limits and obtain
\begin{eqnarray}
\nonumber \prod_{q=1}^rJ(1-G(b_J+a_J(t-U^*_{ij_q})))&\rightarrow&\exp\left\{-rt + \sum_{q=1}^rU^*_{ij_q}\right\}\\
\nonumber J\log G(b_J + a_J(t-U^*_{ij_q}))&\rightarrow&-e^{-t}\exp\{U^*_{ij_q}\}
\end{eqnarray}
Combining the different components, we can take the limit of the integrand in (\ref{ccp_lem_change_vars}),
\begin{eqnarray}\nonumber R_J(t)&:=&\left(\prod_{q=1}^rJ(1-G(b_J + a_J(t-U^*_{ij_q})))\right)J\frac{a_Jg(b_J + a_Jt)}{G(b_J + a_Jt)}\\
\nonumber&&\times\exp\left\{J\log G(b_J + a_Jt) + \frac1J\sum_{q=r+1}^nJ\log G(b_J + a_J(t-U^*_{ij_q}))D_{j_qi}\right\}\\
\label{logit_lem_appx}&=&\exp\left\{-e^{-t}\left(1+\frac1J\sum_{q=r+1}^{n}D_{j_qi}\exp\{U^*_{ij_q}\}\right)-(r+1)t + \sum_{q=1}^rU^*_{ij_q}\right\}
+o(1)\end{eqnarray}
for all $t\in\mathbb{R}$. Using the same argument as in the proof of Lemma B.1 in \cite{Men13}, pointwise convergence and boundedness of the integrand under Assumption \ref{surplus_bd_ass} imply convergence of the integral by dominated convergence, so that we obtain
\begin{eqnarray}\nonumber J^r\Phi(i,j_1,\dots,j_r|\mathbf{z}_i^*)&\rightarrow&\int_{-\infty}^{\infty}\exp\left\{-e^{-t}\left(1+\frac1J\sum_{q=r+1}^{n}
D_{j_qi}\exp\{U^*_{ij_q}\}\right)-(r+1)t + \sum_{q=1}^rU^*_{ij_q}\right\}dt\\
\nonumber&=&\int_{-\infty}^{0}\exp\left\{s\left(1+\frac1J\sum_{q=r+1}^{n}D_{j_qi}\exp\{U^*_{ij_q}\}\right)+ \sum_{q=1}^rU^*_{ij_q}\right\}s^{r}ds\\
\nonumber &=&\frac{r!\exp\{\sum_{q=1}^rU^*_{ij_q}\}}{\left(1+\frac1J\sum_{q=r+1}^nD_{j_qi}\exp\left\{U^*_{ij_q}\right\}\right)^{r+1}}
\end{eqnarray}
where the first step uses a change of variables $s=-e^{-t}$, and the last step can be obtained recursively via integration by parts. Furthermore, if $\frac{r}J\rightarrow0$, boundedness of the systematic parts from Assumption \ref{surplus_bd_ass} implies that \[\left|\frac1J\sum_{j=1}^JD_{ji}\exp\left\{U^*_{ij}\right\}-\frac1J\sum_{q=r+1}^nD_{j_qi}\exp\left\{U^*_{ik_q}\right\}\right|\rightarrow0\]
so that
\[J^r\Phi(i,j_1,\dots,j_r)\rightarrow \frac{r!\prod_{q=0}^r\exp\{U^*_{ij_q}\}}{\left(1+\frac1J\sum_{j=1}^nD_{ji}\exp\left\{U^*_{ij}\right\}\right)^{r+1}}\]
which completes the proof\qed

To establish the leading terms in the asymptotic distribution for generalized subgraph counts below, we also need to show that the limiting distribution conditional on $MC_i$ is strictly decreasing in $MC_i$: Defining
\[\Phi(i,j_1,\dots,j_r|\mu) = \mathbb{P}\left(\left.\mathbf{D_{E_i}}\in\mathcal{D}_{\mathbf{E_i}}^*(\mathbf{s}_{-i})\right|\mathbf{X},\mathbf{s}_{-i}\in\mathcal{S}_{-i}^*(\mathbf{D_{E_i}}),
\mathbf{D_{E_{-i}}}\in\mathcal{D}_{\mathbf{{E_{-i}}}}^*(\mathbf{D_{E_i}}),MC_i-b_J/a_J=\mu\right)\]
we have the following lemma:

\begin{lem}\label{app_ccp_cond_mu_lem}Suppose the assumptions of Lemma \ref{app_logit_ccp_lem}hold. sLet $b_J:=G^{-1}\left(1-\frac1J\right)$ and $a_J=a(b_J)$ and suppose that $r\geq1$. Then for the conditional probability given $MC_i-b_J/a_J=\mu$, we have that
$n^{r/2}\Phi(i,j_1,\dots,j_r|\mu)$ converges to a function of $\mu$ that is strictly decreasing in $\mu$ for $\mu$ large enough.
\end{lem}

\textsc{Proof of Lemma \ref{app_ccp_cond_mu_lem}} As before we let $J=\lceil n^{1/2}\rceil$. Considering the conditional version of (\ref{app_ccp_eqn}),
\begin{eqnarray}
\nonumber J^r\Phi(i,j_1,\dots,j_r|\mu)&=&\left(\prod_{q=1}^rJ(1-G(\mu+b_J/a_J-\sigma^{-1}U^*_{ij_q}))\right)
\left(\prod_{q=r+1}^nG(\mu+b_J/a_J-\sigma^{-1}U^*_{ij_q})^{D_{j_qi}}\right)\\
\nonumber&=&\left(\prod_{q=1}^rJ(1-G(b_J-a_J(U^*_{ij_q}-\mu)))\right)\exp\left\{\frac1J\sum_{q=r+1}^nJ\log G\left(b_J
a_J(U^*_{ij_q}-\mu)D_{j_qi}\right)\right\}
\end{eqnarray}
where the last equality follows from a change of variables $s=a_Jt + b_J$ with $a_J,b_J$ defined as in the proof of Lemma \ref{app_logit_ccp_lem} and Assumption \ref{asy_rate_ass} regarding the asymptotic sequence for $\sigma$. Substituting in the component-wise limits derived in the proof of Lemma \ref{app_logit_ccp_lem}, we obtain
\begin{eqnarray}\nonumber J^r\Phi\left(i,j_1,\dots,j_r\left|\mu\right.\right)&\rightarrow&\exp\left\{\sum_{q=1}^r(U^*_{ij_q}-\mu)
-\frac1J\sum_{q=r+1}^nD_{j_qi}\exp\left\{U^*_{ij_q}-\mu\right\}\right\}
\end{eqnarray}
By inspection, there is a sufficiently large (but finite) value of $\mu_0<\infty$ such that the function on the right-hand side is strictly decreasing in $\mu$ for all $\mu\geq\mu_0$\qed

\subsection*{Proof of Lemma \ref{ev_conv_rate_lem}.}

This result requires only a few modifications to the proof of Lemma \ref{app_logit_ccp_lem} to control the rate at which $J(1-G(b_J+a_Jv))$ converges to $e^{-v}$. For this proof, $b_J$ is again chosen according to
\[1-G(b_J)= \frac1J\]
and $a_J:=\frac{1-G(b_J)}{g(b_J)}$, so that in particular,
\[J(1-G(b_J+a_Jv)) = \frac{1-G(b_J+a_Jv)}{1-G(b_J)}\]

For the main results, we only require rates that are pointwise in $v$, where pointwise convergence rates are invariant under mappings that are continuously differentiable. The mapping $w\mapsto\log(w)$ is continously differentiable for $w\in(0,\infty)$, so it is sufficient to establish rates in levels or logs.

First consider the term
\[\left|\log (1-G(b_J+a_Jv))+\log J +v\right|\rightarrow 0\]
In the following, we let
\[h(u):=\log(1-G(u))\]
By the choice of $b_J$ and the fundamental theorem of calculus, we have
\begin{eqnarray}
\nonumber \log\left(J(1-G(b_J+a_Jv))\right)&=&\log(1-G(b_J+a_Jv) - \log(1-G(b_J))\\
\nonumber&=&\int_{0}^{a_Jv}h'(b_J+u)du
\end{eqnarray}
By the mean-value theorem, for any value of $u\in[0,a_Jv]$, there exists an intermediate value $\bar{u}(u)\in[0,u]$ such that
\[h'(u) = h'(b_J) + h''(b_J+\bar{u}(u))u\]
Furthermore, from the choice of $a_J$ we have
\[a_Jh'(b_J) = -a_J\frac{g(b_J)}{1-G(b_J)}=-1\]
Therefore,
\begin{eqnarray}
\nonumber \left|\log\left(J(1-G(b_J+a_Jv))\right)+v\right|&=& \left|h'(b_J)a_Jv + \int_{0}^{a_Jv}a_Jvh''(b_J+\bar{u}(u))udu +v\right|\\
\nonumber&=&\left|\int_{b_J}^{b_J+a_Jv}h''(b_J + \bar{u}(u))udu\right|\\
\nonumber&\leq&\int_{b_J}^{b_J+a_Jv}|h''(\bar{u}(u))||a_Jv|du\\
\nonumber&=&v\int_{b_J}^{b_J+a_Jv}\frac{|h''(\bar{u}(u))|}{|h'(b_J)|}du
\end{eqnarray}
Since $a_J/b_J\rightarrow0$, we can bound $|h''(\bar{u})|\leq 2|h''(b_J(1-\varepsilon))|$ asymptotically. From Assumption \ref{set_unobs_ass} (ii), it then follows that the right-hand side of this expression is of the order $o\left(\exp\left\{-\frac12b_J\right\}\right)=o(J^{-1})$.

Finally we need to establish the rate of convergence for $J\log G(b_J+a_Jv)+e^{-v}$. The argument is analogous to the previous case - noting that we can rewrite the logarithm via a mean-value expansion around $G=1$,
\[J\log G(b_J+a_Jv)=J\log(1) + \frac{J}1(G(b_J+a_Jv)-1) + \frac{J}{\bar{G}}(G(b_J+a_Jv)-1)^2\]
so that by the same arguments as before,
\[J\log G(b_J+a_Jv)=-\exp\{-v\} + o(J^{-1})\]

Since all multiplicative terms of the integrand in (\ref{app_ccp_eqn}) converge to finite limits, the rate of convergence of the product is governed by the term that converges at the slowest rate, which is also $o(J^{-1})$. This establishes the point-wise convergence rate of the integrand, so that Lemma \ref{ev_conv_rate_lem} follows by dominated convergence \qed

\subsection*{Proof of Lemma \ref{app_ref_fp_lem}}

We first establish a law of large numbers for the bound $\hat{\Psi}_{1n}$ in equation (\ref{M_hat_upper_bound}), showing that \[\left|\hat{\Psi}_{1n}(x,s|R_0)-\frac{\frac1n\sum_{i=1}^n\mathbb{E}_{\Eta}[\tilde{\psi}_{1ni}[\hat{M}_n]](s,R_0,x)}
{\frac1{n^2}\sum_{i=1}^n\mathbb{E}_{\Eta}[\tilde{\psi}_{1ni}[\hat{M}_n]](R_0)}\right|\stackrel{p}{\rightarrow}0,\]
for each $s,R_0,x$. We then establish that $\bar{\psi}_{1ni}[M,\Eta]:=\mathbb{E}_{\Eta}[\tilde{\psi}_{1ni}[M]]$, allowing us to express the fixed point condition as a known expression in terms of the aggregate state variables $M,\Eta$ and economic primitives. To simplify notation, we also let $L_{ij0}:=D_{ij0}D_{ji0}$.

For a point-wise law of large numbers for $\frac1{nq_n}\sum_{i=1}^n\psi_{1ni}^*(s,R_0,x)$ it is then sufficient to show that its variance goes to zero as $n$ grows. Now, by Lemma \ref{app_logit_ccp_lem},
\[n\mathbb{P}(L_{ik0}=1)\leq \left(\frac{\exp\{\bar{U}\}}{1+\exp\{\bar{U}\}}\right)^2=:\bar{C}^2\]
We also let $q_n:=\mathbb{P}(R_{ik}=R_0|L_{ik0}=1)$, where by assumption $nq_n\rightarrow\infty$. We can then bound
\begin{eqnarray}\nonumber\var\left(\frac1{nq_n}\sum_{i=1}^n\sum_{k\neq i}\dum_{A_{k,i}(s,R_0,x)}\right)&=&
\frac{n(n-1)}{n^2q_n^2}\var\left(\dum_{A_{1,2}(s,R_0,x)}\right)\\
\nonumber&&+\frac{n(n-1)(n-2)}{n^2q_n^2}\cov(\dum_{A_{1,2}(s,R_0,x)},\dum_{A_{1,3}(s,R_0,x)})\\
\nonumber&&+\frac{n(n-1)(n-2)}{n^2q_n^2}\cov(\dum_{A_{1,3}(s,R_0,x)},\dum_{A_{2,3}(s,R_0,x)})\\
\nonumber&&+\frac{n(n-1)(n-2)(n-3)}{n^2q_n^2}\cov\left(\dum_{A_{1,2}(s,R_0,x)},\dum_{A_{3,4}(s,R_0,x)}\right)\\
\nonumber&\leq&\frac{\bar{C}^2}{nq_n}\var\left(\dum_{A_{1,2}(s,R_0,x)}\left|L_{210}=1,R_{21}=R_0\right.\right)\\
\nonumber&&+\frac{\bar{C}^4}{nq_n}\cov\left(\dum_{A_{1,2}(s,R_0,x)},\dum_{A_{1,3}(s,R_0,x)}\left|L_{210}=L_{310}=1,R_{21}=R_0\right.\right)\\
\nonumber&&+\frac{\bar{C}^4}{nq_n}
\cov\left(\dum_{A_{1,3}(s,R_0,x)},\dum_{A_{2,3}(s,R_0,x)}\left|L_{310}=L_{320}=1,R_{31}=R_0\right.\right)\\
\nonumber&&+\bar{C}^4\cov\left(\dum_{A_{1,2}(s,R_0,x)},\dum_{A_{3,4}(s,R_0,x)}\left|L_{210}=L_{430}=1,R_{21}
=R_{43}=R_0\right.\right)
\end{eqnarray}

Since $nq_n\rightarrow\infty$, the variance therefore vanishes whenever
\[\cov\left(\dum_{A_{1,2}(s,R_0,x)},\dum_{A_{3,4}(s,R_0,x)}\left|L_{210}=L_{430}=1,R_{21}
=R_{43}=R_0\right.\right)\hspace*{5cm}\]
\begin{eqnarray}
\nonumber&\equiv& \mathbb{P}\left(A_{1,2}(s,R_0,x)\cap A_{3,4}(s,R_0,x)\left|L_{210}=L_{430}=1,R_{21}
=R_{43}=R_0\right.\right)\\
\nonumber&&-\mathbb{P}\left(A_{1,2}(s,R_0,x)\right)\mathbb{P}\left(A_{3,4}(s,R_0,x)\left|L_{210}=L_{430}=1,R_{21}
=R_{43}=R_0\right.\right)\\
\nonumber&\rightarrow&0
\end{eqnarray}
To keep notation manageable, we will take probabilities and expectations for the remainder of this step to be conditional on $L_{210}=L_{430}=1$ and $R_{21}
=R_{43}=R_0$, noting also that the event $A_{3,4}(s,R_0,x)$ does not involve $L_{21}$ and $R_{21}$, and vice versa.

%

To bound $\cov\left(\dum_{A_{1,2}(s,R_0,x)},\dum_{A_{3,4}(s,R_0,x)}\right)$, we partition the event $A_{1,2}(s,R_0,x)\cap A_{3,4}(s,R_0,x)$ according to whether both events are supported by common non-zero edges in $\mathbf{E}_1$ and $\mathbf{E}_2$. That is, by the law of total probability we can write
\begin{eqnarray}
\nonumber \mathbb{P}\left(A_{1,2}(s,R_0,x)\cap A_{3,4}(s,R_0,x)\right)&=&
\mathbb{P}\left(A_{1,2}(s,R_0,x)\cap A_{3,4}(s,R_0,x)\left|\mathbf{L_{E_1\cap E_3}}\neq\mathbf{0}\right.\right)\mathbb{P}\left(\mathbf{L_{E_1\cap E_3}}\neq\mathbf{0}\right)\\
\nonumber &&+\mathbb{P}\left(A_{1,2}(s,R_0,x)\cap A_{3,4}(s,R_0,x)\left|\mathbf{L_{E_1\cap E_3}}=\mathbf{0}\right.\right)\mathbb{P}\left(\mathbf{L_{E_1\cap E_3}}=\mathbf{0}\right)
\end{eqnarray}
Now since the number of edges in $\mathbf{E}_1\cap\mathbf{E}_2$ is fixed, we have by Assumption \ref{asy_rate_ass} that
\[\mathbb{P}\left(\mathbf{L_{E_1\cap E_3}}\neq\mathbf{0}\right) = O(n^{-1})\]
Therefore we have
\[\left|\mathbb{P}\left(\left.A_{1,2}(s,R_0,x)\right|\mathbf{L_{E_1\cap E_3}}=\mathbf{0}\right)\right.-\left.\mathbb{P}\left(A_{1,2}(s,R_0,x)\right)\right|\hspace{8cm}\textnormal{ }\]
\begin{eqnarray}
\nonumber&=&\left|\mathbb{P}\left(\left.A_{1,2}(s,R_0,x)\right|\mathbf{L_{E_1\cap E_3}}\neq\mathbf{0}\right)-\mathbb{P}\left(\left.A_{1,2}(s,R_0,x)\right|\mathbf{L_{E_1\cap E_3}}=\mathbf{0}\right)\right|
\mathbb{P}\left(\mathbf{L_{E_1\cap E_3}}\neq\mathbf{0}\right)\\
\nonumber&\leq&\mathbb{P}\left(\mathbf{L_{E_1\cap E_3}}\neq\mathbf{0}\right) = O(n^{-1})
\end{eqnarray}
Furthermore, by Lemma \ref{app_sampling_rep_lem},
\[\mathbb{P}\left(\left.A_{1,2}(s,R_0,x)\cap A_{3,4}(s,R_0,x)\right|\mathbf{L_{E_1\cap E_3}}=\mathbf{0}\right)
=\mathbb{P}\left(\left.A_{1,2}(s,R_0,x)\right|\mathbf{L_{E_1\cap E_3}}=\mathbf{0}\right)\mathbb{P}\left(\left.A_{3,4}(s,R_0,x)\right|\mathbf{L_{E_1\cap E_3}}=\mathbf{0}\right)\]
Hence, we can bound
\begin{eqnarray}
\nonumber \left|\cov\left(\dum_{A_{1,2}(s,R_0,x)},\dum_{A_{3,4}(s,R_0,x)}\right)\right|&\leq&
\left|\mathbb{P}\left(\left.A_{1,2}(s,R_0,x)\right|\mathbf{L_{E_1\cap E_3}}=\mathbf{0}\right)\mathbb{P}\left(\left.A_{3,4}(s,R_0,x)\right|\mathbf{L_{E_1\cap E_3}}=\mathbf{0}\right)\right.\\
\nonumber&&\left.-\mathbb{P}\left(A_{1,2}(s,R_0,x)\right)\mathbb{P}\left(A_{1,2}(s,R_0,x)\right)\right|+O(n^{-1})\\
\nonumber&\leq&4\mathbb{P}\left(\mathbf{L_{E_1\cap E_3}}\neq\mathbf{0}\right)+O(n^{-1})=O(n^{-1})
\end{eqnarray}
Hence, the variance of the sample mean of $\dum_{A_{1,2}(s,R_0,x)}$ is of the order $O(n^{-1})$ so that by Chebyshev's inequality $\frac1{n}\sum_{i=1}^n\sum_{k\neq i}\psi^*_{1ni}(s,R_0,x)$ converges in probability to its expectation given $\hat{M}_n$. Since Lemma \ref{app_sampling_rep_lem} also implies that
\[\mathbb{E}_{\hat{M}_n}[\psi_{1ni}^*(s,R_0,x)] = \mathbb{E}[\tilde{\psi}_{1ni}[\hat{M}_n]](s,R_0,x) + O\left(\frac1n\right),\]
Similarly,
\[\frac1{n^2q_n}\sum_{i=1}^n\sum_{k\neq i}\psi^*_{1ni}(R_0)-\frac1{n^2}\sum_{i=1}^n\sum_{k\neq i}\mathbb{E}[\tilde{\psi}_{1ni}[\hat{M}_n]](R_0)\stackrel{p}{\rightarrow}0\]
where $\frac1{n^2q_n}\sum_{i=1}^n\sum_{k\neq i}\mathbb{E}[\tilde{\psi}_{1ni}[\hat{M}_n]](R_0)$ is bounded away from zero. Convergence of $\hat{\Psi}_{1n}(s,x|R_0)$ then follows from the continuous mapping theorem.

For the second step, we can evaluate the expectation of $\tilde{\psi}_{1ni}[\hat{M}]$ under the sampling representation in Lemma \ref{app_sampling_rep_lem}. Specifically, by the law of total probability the probability that $s_i^*=s$ and $R_i^*=R_0$ are supported by a pairwise stable network, and $D_{ik0}=1$ for an arbitrary node $k\neq i$,
\begin{eqnarray}
\nonumber\mathbb{E}[\tilde{\psi}_{1ni}[M]]&=&\frac1{n}\sum_{k\neq i}^n\mathbb{P}\left(\mathcal{L}_{k,i,\{j_1,\dots,j_r\}}^*\cap
\mathcal{L}(s,R_0;k,i,\{j_1,\dots,j_r\})\neq\emptyset,x_i= x,\tilde{L}_{ik0}=1|M\right)\\
\nonumber&=&\frac1{n}\sum_{k\neq i}\sum_{r\geq0}\sum_{N_{i,\{j_1,\dots,j_r\}}}
\mathbb{P}\left(\mathcal{L}_{k,i,\{j_1,\dots,j_r\}}^*\cap
\mathcal{L}(s,R_0;k,i,\{j_1,\dots,j_r\})\neq\emptyset,x_i= x,\tilde{D}_{ik0}=1|
\tilde{\mathcal{N}}_i=N_{i,\{j_1,\dots,j_r\}}\right)\\
\nonumber&&\times\mathbb{P}(\tilde{\mathcal{N}}_i=N_{i,\{j_1,\dots,j_r\}}|M)\\
\label{ref_dist_expect}&=&\sum_{r\geq0}\sum_{N_{i,\{j_1,\dots,j_r\}}}\sum_{\mathbf{L}\in\{0,1\}^r} \bar{\psi}_{1ni,j_1\dots j_r}[M,\Eta](\mathbf{L},N_{i,\{j_1,\dots,j_r\}},x)\dum_{\mathbf{L}\in\mathcal{L}(s,R_0;i,N_{i,\{j_1,\dots,j_r\}})}
\end{eqnarray}
where the second equality uses the law of total probability and that by Lemma \ref{app_sampling_rep_lem}, $\tilde{\mathcal{N}}_i$ and $\hat{M}_n$ are independent. For the third equality, we used Lemma \ref{app_logit_ccp_lem} and (\ref{ubar_shift}). Furthermore, by standard arguments, drawing from $\hat{M}_n$ independently with replacement differs from drawing from the underlying population without replacement, and leaving out $i$ by a factor that is at most of the magnitude $O\left(\frac1n\right)$ as $n$ grows.

Averaging over all nodes, we obtain
\begin{eqnarray}
\nonumber\bar{\Psi}_{1n}[M,\Eta](x,s|R_0)&:=&\frac{\frac1n\sum_{i=1}^n\mathbb{E}_{\Eta}\left[\tilde{\psi}_{1ni}[M](s,R_0,x)\right]}
{\frac1n\sum_{i=1}^n\mathbb{E}_{\Eta}\left[\tilde{\psi}_{1ni}[M](R_0)\right]}\\
\nonumber&=:&\frac{\frac1n\sum_{i=1}^n\bar{\psi}_{1ni}[M,\Eta](s,R_0,x)}{\frac1n\sum_{i=1}^n\bar{\psi}_{1ni}[M,\Eta](R_0)}
+o(1)
\end{eqnarray}
as claimed, where $\mathcal{L}_{i}^*$ is the set of subnetworks $\mathbf{L_{E_i}}^*$ that are supported by a pairwise stable subnetwork on $\mathbf{E}_i$.

Combining this with convergence of $\hat{\Psi}_{1n}$ to $\bar{\Psi}_{1n}[\hat{M}_n,H]$ we obtain the conclusion of the Lemma for a singleton argument $s,R_0$. The analogous conclusion for sets $\{(s_1,R_1),\dots,(s_K,R_K)\}$ follows in a completely analogous fashion\qed

\section{Proofs for Section \ref{sec:struct_limit}}

\label{sec:gen_limit_theory}

The main result in this paper is a central limit theorem for moments of the form (\ref{moment_fct}) for pairwise stable networks among finitely many agents. The main argument proceeds along the following steps:
\begin{itemize}
\item For the $D$-fold array of contributions $\left(m_{i_1\dots i_D}\right)_{i_1\dots i_D}$,  we first develop an approximate sampling representation in terms of jointly exchangeable arrays (defined below).
\item Using results by \cite{Ald81} and \cite{Hoo79}, we show that these approximating arrays can be characterized by a function of (vector-valued) i.i.d. factors $\alpha_{i_1}^{(1)},\dots,\alpha_{i_1\dots i_D}^{(D)}$ that are indexed by subsets of $\{i_1,\dots,i_D\}$.
\item This characterization allows us to decompose the D-adic average in (\ref{moment_fct}) of, generally dependent, contributions to the network moment, into a sum of multi-linear forms of random vectors whose entries are independent. We can then use results by \cite{dJo90} to establish CLT-type results to these multi-linear forms.
\item Having established that the network moment is asymptotically normal, we can then characterize the first two moments of its asymptotic distribution in terms of model primitives.
\end{itemize}

This appendix establishes a central limit theorem in terms of high-level conditions, which will then be used to prove Theorem \ref{asy_Gauss_thm}.

\subsection{Coupling to Exchangeable Array}

We start out by approximating the array $\left(Y_{i_1,\dots,i_{D+1}}\right)_{i_1\dots i_{D+1}}$ with elements
\[Y_{i_1,\dots,i_{D+1}}:=\left(m_{i_1\dots i_D}',\psi_{i_{D+1}}\right)_{i_1\dots i_{D+1}}\]
determining the joint distribution of the network moment $\hat{m}_n(\theta)$ and the fixed point conditions for the aggregate state $\eta=[M,\Eta]$, with another array $\left(\tilde{Y}_{i_1,\dots,i_{D+1}}\right)_{i_1\dots i_{D+1}}$ consisting of entries
\[\tilde{Y}_{i_1,\dots,i_{D+1}}[\hat{\eta}_n]:=\left(\tilde{m}_{i_1\dots i_D}',\tilde{\psi}_{i_{D+1}}\right)_{i_1\dots i_{D+1}}\]
constructed under the sampling representation in Lemma \ref{app_sampling_rep_lem}, where the reference distribution $\hat{M}_n$ is given by the first component of $\hat{\eta}_n = [\hat{M}_n,\hat{\Eta}_n]$.

Specifically, we let $\hat{M}_n$ be the reference distribution corresponding to the realization of $(Y_{\mathbf{i}})_{\mathbf{i}}$, and generate $(\tilde{Y}_{\mathbf{i}})_{\mathbf{i}}$ by drawing node attributes, network neighborhoods, and taste shocks for each node $i$ at random from $\hat{M}_n$, independently from the original array. We also denote the properly normalized sample averages with
\[\bar{Y}_n:=\left(\left[\binom{n}{D}p_n\right]^{-1}\sum_{\mathbf{i}}m_{\mathbf{i}},\frac1n\sum_{i=1}^n\psi_{i}\right)\]
and
\[\overline{\tilde{Y}}_n[\hat{\eta}_n]:=\left(\left[\binom{n}{D}p_n\right]^{-1}\sum_{\mathbf{i}}\tilde{m}_{\mathbf{i}},\frac1n\sum_{i=1}^n\tilde{\psi}_{i}\right)\]
respectively. We first note that the array $\tilde{Y}_{i_1\dots i_{D+1}}$ is jointly exchangeable:

\begin{dfn} A \emph{jointly exchangeable array} is an infinite array $(Y_{i_1\dots i_D})_{i_1\dots i_D}$ such that for any integer $\tilde{N}<\infty$ and permutation $\tau:\{1,\dots,\tilde{N}\}\rightarrow\{1,\dots,\tilde{N}\}$, we have \[\left(Y_{\tau(i_1)\dots\tau(i_D)}\right)_{i_1\dots i_D}\stackrel{d}{=}\left(Y_{i_1\dots i_D}\right)_{i_1\dots i_D},\] where ``$\stackrel{d}{=}$" denotes equality in distribution.
\end{dfn}

By construction of either component potential values for the network statistics were drawn independently from the reference distribution so that $\left(\tilde{Y}_{i_1,\dots,i_{D+1}}\right)_{i_1\dots i_{D+1}}$ is in fact jointly exchangeable. It remains to be shown that this array approximates its finite-network analog under the appropriate metric. Specifically, we establish the following:

\begin{lem}\label{sampling_array_appx_lem} Suppose Assumptions \ref{moment_ass}-\ref{asy_rate_ass} hold. Then we have
\[\overline{\tilde{Y}}_n[\hat{\eta}_n]\stackrel{d}{=}\bar{Y}_n + O_p\left(\frac1n\right)\]
conditional on $\hat{\eta}_n=\hat{\Psi}_n[\hat{\eta}_n]=\hat{\tilde{\Psi}}_n[\hat{\eta}_n]$.
\end{lem}


\textsc{Proof:} Note first that $\hat{M}_n$ is the conditional empirical distribution given relevant overlap of draws $(x_k,s_k^*)$ such that $D_{ik0}=D_{ki0}=1$ for some $i\neq k$. As an intermediate step for the approximation, we now consider an array $\left(\ddot{Y}_{\mathbf{i}}[M]\right)_{\mathbf{i}}$ for a fixed value of $M$ which results from permuting the pre-network at random, specifically for a permutation $\tau$ on $\{1,\dots,n\}$ that is generated uniformly at random we let $\ddot{D}_{\tau(i)k0}=\ddot{D}_{k\tau(i)0}=1$ iff $D_{ik0}=D_{ki0}=1$, while leaving $(\ddot{x}_k,\ddot{s}_k^*)=(x_k,s_k^*)$ unchanged. $\ddot{Y}_{\mathbf{i}}[M]$ is then defined in analogy $\tilde{Y}_{\mathbf{i}}[M]$ where network neighborhoods $\ddot{\mathcal{N}}_i$ are defined by the pre-network $\ddot{D}_{ik0}=\ddot{D}_{ki0}=1$. We also denote the polyadic average with
\[\overline{\ddot{Y}}_n:=\left(\left[\binom{n}{D}p_n\right]^{-1}\sum_{\mathbf{i}}\ddot{m}_{\mathbf{i}},\frac1n\sum_{i=1}^n\ddot{\psi}_{i}\right)
=:\left(\hat{\ddot{m}}_n,\hat{\ddot{\Psi}}_n\right)\]


Since for the original array the joint distribution of $\varepsilon_{ik}$ and $MC_i$ is i.i.d. and therefore in particular invariant to permutations of this kind, and polyadic averages are also a symmetric function of node identifiers,  we have that \[\overline{\ddot{Y}}_n[\hat{\eta}_n]\stackrel{d}{=}\bar{Y}_n\]
conditional on $\hat{\eta}_n\in\hat{\ddot{\Psi}}_n[\hat{\eta}_n]$ and $\hat{\eta}_n\in\hat{\Psi}_n[\hat{\eta}_n]$. The most important implication of this step is that for the approximating array the pairwise stability conditions only need to hold at the level of the aggregate fixed point condition $\hat{\eta}_n\in\hat{\Psi}_n[\hat{\eta}_n]$ rather than for each node.

To complete the argument, it remains to be shown that $\overline{\tilde{Y}}_n[\hat{\eta}_n]-\overline{\ddot{Y}}_n[\hat{\eta}_n]=O_P\left(\frac1n\right)$. Here the key difference between the two arrays is that the array $\left(\ddot{Y}_{\mathbf{i}}\right)_{\mathbf{i}}$ can be thought of drawing the pre-network around a node $k$ without replacement, whereas the array $\left(\ddot{Y}_{\mathbf{i}}\right)_{\mathbf{i}}$ is generated by drawing with replacement with probabilities equal to the corresponding frequencies for the former array.

We bound the approximation error via a coupling between the two arrays that is constructed as follows: suppose that for a given node $i$ we have $\ddot{D}_{ik_q0}\ddot{D}_{k_qi0}=1$ for $q=1,\dots,Q(i)$, and
$\ddot{D}_{ik_q0}\ddot{D}_{k_qi0}=0$ for all $k\notin\{k_1,\dots,k_{Q(i)}\}$. We then draw $l_1,\dots,l_{Q(i)}$ uniformly at random and with replacement from $1,\dots,Q(i)$, resulting in the network neighborhood $\tilde{\mathcal{N}}_i$ consisting of nodes $k_{l_1},\dots,k_{l_{Q(i)}}$. This construction $\tilde{\mathcal{N}}_i$ is equivalent to the sampling representation in Lemma \ref{app_sampling_rep_lem}.


In particular it allows for the neighborhood to contain replicates of the same node, whereas in the absence of such ties, $\tilde{\mathcal{N}}_i=\ddot{\mathcal{N}}_i$. Comparing a sequence of $Q<\infty$ draws, the probability of such a tie is bounded by $\frac{Q}n$. Now, a given entry of the array $\tilde{Y}_{\mathbf{i}}\neq \ddot{Y}_{\mathbf{i}}$ only if $\tilde{\mathcal{N}}_{i_d}\neq\ddot{\mathcal{N}}_{i_d}$ for at least one component $i_d$ of the multi-index $\mathbf{i}=(i_1,\dots,i_D)$. However since the pre-network degree $|\mathcal{N}_i|$ is stochastically bounded and $D$ is fixed, the probability for that event is also bounded by a nonstochastic sequence at the rate $\frac1n$. Since the event $\tilde{\mathcal{N}}_i\neq\ddot{\mathcal{N}}_i$ is furthermore independent across $i=1,\dots,n$, this establishes the desired conclusion \qed

\subsection{Aldous-Hoover Representation}

By construction, $\tilde{Y}_{i_1\dots i_D}$ is determined by the draws for marginal costs, taste shocks, and local neighborhoods $\tilde{\mathcal{N}}_{i_1},\dots,\tilde{\mathcal{N}}_{i_D}$, which are generated as i.i.d. random draws from their respective distributions. In particular, $\tilde{Y}_{i_1\dots i_D}$ is jointly exchangeable in $i_1,\dots,i_D$.

From an Aldous-Hoover representation (Theorem 7.22 in \cite{Kal05}), we can represent such a jointly exchangeable $D$-fold array on a Borel space $\mathcal{B}$ as
\[\tilde{Y}_{i_1\dots i_D} = h\left(\alpha^{(0)},\alpha_{i_1}^{(1)},\dots,\alpha_{i_D}^{(1)},\alpha_{\{i_1,i_2\}}^{(2)},\dots,\alpha_{\{i_1,\dots, i_D\}}^{(D)}\right)\]
for some measurable function $h:[0,1]^{2^D}\rightarrow\mathcal{B}$, and $(\alpha^{(0)},\alpha_{i_1}^{(1)},\dots,\alpha_{i_D}^{(1)},\alpha_{i_1i_2}^{(2)},\dots,\alpha_{i_1\dots i_D}^{(D)})$ is a uniform random array whose $2^D$ components are random variables that are independently and uniformly distributed on the unit interval, and which are indexed by the subsets of $\{i_1,\dots,i_D\}$.

\subsection{Decomposition for CLT}

In what follows we therefore restrict our attention to the term
\[\tilde{Z}_n=\binom{n}{D}^{-1}\sum_{i_1<\dots<i_D}(\tilde{Y}_{i_1\dots i_D}-\mathbb{E}[Y_{i_1\dots i_D}|\alpha^{(0)}])\]

For a decomposition of $\tilde{Z}_n$, we now denote the conditional expectations of $Y_{i_1\dots i_D}$ with
\begin{eqnarray}
\nonumber y_{j_1\dots j_D}^{(1)}&:=& \mathbb{E}[\tilde{Z}_n|\alpha^{(0)},\alpha_{j_1}^{(1)},\dots,\alpha_{j_D}^{(1)}]\\
\nonumber y_{j_1\dots j_D}^{(2)}&:=& \mathbb{E}[\tilde{Z}_n|\alpha^{(0)},\alpha_{j_1}^{(1)},\dots,\alpha_{j_2}^{(D)},\alpha_{j_1j_2}^{(2)},\dots,\alpha_{j_{D-1}j_D}^{(2)}]\\
\nonumber \vdots&&\vdots\\
\nonumber y_{j_1\dots j_D}^{(d)}&:=& \mathbb{E}[\tilde{Z}_n|\alpha^{(0)},\alpha_{j_1}^{(1)},\dots,\alpha_{j_{1}\dots j_d}^{(d)}\dots,\alpha_{j_{D-d}\dots j_D}^{(d)}]
\end{eqnarray}
for $d=1,\dots,D$, noting that $y_{j_1\dots j_D}^{(D)}=Y_{j_1\dots j_D}$. We then form recursive projection residuals
\begin{eqnarray}
\nonumber \dot{y}_{j_1\dots j_D}^{(1)}&:=& y_{j_1\dots j_D}^{(1)}\\
\nonumber \dot{y}_{j_1\dots j_D}^{(2)}&:=&y_{j_1\dots j_D}^{(2)} - y_{j_1\dots j_D}^{(1)}\\
\nonumber \vdots&&\vdots\\
\nonumber \dot{y}_{j_1\dots j_D}^{(d)}&:=& y_{j_1\dots j_D}^{(d)} - y_{j_1\dots j_D}^{(d-1)}\\
\nonumber \vdots&&\vdots\\
\nonumber \dot{y}_{j_1\dots j_D}^{(D)}&:=& y_{j_1\dots j_D}^{(D)} - y_{j_1\dots j_D}^{(D-1)}
\end{eqnarray}
By construction, these projection residuals are uncorrelated within and across index sets.

Given these definitions, we can decompose the sum 
\begin{equation}\label{app_z_tilde_decomp}\tilde{Z}_n  = \tilde{Z}_{1n} + \dots + \tilde{Z}_{Dn}\end{equation}
where
\begin{eqnarray}
\nonumber \tilde{Z}_{1n}&:=&\binom{n}{D}^{-1}\sum_{i_1<\dots<i_D}\dot{y}_{i_1\dots i_D}^{(1)}\\
\nonumber \vdots&&\vdots\\
\nonumber \tilde{Z}_{Dn}&:=&\binom{n}{D}^{-1}\sum_{i_1<\dots<i_D}\dot{y}_{i_1\dots i_D}^{(D)}
\end{eqnarray}
By construction, each terms $\tilde{Z}_{dn}$ is mean-independent of its precursors $\tilde{Z}_{1n},\dots,\tilde{Z}_{(d-1)n}$.

We can now analyze the asymptotic distribution of each term $\tilde{Z}_{1n},\tilde{Z}_{2n},\dots$, noting that our analysis is conditional on $\alpha^{(0)}$ so that $\tilde{Z}_{0n}$ is regarded as fixed. We can rewrite
\begin{eqnarray}
\nonumber \dot{y}_{i_1\dots i_D}^{(1)}&=&\sum_{r=1}^D\sum_{q\in Q_1(r,D)} e_{i_1\dots i_r;d_{q(1)}\dots d_{q(r)}}^{(1)}
\end{eqnarray}
where $Q_1(r,d)$ denotes the set of all strictly increasing mappings $\{1,\dots,r\}\rightarrow\{1,\dots,d\}$, and we define
\begin{eqnarray}
\nonumber e_{j_1;d_1}^{(1)}&:=&\mathbb{E}[\dot{y}_{i_1\dots i_D}^{(1)}|\alpha^{(0)},\alpha_{i_{d_1}}^{(1)}=\alpha_{j_1}^{(1)}]\\
\nonumber e_{j_1,j_2;d_1,d_2}^{(1)}&:=&\mathbb{E}\left[\left.
\dot{y}_{i_1\dots i_D}^{(1)}\right|\alpha^{(0)},\alpha_{i_{d_1}}^{(1)}=\alpha_{j_1}^{(1)},\alpha_{i_{d_2}}^{(1)}=\alpha_{j_2}^{(1)}\right]\\
\nonumber&&-e_{j_1;d_1}^{(1)} - e_{j_2;d_2}^{(1)}\\
\nonumber e_{j_1\dots j_r;d_1\dots d_r}^{(1)}&:=&\mathbb{E}\left[\left.
\dot{y}_{i_1\dots i_D}^{(1)}\right|\alpha^{(0)},\alpha_{i_{d_1}}^{(1)}=\alpha_{j_1}^{(1)},\dots,\alpha_{i_{d_r}}^{(1)}=\alpha_{j_r}^{(1)}\right]\\
\nonumber&&-\sum_{s=1}^{r-1}\sum_{q\in Q_1(s,r-1)}e_{j_1\dots j_s;d_{q(1)}\dots d_{q(s)}}^{(1)}
\end{eqnarray}
recursively. This expansion can be viewed as a generalized Hoeffding composition, whose individual components are mean-independent from their respective precursors, again by construction.

Now let $Q_2(r,d)$ denotes the set of all nondecreasing mappings $\mathbf{q}:\{1,\dots,r\}\rightarrow\{1,\dots,d\}^2$ such that the second component is larger than the first, and for each $s=1,\dots,r$, at least one component of $\mathbf{q}(s)$ is strictly greater than the corresponding component of $\mathbf{q}(s-1)$. We then write
\begin{eqnarray}
\nonumber \dot{y}_{i_1\dots i_D}^{(2)}&=&\sum_{r=2}^D\sum_{q\in Q_2(r,D)} e_{\mathbf{i}_1\dots \mathbf{i}_r;\mathbf{d}_{q(1)}\dots \mathbf{d}_{q(r)}}^{(2)}
\end{eqnarray}
where with some abuse of notation we let $\mathbf{i}_{q_1q_2}:=(i_{q_1}i_{q_2})$, $\mathbf{d}_{q_1q_1}:=(d_{q_1}d_{q_2})$, and we define
\begin{eqnarray}
\nonumber e_{\mathbf{j}_1;\mathbf{d}_1}^{(2)}&:=&\mathbb{E}\left[\left.
\dot{y}_{i_1\dots i_D}^{(2)}\right|\alpha^{(0)},\alpha_{i_1}^{(1)},\dots,\alpha_{i_D}^{(1)},\alpha_{i_{\mathbf{d}_1}}^{(2)}=\alpha_{\mathbf{j}_1}^{(2)}\right]\\
\nonumber e_{\mathbf{j}_1\dots\mathbf{j}_r;\mathbf{d}_1\dots\mathbf{d}_r}^{(1)}&:=&\mathbb{E}\left[\left.
\dot{y}_{i_1\dots i_D}^{(2)}\right|\alpha^{(0)},\alpha_{i_1}^{(1)},\dots,\alpha_{i_D}^{(1)},\alpha_{i_{\mathbf{d}_1}}^{(2)}=\alpha_{\mathbf{j}_1}^{(2)},
\dots,\alpha_{i_{\mathbf{d}_r}}^{(2)}=\alpha_{\mathbf{j}_r}^{(2)}\right]\\
\nonumber&&-\sum_{s=2}^{r-1}\sum_{q\in Q_2(s,r-1)}e_{\mathbf{j}_1,\dots \mathbf{j}_s;\mathbf{d}_{q(1)}\dots\mathbf{d}_{q(s)}}^{(2)}
\end{eqnarray}
recursively.

More generally, for $k=1,\dots,D$, we let $Q_d(r,k)$ denote the set of all nondecreasing mappings $\mathbf{q}:\{1,\dots,r\}\rightarrow\{1,\dots,k\}^d$ such that for each $s=1,\dots,r$, at least one component of $\mathbf{q}(s)$ is strictly greater than the corresponding component of $\mathbf{q}(s-1)$, and each component $q_l(s)<q_{l+1}(s)$. We then write
\begin{eqnarray}
\label{y_dot_k_decomp} \dot{y}_{i_1\dots i_D}^{(k)}&=&\sum_{r=k}^D\sum_{q\in Q_k(r,D)} e_{\mathbf{i}_1\dots \mathbf{i}_r;\mathbf{d}_{q(1)}\dots \mathbf{d}_{q(r)}}^{(k)}
\end{eqnarray}
where $\mathbf{i}_{q_1\dots q_r}:=(i_{q_1}\dots i_{q_r})$, $\mathbf{d}_{q_1\dots q_r}:=(d_{q_1}\dots d_{q_r})$, and
\begin{eqnarray}
\nonumber e_{\mathbf{j}_1;\mathbf{d}_1}^{(k)}&:=&\mathbb{E}\left[\left.
\dot{y}_{i_1\dots i_D}^{(k)}\right|\alpha^{(0)},\alpha_{i_1}^{(1)},\dots,\alpha_{i_D}^{(1)},\dots,\alpha_{i_{D-k+1}\dots i_D}^{(k-1)},\alpha_{i_{\mathbf{d}_1}}^{(k)}=\alpha_{\mathbf{j}_1}^{(k)}\right]\\
\nonumber e_{\mathbf{j}_1\dots\mathbf{j}_r;\mathbf{d}_1\dots\mathbf{d}_r}^{(k)}&:=&\mathbb{E}\left[\left.
\dot{y}_{i_1\dots i_D}^{(k)}\right|\alpha^{(0)},\alpha_{i_1}^{(1)},\dots,\alpha_{i_D}^{(1)},,\dots,\alpha_{i_{D-k+1}\dots i_D}^{(k-1)},\alpha_{i_{\mathbf{d}_1}}^{(k)}=\alpha_{\mathbf{j}_1}^{(k)},
\dots,\alpha_{i_{\mathbf{d}_r}}^{(k)}=\alpha_{\mathbf{j}_r}^{(k)}\right]\\
\nonumber&&-\sum_{s=k}^{r-1}\sum_{q\in Q_k(s,r-1)}e_{\mathbf{j}_1\dots \mathbf{j}_s;\mathbf{d}_{q(1)}\dots\mathbf{d}_{q(s)}}^{(k)}
\end{eqnarray}
are again defined recursively.

\subsection{Rate of Convergence}

\label{subsec:roc_subsec}
In the next subsections \ref{subsec:roc_subsec}-\ref{subsec:generic_clt_subsec}, we present generic results for a scalar component of the multivariate exchangeable array $\tilde{\mathbf{Y}}{\mathbf{i}}$, which will subsequently be extended to the full array using the Cram\'er-Wold device.

For the scalar case, we can write
\[\sigma_{k,\mathbf{d}_1\dots\mathbf{d}_r}^2:=\var(e_{\mathbf{j}_1\dots\mathbf{j}_r;\mathbf{d}_1\dots\mathbf{d}_r}^{(k)})  \]
for any $k=1,\dots,D$, where each $\mathbf{d}_s\in\{1,\dots,D\}^k$ and $\mathbf{j}_s\in\{1,\dots,D\}^k$. Under sparse asymptotics, $\sigma_{k,\mathbf{d}_1\dots\mathbf{d}_r}^2$ will in general not be constant, but some components will go to zero at some rate as $n$ increases. We first establish the stochastic order of the terms the expansion (\ref{app_z_tilde_decomp}) and (\ref{y_dot_k_decomp}) corresponding to projections of degree $r=1$.

\begin{lem}\label{app_lead_terms_lem} Suppose that $\tilde{Y}_{i_1\dots i_D}$ is a jointly exchangeable array with scalar entries which are symmetric across dimensions $d=1,\dots,D$ and satisfy the conditions of Assumption \ref{moment_ass}. Then for any $k=1,\dots, D$ such that $\binom{n}{k}\sigma_{k,\mathbf{d}_1}^2\rightarrow\infty$, the terms
\[\sum_{1\leq i_1<\dots<i_k\leq n}\sum_{1\leq d_1<\dots<d_k\leq D}e_{i_1\dots i_k,d_1\dots d_k}^{(k)} = O_P(r_{n,k}^{-1/2})\]
where
\[r_{n,k}^{-1}:=\binom{n}{k}^{-1}\sigma_{k,\mathbf{d}_1}^2\]
and $\mathbf{d}_1:=(1,\dots,k)'$.
\end{lem}

\textsc{Proof:} From the recursive construction of $e_{i_1\dots i_{k},d_1\dots d_{k}}^{(k)}$ and independence of $\{\alpha_{i_{d_1},\dots,i_{d_s}}^{(s)}:s=1,\dots,D\}$, we have that conditional on $\{\alpha_{i_{d_1},\dots,i_{d_s}}^{(s)}:s=1,\dots,k-1\}$, $e_{i_1\dots i_{k},d_1\dots d_{k}}^{(k)}$ are mean-independent, although not necessarily i.i.d.. Furthermore, $Y_{i_1\dots i_D}$ are bounded by Assumption \ref{moment_ass} (a), so that the conclusion follows from a martingale difference central limit theorem for triangular arrays, see e.g. the main theorem in \cite{McL74}. The unconditional statement follows from the law of iterated expectations and the fact that $\{\alpha_{i_{d_1},\dots,i_{d_s}}^{(s)}:s=1,\dots,k-1\}$ are also i.i.d.\qed

The next lemma establishes that under regularity conditions, the projections of degree $r=1$ do in fact provide the leading terms in that expansion, possibly for multiple values of $k=1,\dots,D$, whereas terms corresponding to $r>1$ are generally dominated.

\begin{lem}\label{app_dom_terms_lem} Suppose that $\tilde{Y}_{i_1\dots i_D}$ is a jointly exchangeable array with scalar entries which are symmetric across dimensions $d=1,\dots,D$ and satisfy the conditions of Assumption \ref{moment_ass}. Furthermore assume that the component variances satisfy $\sigma_{k,\mathbf{d}_1}^2/\sigma_{k,\mathbf{d}_1\dots\mathbf{d}_r}^2\geq B$ for some lower bound $B>0$. Then for any $k=1,\dots, D$ and $r=2,\dots,k$, the terms
\[\sum_{1\leq i_1<\dots<i_D\leq n}\sum_{q\in Q_k(r,D)}e_{\mathbf{i}_{q_1}\dots \mathbf{i}_{q_r},\mathbf{d}_{q_1}\dots \mathbf{d}_{q_r}}^{(k)} = o_P(r_{n,k}^{-1/2})\]
\end{lem}

\textsc{Proof:} We first notice that by Assumption \ref{moment_ass} and the triangle inequality all moments of $e_{i_1\dots i_{r},d_1\dots d_{r}}$ exist. Also by construction, $e_{i_1\dots i_{r},d_1\dots d_{r}}$ are mean-independent and therefore uncorrelated. Since there are $\binom{\binom{n}{k}}{r}$ of these terms, the variance of the $r$-fold mean equals $\left[\binom{\binom{n}{k}}{r}\sigma_{k,\mathbf{d}_1\dots\mathbf{d}_r}^2\right]^{-1}\leq \left[\binom{\binom{n}{k}}{r}B\sigma_{K,\mathbf{d}_1}^2\right]^{-1}$, which converges to zero at a rate faster than $r_{n,K}^{-1}$, so that the conclusion follows directly from Chebyshev's inequality\qed

\subsection{Generic LLN and CLT}

\label{subsec:generic_clt_subsec}

Given the previous two lemmata, we can establish a Law of Large Numbers for the average $\tilde{Z}_n$.

\begin{lem}\label{app_generic_lln_lem} Suppose that $\tilde{Y}_{i_1\dots i_D}$ is a jointly exchangeable array with scalar entries which are symmetric across dimensions $d=1,\dots,D$ and satisfy the conditions of Assumption \ref{moment_ass}. Then for any sequence $q_n$ such that for each $k=1,\dots,D$, $\sigma_{n,k}^2/p_n$ is bounded and $\binom{N}{D}p_n\rightarrow\infty$, we have
\[p_n^{-1}\tilde{Z}_{n}\stackrel{p}{\rightarrow}0\]
\end{lem}

This lemma is an immediate consequence of Lemmas \ref{app_lead_terms_lem} and \ref{app_dom_terms_lem} and Chebyshev's Inequality. These lemmas furthermore identify the leading terms in the expansion for $\tilde{Z}_n$, allowing us to establish a generic central limit theorem. Let $r_n:=\min\{r_{n,1},\dots,r_{n,D}\}$ and define
\[V_n:=r_n\sum_{k=1}^D\binom{n}{k}\binom{D}{k}\sigma_{k,\mathbf{d}_1}^2\]
We then have the following:

\begin{lem}\label{app_generic_clt_lem} Suppose that $\tilde{Y}_{i_1\dots i_D}$ is a jointly exchangeable array with scalar entries which are symmetric across dimensions $d=1,\dots,D$ and satisfy the conditions of Assumption \ref{moment_ass}. If furthermore $V_n\rightarrow V$ for some $\mathcal{F}$-measurable positive definite matrix $V$ almost surely, we have
\[r_n^{1/2}V^{-1/2}\tilde{Z}_{n}\stackrel{d}{\rightarrow}N(0,\mathbf{I}),\]
mixing with respect to $\mathcal{F}$.
\end{lem}

Note that the asymptotic variance $V$ is defined only implicitly as a limit in order to allow for drifting sequences to accommodate sparse network asymptotics. We analyze this expression in terms of the primitives of the economic model further below.\\

\textsc{Proof:} As before, we decompose $\tilde{Z}_n$ into
\[\tilde{Z}_n = \tilde{Z}_{1n} + \tilde{Z}_{2n} + \dots +\tilde{Z}_{Dn}\]
with $\tilde{Z}_{kn}$ defined in (\ref{app_z_tilde_decomp}). It follows from Lemmas \ref{app_lead_terms_lem} and \ref{app_dom_terms_lem} that
\[\tilde{Z}_{kn} = \sum_{i_1<\dots<i_D}\sum_{q\in Q_k(D,k)} e_{i_1\dots i_k;d_{q_1}\dots d_{q_k}}^{(k)}+o_P(r_{nk}^{-1/2})\]
so that in the following we can focus on the leading term of each component $\tilde{Z}_{kn}$.

We now define the filtration
\[\mathcal{G}_i:=\sigma\left(\left\{\alpha^{(0)},\alpha_{i_1}^{(1)},\dots,\alpha_{i_1\dots i_D}^{(D)}:i_1\leq i\right\}\right)\]
By definition of $e_{\mathbf{i}_1\dots\mathbf{i}_r;\mathbf{d}_1\dots\mathbf{d}_r}^{(k)}$,
\[\mathbb{E}[e_{i_1\dots i_k;d_1\dots d_k}^{(k)}|\mathcal{G}_i]=\left\{\begin{array}{lcl}
e_{i_1\dots i_k;d_1\dots d_k}^{(k)}&\hspace{0.3cm}&\textnormal{if }i_1\leq i\\0&&\textnormal{otherwise}\end{array}\right.\]

Hence, $\tilde{Z}_{kn}$ is a martingale adapted to $\mathcal{G}_I$ with the martingale differences
\begin{eqnarray}
\nonumber X_{kni}&=&\mathbb{E}[Z_{kn}|\mathcal{G}_i] - \mathbb{E}[Z_{kn}|\mathcal{G}_{i-1}]\\
\nonumber &=&\sum_{i=i_1<i_2<\dots<i_D}\sum_{q\in Q_k(D,k)} e_{i_1\dots i_k;d_{q_1}\dots d_{q_k}}^{(k)}+o_P(r_{nk}^{-1/2}/n)
\end{eqnarray}
Since each component of $e_{i_1\dots i_k;d_{q_1}\dots d_{q_k}}^{(k)}$ is mean-independent of its predecessors $e_{i_1\dots i_{k-s};d_{q_1}\dots d_{q_{k-s}}}^{(k-s)}$, $s=1,\dots, k-1$, the first three conditional moments of $X_n:=X_{n1i}+ \dots + X_{nDi}$ given $\mathcal{G}_{j}$ are equal to the sum of the respective conditional moments of $X_{n1i},\dots,X_{nDi}$.

By construction, $\mathbb{E}[X_{ni}|\mathcal{G}_{i-1}]=0$, and
\[\var(r_n^{1/2}X_{ni}|\mathcal{G}_{i-1}) =r_n\sum_{k=1}^D\binom{n-1}{k-1}\binom{D-1}{k-1}\sigma_{k,\mathbf{d}_1}^2+o_P\left(\frac{r_n}{nr_{n,k}}\right)\]
which is bounded by the definition of $r_n$. Summing over $i=1,\dots,n$ and rearranging terms, we get
\[\var(r_n^{1/2}\tilde{Z}_n)=r_n\sum_{k=1}^D\binom{n}{k}\binom{D}{k}\sigma_{k,\mathbf{d}_1}^2+o_P\left(1\right)\]

Furthermore, the third absolute moments of the joint distribution of $r_n^{1/2}(X_{1ni},\dots,X_{Dni})$ are bounded, so that the third moments of $r_n^{1/2}X_{ni}$ are also bounded. This implies a Lindeberg condition as stated in (2.4) in \cite{McL74}, which ensures that conditions (a) and (b) of Theorem (2.3)in \cite{McL74} are satisfied. Condition (c) follows from similar considerations for the fourth moments and Chebyshev's inequality. Hence, the conclusion of this Lemma follows from Theorem (2.3) in \cite{McL74}\qed



Given these intermediate results, we now conclude with the proofs of the main results for Section \ref{sec:struct_limit}.

\subsection{Proof of Theorem \ref{primitive_lln_thm}} The asymptotic representation of the moment in terms of the aggregate state variables follows from Lemmas \ref{fp_representation_lem} and \ref{app_ref_fp_lem}, and Theorem \ref{fp_existence_thm} ensures that the limiting model is well-defined. The existence a sequence of fixed points
\[\eta_n\in\Psi_0(\eta_n)\]
supporting a converging sequence of the moment follows immediately from Theorem 4.1 in \cite{Men16}. Furthermore Lemma B.3 in \cite{Men16} implies convergence of $\hat{\eta}_n\equiv \hat{\eta}_n(\alpha^{(0)})$ to the set of fixed points of $\Psi_0$ conditional on $\mathcal{F}$. In particular, given the conclusion of Lemma \ref{sampling_array_appx_lem}, a law of large numbers for the mapping $\hat{\Psi}_n$ follows from the Birkhoff LLN for exchangeable sequences together with the continuous mapping theorem and the Glivenko-Cantelli property for $\tilde{\psi}_i(\eta)$. The conclusion then follows from Lemmas \ref{sampling_array_appx_lem} and \ref{app_generic_lln_lem} with $q_n:=p_n$, noting that $\sigma_{k,n}^2/p_n$ is asymptotically bounded and that $m_0(\theta;\eta)$ is continuous in $\eta$ \qed

\subsection{Embedding into Asymptotic Sequence}

For a central limit theorem with mixing we embed the approximating array $\left(\tilde{Y}_{\mathbf{i}}\right)_{\mathbf{i}}$ into an asymptotic sequence along which we then take limits. Since the fixed points for $\hat{\eta}_n$ and $\hat{\tilde{\eta}}_n$ are not necessarily unique but may be selected at random, this requires that the approximating sequence is defined on a common probability space such that $\hat{\tilde{\eta}}_n$ converges to a well-defined limit $\eta_0^*$ which may be stochastic. Specifically we have the following:

\begin{lem}\label{sampling_array_coupling_lem}Suppose Assumptions \ref{surplus_bd_ass}-\ref{set_unobs_ass}, \ref{asy_rate_ass}, and \ref{fp_map_ass} hold. Then for any regular fixed point of the limiting mapping $\eta_0^*\in\Psi_0[\eta_0^*]$ we can construct a sequence of exchangeable arrays $\left(\tilde{Y}_{\mathbf{i}n}\right)_{\mathbf{i}}$, $n=1,2,\dots$ defined on a common probability space according to the sampling representation in Lemma \ref{app_sampling_rep_lem} such that $\hat{\tilde{\eta}}_n\in\overline{\tilde{\Psi}}_n[\hat{\tilde{\eta}}_n]$ and $\hat{\tilde{\eta}}_n\rightarrow\eta_0^*$ almost surely.
\end{lem}

\textsc{Proof:} From the law of large numbers in Theorem \ref{primitive_lln_thm} it follows that for the set $\mathcal{H}_0^*:=\left\{\eta_0^*:\eta_0^* = \Psi_0[\eta_0^*]\right\}$ we have $d(\hat{\tilde{\eta}}_n,\mathcal{H}_0^*)\stackrel{p}{\rightarrow}0$. Hence we can choose $\eta_0^*$ as the element of $\mathcal{H}_0^*$ that is closest to $\hat{\eta}_n$. Recall that Assumption \ref{fp_map_ass} (iv) states that the fixed point $\eta_0^*$ must be a regular point of the mapping $\Psi_0[\eta]$. Local existence of a fixed point $\hat{\tilde{\eta}}_{n'}\in\overline{\tilde{\Psi}}_n[\hat{\tilde{\eta}}_{n'}]$ for each $n'$ and $\hat{\tilde{\eta}}_{n'}\stackrel{a.s.}{\rightarrow}\eta_0^*$ then follows from Theorem 3.1(b) in \cite{Men12}\qed

Since we are coupling realizations of $\left(\tilde{Y}_{\mathbf{i}}\right)_{\mathbf{i}}$ rather than the initial array $\left(Y_{\mathbf{i}}\right)_{\mathbf{i}}$, this does not require existence of a pairwise stable network supporting values $\hat{\eta}_n\rightarrow\eta_0^*$ for any given realization of payoffs, which would be more difficult to establish. For the remainder of the argument, we take the approximating sequence to be defined on a common probability space according to the coupling in Lemma \ref{sampling_array_coupling_lem}.


Given the generic CLT in Lemma \ref{app_generic_clt_lem} in the appendix, we now turn to our main results which will be stated in terms of primitives of the network formation model. We first note that the network moments have the specific structure
\[m_{\mathbf{i}}(\mathbf{L},\mathbf{X};\theta) = \dum\left\{\mathbf{L}_{|\mathbf{i}}\in\mathcal{A}_{\mathbf{i}}\right\}h_{\mathcal{A}_i}(\mathbf{x}_{\mathbf{i}};\theta)\]

For simple (i.e. singleton) events $\mathcal{A}_{\mathbf{i}}$, the corresponding indicator variable is the product of edge-specific indicators $L_{i_qi_r},(1-L_{i_qi_r})$ for $q,r\in\{1,\dots,D\}$. If the two-dimensional approximating array $\left(\tilde{L}_{ij}\right)$ is jointly exchangeable, it follows that it can be represented as a function
\[\tilde{L}_{i_1i_2} = g(\xi^{(0)},\xi_{i_1}^{(1)},
\xi_{i_2}^{(1)},\xi_{i_1i_2}^{(2)})\]
Since attributes $x_{i_1}$ are i.i.d. by assumption, it follows that we can represent
\begin{equation}\label{app_2nd_order_rep_struc}m_{\mathbf{i}}(\tilde{\mathbf{L}},\mathbf{X};\theta) = \tilde{m}(\xi^{(0)},\xi_{i_1}^{(1)},\dots,\xi_{i_D}^{(1)},\xi_{i_1i_2}^{(2)},\dots,\xi_{i_{D-1}i_D}^{(2)},x_{i_1},\dots,x_{i_D};\theta)
\end{equation}
as a function of i.i.d. random variables at the node or edge level alone, which does not depend on any shocks at the level of $d$-ads of any order higher than $d=2$.

For composite events, corresponding to the case where $\mathcal{A}_{\mathbf{i}}$ is a set of size greater than one, the same argument holds, noting that the indicator for the composite event is the maximum over the indicators for each of the simple events contained in $\mathcal{A}_{\mathbf{i}}$. Lemma \ref{app_prim_rate_lem} in the appendix then determines convergence rates given the network formation model in Assumptions \ref{surplus_bd_ass}-\ref{set_unobs_ass}.

Given the projection representation of the network moment introduced before, we next establish the convergence rate given the network formation model:

\begin{lem}\label{app_prim_rate_lem} Assume the network formation model in Assumptions \ref{surplus_bd_ass}-\ref{set_unobs_ass}, and \ref{asy_rate_ass}. Furthermore suppose that Assumption \ref{moment_ass} holds and that the network event $\mathcal{A}_{i_1\dots i_D}$ implies that $L_{i_1i_s}=1$ for at least one $s\in\{2,\dots,D\}$. We then have that
$\sigma_1^2/p_n^2$ converges almost surely to a limit that is finite and bounded away from zero, and $\sigma_2^2/p_n^2 = O\left(1\right)$ so that $r_n = \left(\binom{n}{1}p_n^2\right)^{-1}$. Furthermore,
\[r_n\sigma_{k,\mathbf{d}_1}^2\rightarrow0\]
almost surely for all $k>2$.
\end{lem}

\textsc{Proof:} Note first that by Lemma \ref{sampling_array_coupling_lem}, the second moments converge almost surely to their $\mathcal{F}$-measurable limits. For the first claim, consider the conditional probability of $\mathcal{A}_i$ given $MC_i$,
\[\pi_{n1}(\mu):=\mathbb{P}\left(\left.\mathbf{L}_{|\mathbf{i}}\in\mathcal{A}_{\mathbf{i}}\frac{}{}\right|MC_{i_1}-\frac12\log n=\mu\right)\] where by the law of iterated expectations, $\mathbb{E}[\pi_{n1}(MC_{i_1})]=p_n$. Since
\[e_{i_1;1}^1:=(\pi_{n1}(MC_{i_1}) - p_n)\mathbb{E}[h(x_{i_1},\dots,x_{i_D})|x_{i_1}]\]
it that it will be sufficient to show that $\var\left(\frac{\pi_{n1}(MC_{i_1})-p_n}{p_n}\right)$ remains bounded away from zero.

In order to evaluate $\pi_{n1}(MC_{i_1})$, we make use of the conditional independence result in Lemma \ref{app_independence_lem}. As before, we define $D_{ij}:=\dum\left\{U_{ij}(\mathbf{L},\mathbf{X})\geq MC_i\right\}$ as an indicator whether $i$ would agree to form a link to $j$ given that network $\mathbf{L}$, and let $D_{ij}^*:=\dum\left\{U_{ij}(\mathbf{L}^*,\mathbf{X})\geq MC_i\right\}$ denote the corresponding indicator given the pairwise stable network $\mathbf{L}^*$.

Now for every network $\bar{\mathbf{L}}_{|\mathbf{i}}\in\mathcal{A}_{\mathbf{i}}$, we let $\mathcal{A}_{Di_1;i_2\dots i_D}(\bar{\mathbf{L}}_{|\mathbf{i}})$ denote the event that $D_{i_1i_s}=1$ for each $i_s\in\{i_2,\dots, i_D\}$ with $\bar{L}_{i_1i_s}=1$. Since it was assumed that $\mathcal{A}_{i_1\dots i_D}$ contains at least one non-zero link between $i_1$ and another node $i_s\in\{i_2,\dots, i_D\}$, there exists at least one such $i_s$. We also let $A_{Di_1;i_2\dots i_D}(\bar{\mathbf{L}}_{|\mathbf{i}})\in\mathcal{A}_{Di_1;i_2\dots i_D}(\bar{\mathbf{L}}_{|\mathbf{i}})$ denote the (singleton) event $D_{i_1i_s}=\bar{L}_{i_1i_s}$ for all $s=2,\dots,D$. Since under sparse asymptotics in Assumption \ref{asy_rate_ass}, $\mathbb{P}(D_{ij}=1)$ goes to zero, $\frac{\mathbb{P}(A_{Di_1;i_2\dots i_D}(\bar{\mathbf{L}}_{|\mathbf{i}}))}{\mathbb{P}(\mathcal{A}_{Di_1;i_2\dots i_D}(\bar{\mathbf{L}}_{|\mathbf{i}}))}\rightarrow1$ so that we can without loss of generality restrict our attention to the singleton event $A_{Di_1;i_2\dots i_D}(\bar{\mathbf{L}}_{|\mathbf{i}})$.

We then have by the law of iterated expectations that
\begin{eqnarray}
\nonumber \pi_{n1}(\mu)&:=&\mathbb{P}\left(\left.\mathbf{L}_{|\mathbf{i}}\in\mathcal{A}_{\mathbf{i}}\right|\mathcal{A}_{Di_1;i_2\dots i_D},MC_{i_1}-\frac12\log n=\mu\right)\mathbb{P}(\mathcal{A}_{Di_1;i_2\dots i_D}|MC_{i_1}-\frac12\log n=\mu)\\
\nonumber&=&\sum_{\bar{\mathbf{L}}_{|\mathbf{i}}\in\mathcal{A}_{\mathbf{i}}}
\mathbb{P}\left(\left.\mathbf{L}_{|\mathbf{i}}\right|A_{Di_1;i_2\dots i_D}(\bar{\mathbf{L}}_{|\mathbf{i}}),MC_{i_1}-\frac12\log n=\mu\right)\mathbb{P}(\mathcal{A}_{Di_1;i_2\dots i_D}(\bar{\mathbf{L}}_{|\mathbf{i}})|MC_{i_1}-\frac12\log n=\mu) + o(p_n)\\
\nonumber&=&\sum_{\bar{\mathbf{L}}_{|\mathbf{i}}\in\mathcal{A}_{\mathbf{i}}}
\mathbb{P}\left(\left.\mathbf{L}_{|\mathbf{i}}\right|A_{Di_1;i_2\dots i_D}(\bar{\mathbf{L}}_{|\mathbf{i}})\right)\mathbb{P}(\mathcal{A}_{Di_1;i_2\dots i_D}(\bar{\mathbf{L}}_{|\mathbf{i}})|MC_{i_1}-\frac12\log n=\mu) + o(p_n)
\end{eqnarray}
where the last step follows from Lemma \ref{app_independence_lem}.

Since $\eta_{i_1i_s}$ is continuously distributed with full support, $\mathbb{P}(\mathcal{A}_{Di_1;i_2\dots i_D}(\bar{\mathbf{L}}_{|\mathbf{i}})|MC_{i_1}-\frac12\log n=\mu)\mathbb{P}(\mathcal{A}_{Di_1;i_2\dots i_D}(\bar{\mathbf{L}}_{|\mathbf{i}}))$ is strictly increasing in $\mu$ for any $n$, and by Lemma \ref{app_ccp_cond_mu_lem}, its limit is also strictly decreasing in $\mu$. Since $MC_i$ is continuously distributed with full support on $\mathbb{R}$, it follows that $\var\left(\frac{\pi_{n1}(MC_{i_1})-p_n}{p_n}\right)$ converges to a strictly positive limit, establishing the first claim.

From analogous arguments, we can show that $\sigma_2^2 = O\left(p_n\right)$ and $\sigma_{k,\mathbf{d}_1\dots\mathbf{d}_r}^2=O\left(p_n\right)$ for every $k\geq2$ and $r=1,\dots,k$. By Lemma \ref{app_dom_terms_lem}, this establishes the rate $r_n = \left(\binom{n}{D}p_n\right)^{-1}$. Finally (\ref{app_2nd_order_rep_struc}) implies that jointly exchangeable approximation to the network moment can be represented in terms of Aldous-Hoover factors of degree less than 3, establishing the last claim of this Lemma\qed

We can now state the proofs for the remaining results in Section \ref{sec:struct_limit}.

\subsection{Proof of Lemma \ref{asy_bias_lem}} Without loss of generality, we assume that the approximating sequence of arrays $\left(\tilde{Y}_{\mathbf{i}n}\right)_{\mathbf{i}}$ is defined on a common probability space using the construction in Lemma \ref{sampling_array_coupling_lem}.

Using the notation in Lemma \ref{psn_potential_val_lem}, we let $\mathcal{J}_i$ denote the set of nodes available to $i$ given the selected potential values of network statistics, $\mathbf{s_{E_{i,2}}}$,
\[\mathcal{J}_i^*\equiv :=\mathcal{J}_i^*(\mathbf{D_{E_i}}):=\{j: D_{ji}^*=1\}\subset\{1,\dots,n\}\]
where $\mathbf{D_{E_i}}\in\mathcal{D}_{\mathbf{E_i}}^*(\mathbf{s_{E_{i,2}}})$. We also let
\[J^*:=|\mathcal{J}_i^*|\]
denote the size of the link opportunity set $\mathcal{J}_i^*$. By Assumption \ref{asy_rate_ass}, $J^*/\sqrt{n}=O_p(1)$ with finite variance.

The expectation of $\hat{m}_n$ and $\hat{\Psi}_n$ will be determined by conditional link acceptance probabilities of the form
\[\Phi(i,j_1,\dots,j_r|\mathbf{z}_i^*) = P\left(U_{ij_1},\dots,U_{ij_r}\geq MC_i >U_{ij'}\textnormal{ all other }j'\in W_i(\mathbf{L}^*)|\mathbf{z}_i^*\right)\]
For a given node set $\{j_1,\dots,j_r\}$ we also denote
\[\bar{\mathcal{J}}_i^*:=\mathcal{J}_i^*\backslash \{j_1,\dots,j_r\}\]

Given the conclusion of Lemma \ref{ev_conv_rate_lem}, conditional link acceptance probabilities $n^{r/2}\Phi(i,j_1,\dots,j_r|\mathbf{z}_i^*)$ can be approximated by a smooth function of $\hat{I}_{i,j_1,\dots,j_r;n}:=\frac1{J^*}\sum_{j\in\bar{\mathcal{J}}_i^*}\exp\{\tilde{U}_{ij}\}$ at a rate faster than $n^{-1/2}$. Specifically we denote
\[g(\hat{I}_{i,j_1,\dots,j_r;n}):=\frac{r!\prod_{s=1}^{r}\exp\{\tilde{U}_{ij_s}\}}
{\left(1+\hat{I}_{i,j_1,\dots,j_r;n}\right)^{r+1}}\]
We also define $\hat{I}_{i;n}:=\frac1{J^*}\sum_{j\in\mathcal{J}_i^*}\exp\{\tilde{U}_{ij}\}$.

First, if $\frac{r}{n^{1/2}}\rightarrow0$, boundedness of the systematic parts from Assumption \ref{surplus_bd_ass} implies that \[n^{1/2}|\hat{I}_{i,j_1,\dots,j_r;n}-\hat{I}_{i,n}|=
n^{1/2}\left|\frac1{J^*}\sum_{j\in\mathcal{J}_i^*}\exp\left\{\tilde{U}_{ij}\right\}-\frac1{J^*}\sum_{j\in\bar{\mathcal{J}}_i^*}
\exp\left\{\tilde{U}_{ij}\right\}\right|\]
as asymptotically bounded. 

The second approximation error arises from the randomness in $\hat{I}_{i,n}-\hat{\Eta}_n(x_i,s_i)$. To this end we establish that $n^{1/4}(\hat{I}_{i,n}-\hat{\Eta}_n(x_i,s_i))$ is asymptotically normal, so that in particular its asymptotic variance converges to a finite limit. Define $X_{kn}:=\left(\dum\{U_{ki}\geq MC_k\}-\pi_{kn}\right)\exp\{U_{ik}^*(L^*)\}$, where $\pi_{kn}:=\mathbb{P}(U_{ki}\geq MC_k|z_{1i}^*,\dots,z_{ni}^*)$. By Lemma \ref{app_independence_lem}, $\varepsilon_{ki}$ and $MC_k$ are drawn without replacement from $\varepsilon_{1i},\dots,\varepsilon_{ni}$ and $MC_1,\dots,MC_n$, independent of $z_{1i}^*,\dots,z_{ni}^*$.

Since $\varepsilon_{1i},\dots,\varepsilon_{ni}$ and $MC_1,\dots,MC_n$ are in turn i.i.d. draws from the respective marginal distribution independent of $z_{1i},z_{2i},\dots$, $X_{kn}$ are conditionally independent given $z_{1i}^*,z_{2i}^*,\dots$. We also define $s_n^2:=\sum_{k\neq i}\pi_{kn}(1-\pi_{kn})\exp\{2U_{ik}^*(L^*)\}$. We then consider convergence of $T_n:=\frac1{s_n}\sum_{k\neq i}X_{kn}$ to a standard normal random variable conditional on $z_{1i}^*,\dots,z_{ni}^*$. A sufficient condition for this is the Lyapounov condition \begin{equation}\label{app_lyapounov_cond}\frac1{s_n^3}\sum_{k\neq i}\mathbb{E}[|X_{kn}|^3]\rightarrow0.\end{equation}
Noting that by Assumption \ref{surplus_bd_ass}, $0<\exp\{-\bar{U}\}\leq \exp\{2U_{ik}^*(L^*)\}\leq \exp\{\bar{U}\}<\infty$ for all $k=1,2,\dots$. Furthermore, the (centered) Bernoulli random variable $\mathbb{E}\left|\dum\{U_{ki}\geq MC_k\}-\pi_{kn}\right|^3=\pi_{kn}(1-\pi_{kn})^3 + (1-\pi_{kn})\pi_{kn}^3\leq 2\pi_{kn}$. Hence, we have that
\[\frac1{s_n^3}\sum_{k\neq i}\mathbb{E}[|X_{kn}|^3]\leq 2\exp\{6\bar{U}\}\frac{\sum_{k\neq i}\pi_{kn}}{\left(\sum_{k\neq i}\frac14\pi_{kn}\right)^{3/2}}=\frac14\exp\{6\bar{U}\}\left(\sum_{k\neq i}\pi_{kn}\right)^{-1/2}\]
for $n$ sufficiently large since for each $k$, $\pi_{kn}$ is bounded by a sequence that converges to zero.

Since $\sqrt{n}p_n$ is bounded from above and away from zero, $\sum_{k\neq i}\pi_{kn}$ diverges to infinity so that the Lyapounov condition (\ref{app_lyapounov_cond}) holds. It then follows from Lyapounov's CLT for triangular arrays that conditional on $z_{1i}^*,z_{2i}^*,\dots$,
\[\frac{\sum_{k\neq i}\left(\dum\{U_{ki}\geq MC_k\}-\pi_{kn}\right)\exp\{U_{ik}^*(L^*)\}}{s_n}\stackrel{d}{\rightarrow}N(0,1)\]
In particular, \[n^{-1/4}\sum_{k\neq i}\left(\dum\{U_{ki}\geq MC_k\}-\pi_{kn}\right)\exp\{U_{ik}^*(L^*)\}\stackrel{d}{\rightarrow}N(0,V)\]
conditional on $z_{1i}^*,z_{2i}^*,\dots$.

We can now combine these findings with a mean-value expansion to conclude that
\begin{eqnarray}
\nonumber \mathbb{E}[\sqrt{n}g(\hat{I}_{i,j_1,\dots,j_r;n})-g(\hat{\Eta}_n(x_i,s_i))] &=& \sqrt{n}g'(\hat{\Eta}_n(x_i,s_i))\mathbb{E}[\hat{I}_{i,j_1,\dots,j_r;n})-\hat{\Eta}_n(x_i,s_i)]\\
\nonumber&&+\frac12g''(\hat{\Eta}_n(x_i,s_i))\mathbb{E}[(\hat{I}_{i,j_1,\dots,j_r;n}-\hat{\Eta}_n(x_i,s_i))^2]+o(1)\\
\nonumber&=&B_{g,n} + o(1)
\end{eqnarray}
conditional on $\hat{\Eta}_n(x_i,s_i))$, which is asymptotically bounded given the bounds on the bias and variance of $\hat{I}_{i,j_1,\dots,j_r;n}-\hat{\Eta}_n(x_i,s_i)$ established before. Finally, the expectations of $\hat{m}_n(\theta;\eta)$ and $\hat{\Psi}_n(\theta;\eta)$ are Lipschitz in the asymptotic conditional link acceptance probabilities, establishing the conclusion of this Lemma \qed

\subsection{Proof of Theorem \ref{asy_Gauss_thm}} Stacking the moments, we can consider linear functionals of the form
\[Z_n:=\mathbf{a}'\left(n^{1/2}\left[\begin{array}{c}\hat{m}_n(\theta;\eta) - m_0(\theta;\eta)\\\hat{\Psi}_n(\theta;\eta)-\Psi_0(\theta;\eta)  \end{array}\right] - B\right)\]
for an arbitrary conformable column vector $\mathbf{a}$, noting that $B$ is well-defined by Lemma \ref{asy_bias_lem}.

Without loss of generality, we assume that the approximating sequence of arrays $\left(\tilde{Y}_{\mathbf{i}n}\right)_{\mathbf{i}}$ is defined on a common probability space using the construction in Lemma \ref{sampling_array_coupling_lem}. Given the convergence rate for $\hat{m}_n$ established in Lemma \ref{app_prim_rate_lem}, we can apply Lemma \ref{app_generic_clt_lem} to establish convergence
\[(a'Va)^{-1/2}Z_n\stackrel{d}{\rightarrow}N(0,1)\]
for each $a$. By the Cram\'er-Wold device we then have that for any $\theta,\eta$,
\[\sqrt{n}V^{-1/2}\left(\begin{array}{c}\hat{m}_n(\theta;\eta) - m_0(\theta;\eta)\\ \hat{\Psi}_n(\theta;\eta)-\Psi_0(\theta;\eta)\end{array}\right) - B\stackrel{d}{\rightarrow} N(0,\mathbf{I})\]
mixing with respect to $\mathcal{F}$.

Since by Assumption \ref{fp_map_ass}, $\eta_0$ is a regular fixed point of $\Psi_0$ the conclusion follows from a mean-value expansion of $[\hat{m}_n(\theta;\hat{\eta}_n),\hat{\Psi}_n(\theta;\hat{\eta}_n)]'$ about $\eta_0$\qed

\end{document}